\documentclass[a4paper,11pt]{article}

\usepackage{jcappub} 

\usepackage{graphicx}
\usepackage{amsmath}
\usepackage{dcolumn}
\usepackage{bm}

\def\k{\kappa}

\def\beq{\begin{eqnarray}}
\def\eeq{\end{eqnarray}}

\def\dd{\mathrm{d}}

\def\Mpl{M_{\rm Pl}}
\def\GeV{ \ {\rm GeV}}

\def\TeV{\ {\rm TeV}}

\def\TR{T_{\rm RH}}

\newcommand{\lmk}{\left(}  
\newcommand{\rmk}{\right)}
\newcommand{\lkk}{\left[}  
\newcommand{\rkk}{\right]}

\newcommand{\del}{\partial}  
\newcommand{\la}{\left\langle} 
\newcommand{\ra}{\right\rangle}

\newcommand{\half}{\frac{1}{2}}

\newcommand{\abs}[1]{\left\vert {#1} \right\vert}
\def\Im{{\rm Im}}
\def\bea{\begin{array}}
\def\eea{\end{array}}
\def\Mpl{M_{\text{Pl}}}
\def\M{M_{\text{Pl}}}

\def\mphi{m_{\phi}}

\def\ij{_{ij}}
\def\k{\lmk {\bf k} \rmk}
\def\tk{\lmk \tau, {\bf k} \rmk}

\def\TT{T_{ij}^{\rm TT}}

\def\Max{\text{Max}}
\def\Kahler{K\"{a}hler }
\def\cphi{\varphi}

\title{
\boldmath
Gravitational~wave~signals from~short-lived~topological~defects in~the~MSSM
}

\author[a,b]{Ayuki Kamada}
\author[a,c]{and Masaki Yamada
}

\affiliation[a]{Kavli IPMU (WPI), UTIAS, The University of Tokyo, Kashiwa, Chiba 277-8583, Japan}
\affiliation[b]{Department of Physics and Astronomy, University of California, Riverside, CA, 92507, USA}
\affiliation[c]{ICRR, The University of Tokyo, Kashiwa, Chiba 277-8582, Japan}

\emailAdd{ayuki.kamada@ucr.edu}
\emailAdd{yamadam@icrr.u-tokyo.ac.jp}

\abstract{
Supersymmetric theories, including the minimal supersymmetric standard model, 
usually contain many scalar fields whose potentials are absent in the exact supersymmetric limit 
and within the renormalizable level. 
Since their potentials 
are vulnerable to the finite energy density of the Universe 
through supergravity effects, 
these flat directions have nontrivial dynamics in the early Universe. 
Recently, we have pointed out that 
a flat direction may have a positive Hubble induced mass term during inflation 
whereas a negative one after inflation. 
In this case, the flat direction stays at the origin of the potential during inflation 
and then obtain a large vacuum expectation value after inflation. 
After that, 
when the Hubble parameter decreases down to the mass of the flat direction, 
it starts to oscillate around the origin of the potential. 
In this paper, we investigate the dynamics of the flat direction 
with and without higher dimensional superpotentials 
and show that topological defects, such as cosmic strings and domain walls, 
form at the end of inflation and disappear at the beginning of oscillation of the flat direction. 
We numerically calculate their gravitational signals 
and find that 
the observation of gravitational signals would give us information of supersymmetric scale, 
the reheating temperature of the Universe, and higher dimensional operators. 
}

\date{\today}

\begin{document}

\begin{flushright}
IPMU 15-0064
\end{flushright}

\maketitle
\flushbottom

\section{\label{introduction}Introduction}

The discovery of the Higgs boson~\cite{Aad:2012tfa, Chatrchyan:2012ufa} makes it clear that 
the Standard Model has a hierarchy problem between the electroweak scale 
and the Planck scale. 
One of the most elegant solutions is 
a symmetry relating fermions and bosons, called a supersymmetry (SUSY), 
which results in cancellations of quadratic divergences in quantum collections of scalar field masses including Higgs mass. 
In addition, 
the gauge couplings are miraculously unified at the grand unification scale 
in the minimal SUSY Standard Model (MSSM). 
This may indicate that the Standard Model gauge interactions are unified as a single 
gauge interaction at that scale. 
In addition to these phenomenological successes in particle physics, 
SUSY theories have plentiful implications in the early Universe.

SUSY theories, including the MSSM, 
usually contain many scalar fields whose potentials are absent in the exact SUSY limit 
and within the renormalizable level 
(see e.g., Ref.~\cite{Gherghetta:1995dv}). 
Since their potentials 
are vulnerable to 
the finite energy density of the Universe 
through supergravity effects, 
these flat directions have nontrivial dynamics in the early Universe. 
One famous example is the Affleck-Dine mechanism, 
in which the baryon asymmetry is generated through a rotational motion of a flat direction in field-value complex plane~\cite{AD, DRT}.
In that mechanism, so-called a negative Hubble induced mass term plays an important role 
for the flat direction 
to have an initial vacuum expectation value (VEV), 
which is closely related to the amount of the 
resulting baryon asymmetry. 
Flat directions can also be considered as inflaton because their potentials are flat. 
However, inflation models in supergravity are usually affected by Hubble induced mass terms 
unless we introduce a shift symmetry~\cite{Kawasaki:2000yn} 
or consider a class of $D$-term inflation~\cite{Binetruy:1996xj, Halyo:1996pp}.

In the literature, it is usually assumed that the coefficient of Hubble induced mass terms 
is constant during and after inflation~\cite{DRT}. 
However, in Ref.~\cite{previous work}, 
we have pointed out a possibility that the coefficient of Hubble induced mass terms changes 
at the end of inflation. 
This is partly because 
in most of the SUSY inflationary models 
the energy density of the Universe is dominated by 
different sources between the era during inflation and the one after inflation. 
In particular, the coefficient of the Hubble induced mass term 
can be positive during inflation and negative after inflation. 
With such a Hubble induced mass term,
the flat direction stays at the origin of the potential during inflation 
and then obtains a large VEV after inflation.%
\footnote{
Note that the flat direction considered in this paper 
does not belong to the inflation sector. 
It is just one of the flat direction in the MSSM sector 
like the one assumed in the context of the Affleck-Dine baryogenesis. 
In addition, we are not restricted to generating the baryon asymmetry
but consider the dynamics of general flat directions in SUSY theories, including the MSSM. 
}

Since the flat direction obtains a large VEV after inflation 
due to the negative Hubble induced mass term, 
we need to consider higher dimensional operators to stabilize its VEV. 
In some cases, 
the higher dimensional operator has $U(1)$ or $Z_n$ symmetry. 
The VEV of the flat direction breaks the symmetry spontaneously and 
results in formations of topological defects, such as cosmic strings and domain walls, 
after the end of inflation. 
Interestingly, a typical width of these topological defects 
is always of the order of (but smaller than) the Hubble length 
because the curvature of the potential for the flat direction is of the order of the Hubble parameter. 
It follows that the energy density of the topological defects never dominate 
that of the Universe, as shown in this paper. 
When 
the Hubble parameter decreases 
down to the soft mass of the flat direction, 
which we expect $\mathcal{O}(1) \TeV$, 
the potential for the flat direction is dominated by its soft mass term 
and it starts to oscillate around the origin of the potential. 
Since the symmetry is restored at the origin of the potential, 
the topological defects disappear at that time. 
Therefore, the topological defects exist only between 
the end of inflation and the beginning of the oscillation, 
so that we can not directly observe them at the present epoch. 
However, 
we have a chance to observe redshifted gravitational waves (GWs) emitted by 
the topological defects.

Stochastic GW signals are generated from the dynamics of cosmic strings~\cite{
Vilenkin:1981bx, Caldwell:1991jj, Vachaspati:1984gt, Olmez:2010bi} 
and domain walls~\cite{Kawasaki:2011vv, Figueroa:2012kw, Hiramatsu:2013qaa}. 
In the present case, emission of GWs begins at the end of inflation and ceases when the Hubble parameter decreases down to 
the soft mass of the flat direction.
Since the emitted GW signals have a peak at the wavenumber corresponding to the 
Hubble length, 
resulting GW signals have a peak at the wavenumber corresponding to the soft mass 
of the flat direction. 
This implies that in principle we can obtain the information of the SUSY scale 
from the observation of GW signals~\cite{previous work}. 
In addition, 
since the energy density of the topological defects is roughly proportional to the square of 
the VEV of the flat direction, which is related to its higher dimensional operators, 
its GW signal also has information of higher dimensional operators. 
Finally, since the GW spectrum is sensitive to what dominates the energy density of the Universe, 
we could obtain information of the reheating temperature 
from the observation of GW signals~\cite{Seto:2003kc, Nakayama:2008ip}.

In this paper, we investigate the dynamics of the flat direction 
which obtains a positive Hubble induced mass term during inflation 
and obtains a negative one after inflation. 
We consider three cases for its higher dimensional operators to stabilize its VEV; 
the one coming from a nonrenormalizable superpotential, 
the one coming from a \Kahler potential with $U(1)$ symmetry, 
and the one coming from \Kahler potentials with $Z_n$ symmetry. 
We numerically follow its dynamics 
and confirm the above qualitative understanding, i.e., topological defects, such as cosmic strings and domain walls, 
form after the end of inflation 
and disappear when the Hubble parameter decreases down to the soft mass of the flat direction. 
From the simulations, we also evaluate its GW signals 
and discuss its detectability by future GW detectors.

In the next section, 
we briefly review flat directions in SUSY theories. 
In Sec.~\ref{potentials}, 
we explain the potentials for flat directions
originating from SUSY breaking 
and higher dimensional operators. 
We consider three important cases for higher dimensional operators 
and show our results of numerical simulations 
for each case 
in Secs.~\ref{W ne 0}, \ref{W=0}, and \ref{W=0 without U(1)}. 
Finally, Sec.~\ref{conclusions} is devoted to the conclusion.

\section{\label{flat directions}Flat directions in the MSSM}

A large number of scalar fields are introduced in SUSY theories 
because SUSY relates fermions and bosons. 
The potentials for scalar fields are severely restricted by SUSY, 
which 
especially results in existence of many flat directions. 
Flat directions are scalar fields whose potentials are absent 
within the renormalizable level
as long as SUSY is unbroken.

Here, we illustrates how the potentials for scalar fields are absent
in SUSY theories by taking a flat direction called the $u^c d^c d^c$ flat direction as an example. 
Let us focus on a scalar field constructed by right-handed squarks 
through the following orthogonal matrix: 
\beq
\lmk
\begin{array}{lll}
\phi \\
\, \cdot \\
\, \cdot
\end{array}\rmk
= \frac{1}{\sqrt{3}} \lmk
\begin{array}{lll}
 1 \ \ & 1 \ \ & 1 \\
  & \cdot & \\
  & \cdot & 
\end{array}\rmk 
\lmk
\begin{array}{lll}
(u^c)_1^R\\
(d^c)_1^G\\
(d^c)_2^B
\end{array}\rmk, \label{phi flat}
\eeq
where the lower and upper indices represent flavor and color, respectively. 
The dots represent other directions, which we are not interested in. 
Since the inverse matrix is given by the transposed matrix, 
we obtain 
\beq
\lmk
\begin{array}{lll}
(u^c)_1^R\\
(d^c)_1^G\\
(d^c)_2^B
\end{array}\rmk
= \frac{1}{\sqrt{3}} \lmk
\begin{array}{lll}
 1 &   &  \\
 1 \ \ & \cdot \ \ & \cdot \ \ \\
 1 &  & 
\end{array}\rmk 
\lmk
\begin{array}{lll}
\phi \\
\, \cdot \\
\, \cdot
\end{array}\rmk. 
\label{decomposition}
\eeq
In the MSSM, the superpotential is given by%
\footnote{
We assume a symmetry, such as R-parity, to forbid baryon and lepton number violating renormalizable terms. 
}
\beq
 W_{\rm MSSM} = y_u H_u Q u^c - y_d H_d Q d^c - y_e H_d L e^c + \mu H_u H_d, 
\eeq
within the renormalizable level. 
Here we have omitted flavor indices for notational simplicity. 
It is easy to see that 
the $F$-term potential for the field $\phi$ is absent 
such as 
\beq
 V_F = 
 \abs{\frac{\del W_{\rm MSSM}}{\del \phi}}^2
 = 
  \frac{1}{3} \lmk \abs{\frac{\del W_{\rm MSSM}}{\del (u^c)^R_1}}^2 
 + \abs{\frac{\del W_{\rm MSSM}}{\del (d^c)^G_1}}^2
 + \abs{\frac{\del W_{\rm MSSM}}{\del (d^c)^B_2}}^2 \rmk = 0. 
\eeq
The $D$-term potential is also absent 
like 
\beq
\abs{D^a_3}^2 &=&  g_3^2 \abs{ (u^{c})^{R*}_1 T^a (u^c)^R_1 + (d^c)^{G*}_1 T^a (d^c)^G_1 + (d^c)^{B*}_2 T^a (d^c)^B_2 }^2 \\
&=&  \frac{g_3^2}{9} |\phi|^4 \abs{ \text{Tr} \lmk T^a \rmk}^2 =0, \\
\abs{D_1}^2 &=&  \abs{ -\frac{2}{3} \abs{(u^c)^R_1}^2 +\frac{1}{3} \abs{(d^c)^G_1} + \frac{1}{3} \abs{(d^c)^B_2} }^2 =0, 
\eeq
where $\abs{D^a_3}^2$ and $\abs{D_1}^2$ 
are $D$-term potentials for $SU(3)$ and $U(1)_Y$, respectively. 
Therefore, the field $\phi$ has a flat potential 
and is called a flat direction. 
The above example consists of right handed squarks of $u^c, d^c, d^c$, 
and is called the $u^c d^c d^c$ flat direction.
It is known that every flat direction is characterized by gauge-invariant monomial in this manner.
Table~\ref{table} is the list of flat directions in the MSSM~\cite{Gherghetta:1995dv}. 
Note that there are many flat directions even in such a simple model.

\begin{table}
\caption{\label{table}
Flat directions in the MSSM and possible values of the power of higher dimensional superpotentials~\cite{Gherghetta:1995dv}. 
}
\begin{center}
\begin{tabular}{ll}
 flat directions & $n$ 
\\
 \hline \hline
$ L H_u $  & 4, 6, 8, \dots \\ \hline
$ H_uH_d $  & 4, 6, 8, \dots \\ \hline
$ u^c d^c d^c $  & 6, 9, 12, \dots \\ \hline
$ LLe^c $  & 6, 9, 12, \dots \\ \hline
$ Qd^c L $  & 6, 9, 12, \dots \\ \hline
$QQQL  $  & 4, 8, 12, \dots \\ \hline
$ Qu^cQd^c $ & 4, 8, 12, \dots  \\ \hline
$ Qu^cLe^c $ & 4, 8, 12, \dots  \\ \hline
$ u^cu^cd^ce^c $ & 4, 8, 12, \dots  \\ \hline
$d^cd^cd^cLL  $ & 5, 10, 15, \dots  \\ \hline
$ u^cu^cu^ce^ce^c $  & 5, 10, 15, \dots \\ \hline
$ Qu^cQu^ce^c $  & 5, 10, 15, \dots \\ \hline
$ QQQQu^c $  & 5, 10, 15, \dots \\ \hline
$ (QQQ)_4 LLLe^c $  & 7, 14, 21, \dots \\ \hline
$ u^cu^cd^c Qd^cQd^c $  & 7, 14, 21, \dots \\ \hline
\end{tabular}
\end{center}
\end{table}

Although flat directions have no potential 
within the exact SUSY limit 
and 
the renormalizable level, 
they obtain nonzero potentials 
through SUSY breaking and nonrenormalizable operators (i.e., underlying higher energy theory). 
These potentials induce non-trivial dynamics of flat directions. 
The next section is devoted to discussing this point.

\section{\label{potentials}Potentials for flat directions}

In this section, we discuss the induced potentials for flat directions
through SUSY breaking and nonrenormalizable operators. 
Flat directions have soft (SUSY breaking) masses of the order of sparticle masses, 
which are subject to collider experiments and should be larger than $\mathcal{O}(10^{2{\mathchar`-}3}) \GeV$\,\cite{atlassusy, cmssusy}. 
In addition, 
the finite energy density of the Universe contributes to potentials for flat directions. 
For instance, 
scalar fields obtain so-called Hubble induced terms through supergravity effects during inflation 
because inflation is driven by a finite energy density~\cite{DRT}. 
This is also the case during the inflaton oscillation dominated era 
as we explain in Sec.~\ref{Hubble induced terms}. 
Finally, when a flat direction has a large VEV, 
nonrenormalizable operators become important.%
\footnote{
Although thermal effects sometimes play an important role 
to investigate the dynamics of flat directions~\cite{DRT, Allahverdi:2000zd, Anisimov:2000wx, Asaka:2000nb}, 
it is irrelevant in this paper (see footnote~\ref{footnote2}). 
}
In this paper, we consider three important cases for nonrenormalizable operators, 
which we specify in Sec.~\ref{higher dimensional terms}.

\subsection{\label{soft terms}Soft terms}

Since we have not discovered SUSY particles yet, 
SUSY particles have to be much heavier than the SM particles. 
This can be achieved by introducing a SUSY breaking hidden sector, 
which induces soft mass terms to SUSY particles 
through some mediation mechanism. 
The potential for a flat direction can be written like
\beq
 V(\phi) = m_\phi^2 \abs{\phi}^2, 
 \label{soft mass}
\eeq
around the weak scale vacuum. 
Note that not only soft masses but $\mu$-term can also contribute to $\mphi$ if the flat direction consists of Higgs field.%
\footnote{
\label{flavor indices}
In this paper, hereafter, 
we neglect the flavor indices. 
This is valid  
when the Hubble induced terms and/or nonrenormalizable terms breaks flavor symmetry completely. 
The case with a flavor symmetry is considered in Appendix~\ref{flavor}. 
} 
Here we neglect $A$ terms
because they are subdominant and irrelevant in the following discussion.%
\footnote{
\label{footnote5}
There might be Hubble induced $A$ terms during inflation, which plays 
important role in the context of the Affleck-Dine baryogenesis. 
In the present case, 
the flat direction stays at the origin during inflation, 
so that such higher dimensional terms are irrelevant. 
In addition, 
Hubble induced $A$ terms are irrelevant after the end of inflation 
because the time-averaged $F$ term $\la F \ra$ vanishes 
(see e.g., Ref.~\cite{Kamada:2008sv}). 
Therefore, in the present scenario, the flat direction can not generate the baryon asymmetry. 
}
In this paper, we assume that 
weak scale SUSY is broken 
at latest earlier than the time at which the Hubble parameter $H(t)$ decreases down to $m_\phi$ 
so that 
the soft mass term of Eq.~(\ref{soft mass}) 
dominates the potential for the flat direction 
after $H(t) \simeq m_\phi$.

\subsection{\label{Hubble induced terms}Hubble induced terms}

The potential for a flat direction $\phi$ is modified by supergravity effects during and after inflation. 
The energy density of the Universe is so large that 
we have to consider its effects on the potential for the flat direction. 
In the supergravity, scalar potentials are written in terms of superpotential, $W$, 
and K\"{a}hler potential, $K$, which are functions of scalar fields $\psi^i$ specified below. 
The potential for scalar fields is given by 
\beq 
 V = e^{K/\M^{2}} \lkk D_i W K^{i \bar{j}} D_{\bar{j}} W^* - 3 \abs{W}^2 /\M^{2} \rkk, 
 \label{SUGRA potential}
\eeq
where $K^{i \bar{j}} \equiv (K_{i \bar{j}})^{-1} \equiv (\del^2 K / \del \psi^i \del \psi^{*j})^{-1}$ 
and $D_i W \equiv \del W/ \del \psi^i + W/\M^{2} \del K / \del \psi^i$. 
Hereafter, we assume $F$-term inflation models (see Appendix~\ref{inflation} for example), in which 
we can neglect the contribution from $D$-term potentials. 
The \Kahler potential also determines kinetic terms such as 
\beq
 \mathcal{L}_{\rm kinetic} = K_{i \bar{j}} \del_\mu \psi^i (\del^\mu \psi^{j})^*. 
 \label{kinetic term}
\eeq
One may assume a minimal \Kahler potential of $K = \abs{\psi^i}^2$, 
which results in the canonical kinetic term. 
However, since we do not have any knowledge 
about the physics above a cutoff scale, such as the Planck scale, 
there are expected to be cutoff suppressed terms in the \Kahler potential (see Eq.~(\ref{kahler})).

The relevant fields are a flat direction $\phi$ 
and two fields in the inflaton sector denoted as $I$ and $X$, 
so that $\psi^i$ are identified as $\phi$, $I$, and $X$ in Eqs.~(\ref{SUGRA potential}) and (\ref{kinetic term}). 
The superfield $I$ contains a field $\chi$ which 
starts to oscillate after inflation 
and whose energy dominates the energy density of the Universe, 
that is, $\chi$ is identified as inflaton in chaotic inflation models~\cite{Kawasaki:2000yn} 
and as water-fall fields in hybrid inflation models~\cite{Copeland:1994vg, Dvali:1994ms}. 
The minimal \Kahler potential for the field $I$ is given by 
$K_{\text{min}} (I) = (I + I^*)^2/2$ for the case of the chaotic inflation model proposed in Ref.~\cite{Kawasaki:2000yn} 
and 
$K_{\text{min}} (I) = \abs{I}^2$ for the case of hybrid inflation models (see Appendix~\ref{inflation}). 
The superfield $X$ represents the field whose $F$-term potential energy drives inflation. 
In other words, 
the $F$ term of $X$ satisfies the relation of 
$\abs{\del W_{\rm inf} / \del X}^2 = 3 H_{\rm inf}^2 \M^2$. 
In Appendix~\ref{inflation}, we briefly explain chaotic and hybrid inflation models 
and identify the fields $I$ and $X$ in those models.

To see how the potential for the flat direction is modified, 
let us consider the following non-minimal K\"{a}helr potential: 
\beq
 K = \abs{\phi}^2 + \abs{X}^2 + \frac{c_{X}}{M_*^2} \abs{X}^2 \abs{\phi}^2 
 + K_{\text{min}} (I) \lmk1 + \frac{c_I}{M_*^2} \abs{\phi}^2 \rmk, 
 \label{kahler}
\eeq
where 
$M_*$ 
is a cutoff scale, 
and $c_a$ ($a=X, I$) are constants of order one. 
We expect that 
$M_*$ is less than or equal to the Planck scale $\M$ ($\simeq 2.4 \times 10^{18} \GeV$). 
We set $M_* = \M$ in Sec.~\ref{W ne 0}, while we assume $M_*$ to be less than $\M$ in Secs.~\ref{W=0} 
and \ref{W=0 without U(1)}. 
Since the $F$ term of $X$ satisfies the relation of 
$\abs{\del W / \del X}^2 = 3 H_{\rm inf}^2 \M^2$, 
the flat direction obtains 
the following mass, called the Hubble induced mass, 
during inflation: 
\beq
 V_H (\phi) &=& c_{H_{\rm inf}} H_{\rm inf}^2 \abs{\phi}^2, \\
 c_{H_{\rm inf}} &=& 3 \left( 1 - c_X \frac{\M^2}{M_*^2} \right). 
 \label{c_H during inflation}
\eeq
Let us note that $c_{H_{\rm inf}}$ is of the order of $\M^2/M_*^2$ when $c_X$ is of order one.
While the coefficient $c_{H_{\rm inf}}$ is assumed to be negative 
in the context of the Affleck-Dine baryogenesis~\cite{AD, DRT}, 
we assume it to be positive in this paper. 
In this case, the flat direction stays at the origin, i.e., $\phi = 0$, 
during inflation.

After inflation ends and before reheating completes, 
the energy density of the Universe is 
dominated by that of the oscillation of a scalar field (denoted by $\chi$) 
and the Hubble parameter decreases with time as $H (t) \propto a^{-3/2} (t)$, 
where $a(t)$ is the scale factor. 
The field $\chi$ oscillates with the frequency of its mass $m_\chi$.
The oscillation amplitude is redshifted as 
\beq
\chi \propto \chi_i \lmk \frac{a(t_i)}{a} \rmk^{3/2}, 
\eeq
where $\chi_i$ is the oscillation amplitude just after the oscillation starts around $t=t_i$. 
Since the oscillation time scale of $\chi$ is much smaller than $H^{-1}$, 
that is, $H \ll m_\chi$, 
we can 
average $\dot{\chi}^2$ and $\chi^2$ over the oscillation time scale like 
\beq 
 \la \dot{\chi}^2 \ra \simeq \la m_\chi^2 \chi^2 \ra \simeq \la \abs{\del W / \del X}^2 \ra \simeq \frac{3}{2} H^2 \M^2. 
\label{chi}
\eeq
The flat direction obtains a Hubble induced mass
through higher dimensional kinetic terms as well as the $F$-term potential of Eq.~(\ref{SUGRA potential}) 
during this oscillation dominated era. 
From Eq.~(\ref{kinetic term}), 
we can see that 
the relevant term comes from 
\beq
 K_{I \bar{I}} \abs{\del_\mu I}^2 \supset 
 \frac{c_I}{M_*^2}  \dot{\chi}^2 \abs{\phi}^2, 
\eeq
which, together with the one coming from the $F$-term potentials, 
results in 
the Hubble induced mass term of $c_H H^2 (t) \abs{\phi}^2$ with 
\beq
 c_{H_{\rm osc}} = \frac{3}{2} \lmk 1-c_X \frac{\M^2}{M_*^2} \rmk -  \frac{3}{2} c_I \frac{\M^2}{M_*^2}, 
 \label{c_H}
\eeq
during the oscillation dominated era. 
Let us note that $c_{H_{\rm osc}}$ is of the order of $\M^2/M_*^2$ when $c_X$ and $c_I$ are of order one.

As we can see from Eqs.~(\ref{c_H during inflation}) and (\ref{c_H}), 
the coefficient of the Hubble induced mass term during inflation, 
$c_{H_{\rm inf}}$, is generally different from the one 
during the oscillation dominated era, $c_{H_{\rm osc}}$~\cite{previous work}. 
This is because 
the field $X$, whose $F$ term drives inflation, 
is generally different from 
the field $I$, 
whose oscillation energy dominates the energy density of the Universe 
after inflation. 
In Appendix~\ref{inflation}, 
by taking the models of chaotic inflation 
and hybrid inflation in supergravity as examples, 
we see that
the field $X$ 
is indeed different from the field $I$. 
While the nonrenormalizable terms are heavily dependent on 
the high energy physics beyond the cutoff scale $M_*$, 
we assume that among many flat directions in SUSY theories (at least) 
one flat direction has 
$c_{H_{\rm inf}} >0$ and 
$c_{H_{\rm osc}} < 0$. 
In this case, 
the flat direction 
stays at the origin ($\phi=0$) during inflation, and then, 
it starts to roll down to a large VEV due to the negative curvature after inflation.

Here we comment on how natural to consider the above flat direction. 
There is no guiding principle to determine the signs of Hubble induced mass terms 
unless we specify the underlying high energy model beyond cutoff scale. 
This is what the well-known Affleck-Dine mechanism is also based on, 
while it assumes a negative coefficient during inflation. 
As shown above, 
the coefficient during inflation and that after inflation do not have the common origin 
and hence can be different in general. 
We consider one out of four possible combinations of the signs of the two coefficients. 
It is natural that one or more flat directions have such a combination 
because there are a lot of flat directions in the MSSM (see Table~\ref{table}). 
This is why our work can find its application in not all but wide class of supersymmetric models.

\subsection{\label{higher dimensional terms}Higher dimensional terms}

Since the flat direction obtains a large VEV due to the negative Hubble induced mass term 
after inflation, 
we need to consider its higher dimensional terms to stabilize the VEV of the flat direction. 
Depending on symmetries that the flat direction possesses, 
three important cases are considered in this paper.

In the next section, 
we consider the case that the flat direction has a nonrenormalizable superpotential 
like 
\beq
 W = \lambda \frac{\phi^n}{n \M^{n-3}}, 
 \label{W}
\eeq
where $n$ ($\ge 4$) is an integer dependent on each flat direction and $\lambda$ is an positive integer. 
Let us stress that among possible higher dimensional terms, 
only the lowest dimensional one is relevant to the dynamics of the flat direction.
In the case of the $u^c d^c d^c$ flat direction, for instance, 
it may come from a term like (see Eq.~(\ref{decomposition})) 
\beq
 W = \lambda \frac{\lmk u^c d^c d^c \rmk^2}{6 \M^{3}} = \lambda \frac{\phi^6}{6 \M^{3}}. 
 \label{W for udd}
\eeq
In this case, the power of superpotential $n$ is $6$. 
Note that $n=3$ superpotential is forbidden due to R-parity. 
When we extend this symmetry to a discrete R-symmetry, 
$n$ can be $6, 9, 12, \dots$ for the $u^c d^c d^c$ flat direction. 
For each flat direction, 
possible values of $n$ are shown in Table~\ref{table}~\cite{Gherghetta:1995dv}. 
The $F$-term potential for the flat direction comes from the superpotential and 
is given by (see Eq.~(\ref{SUGRA potential})) 
\beq
 V_{\rm NR} \simeq \abs{\frac{\del W}{\del \phi} }^2
 = \lambda^2 \frac{\abs{\phi}^{2n-2}}{\M^{2n-6}}, 
 \label{potential1}
\eeq
where we neglect irrelevant higher dimensional terms. 
Note that 
the potential of Eq.~(\ref{potential1}) possesses a global $U(1)$ symmetry. 
This is related to the baryon charge (or the lepton charge or a combination between them), 
which plays an important role in the context of the Affleck-Dine baryogenesis. 
Let us emphasize that 
the symmetry is a global symmetry 
independent from the local symmetries of the flat direction. 
The symmetry can be explicitly described as (see Eq.~(\ref{decomposition})) 
\beq
 \phi \to e^{i \theta} \phi 
 \quad
 \Leftrightarrow 
 \quad
 \left\{
 \bea{ll} 
 (u^c)^R_1 \to e^{i \theta} (u^c)^R_1 \\
 (d^c)^G_1 \to e^{i \theta} (d^c)^G_1 \\
 (d^c)^B_2 \to e^{i \theta} (d^c)^B_2 
 \eea
 \right., 
\eeq
for the case of the $u^c d^c d^c$ flat direction. 
One can see that this transformation can not be described by the gauge symmetries of 
$SU(3) \times U(1)_Y$, 
so that the $U(1)$ symmetry is an additional global symmetry.

If $\lambda$ in Eq.~(\ref{W}) 
is sufficiently small or superpotential is absent 
due to some symmetry, 
the VEV of the flat direction can be as large as the cutoff scale $M_*$.
For example, if we introduce $B-L$ symmetry, it forbids superpotentials for 
the $u^c d^c d^c$ flat direction like Eq.~(\ref{W for udd}). 
If we assign R-charges as, for example, $R(Q) = R(L) = R(e^c) = R(u^c) = R(d^c) = 1$ and $R(H_u) = R(H_d) = 0$, 
it also forbids superpotentials for the $u^c d^c d^c$ flat direction.%
\footnote{
\label{R-symmetry breaking}
The R-symmetry has to be broken to obtain nonzero gaugino and higgsino masses and to set the vacuum energy (almost) zero. 
Its symmetry breaking order parameter is proportional to the SUSY breaking scale, 
so that R-symmetry breaking terms are suppressed by a factor of $m_\phi^2 / M_*^2$ . 
These terms may be comparable to the other terms, such as the soft mass term, 
around the time when the Hubble parameter decreases down to the soft mass scale. 
This implies that there is $\mathcal{O}(1)$ uncertainty in our result when we omit those R-symmetry breaking terms. 
On the other hand, it might be possible that R-symmetry breaking terms are suppressed by a factor of $m_\phi^2 / \Mpl^2$. 
In this case, one can easily find that they can be neglected even when the Hubble parameter decreases down to the soft mass scale. After that, higher dimensional terms are irrelevant because the potential for the flat direction is dominated by the soft mass term. 
Therefore, it is reasonable to consider that $R$-symmetry may suppress the superpotential for the flat direction. 
}
In such cases, 
the potential for the flat direction may be dominated by 
higher dimensional terms 
coming from 
nonrenormalizable K\"{a}hler potential~\cite{Fujii:2002kr}. 
The induced potential is proportional to $H^2$ like the Hubble induced mass term described above. 
In Sec.~\ref{W=0}, 
we consider a flat direction with a potential given by 
\beq
 V_{\rm NR} (\phi) = 
 a_{H} H^2(t) \frac{\abs{\phi}^{2m-2}}{\M^{2m-4}}, 
 \label{potential2}
\eeq
where 
$m$ ($\ge 3$) is a certain integer.
\footnote{
In general, two or more terms can give comparable contributions to the potential for flat direction.
While it can change our results by a factor of $\mathcal{O}(1)$, in this paper,
we consider a nonrenormalizable potential with a single term for simplicity.
}
The coefficient $a_{H}$ is given by 
\beq
 a_H = a_H' \lmk \frac{\M}{M_*} \rmk^{2m-2}, 
 \label{a_H0}
\eeq
where we assume $a_H'$ is positive and of order one. 
Note that the potential of Eq.~(\ref{potential2}) respects $U(1)$ symmetry.

On the other hand, if we assign R-charges as, for example, 
$R(Q) = R(L) = R(e^c) = R(u^c) = R(d^c) = 0$ and $R(H_u) = R(H_d) = 2$, 
nonrenormalizable \Kahler potentials may not respect $U(1)$ symmetry. 
In Sec.~\ref{W=0 without U(1)}, 
we consider the same potential with 
Eq.~(\ref{potential2}) 
except for an additional $U(1)$ breaking term 
coming from higher dimensional \Kahler potentials: 
\beq
 V_{\rm NR} (\phi) = 
 a_{H} H^2(t) \frac{\abs{\phi}^{2m-2}}{\M^{2m-4}} 
 -  b_H H^2(t) \lmk \frac{\phi^{m'}}{\M^{m'-2}} + {\rm c.c.} \rmk , 
 \label{potential3}
\eeq
where $m' = 2, 3, 4, \dots,$ is an integer. 
We assume $m' < 2m-2$ so that 
the potential for the flat direction is bounded from below. 
The coefficient $b_H$ is given by 
\beq
 b_H = b'_H \lmk \frac{\Mpl}{M_*} \rmk^{m'}, 
 \label{b_H0}
\eeq
where we assume $b'_H = \mathcal{O}(1)$. 
Here we have redefined the field $\phi$ such that $b'_H$ is postitive.
This additional term explicitly breaks the $U(1)$ symmetry, 
but there remains $Z_{m'}$ symmetry in that potential.

\subsection{\label{summary in Sec3}Summary of this section
}

Here we summarize the potential for the flat direction. 
It is given by 
\beq
V (\phi) = m_\phi^2 \abs{\phi}^2 + c_H H^2 (t) \abs{\phi}^2 + V_{\rm NR}, 
\eeq
where 
the higher dimensional terms $V_{\rm NR}$ is given by 
Eqs.~(\ref{potential1}), (\ref{potential2}), or (\ref{potential3}). 
Hereafter, $c_{H_{\rm inf}}$ and $c_{H_{\rm osc}}$ are collectively denoted as $c_H$. 
We consider a flat direction which has a positive Hubble induced mass term ($c_H = c_{H_{\rm inf}} >0$) during inflation 
and a negative one ($c_H = c_{H_{\rm osc}} < 0$) after inflation.

In the following sections, 
we investigate the dynamics of the flat direction 
and show that topological defects, such as cosmic strings and domain walls, 
form after the end of inflation. 
We numerically follow the dynamics of the flat direction 
and calculate its GW signal. 
We also discuss the detectability of GW signals 
and explain that 
the information of the mass of the flat direction, the reheating temperature of the Universe, and 
higher dimensional potentials are imprinted on GW signals. 
In Secs.~\ref{W ne 0}, \ref{W=0}, and \ref{W=0 without U(1)}, 
we consider the case that 
the flat direction has a higher dimensional potential of 
Eqs.~(\ref{potential1}), (\ref{potential2}), and (\ref{potential3}), respectively.

\section{\label{W ne 0}Case A: $W \ne 0$}

In this section, 
we consider the case that the flat direction has a nonrenormalizable superpotential 
of Eq.~(\ref{W}). 
The $F$-term potential for the flat direction 
is given by 
\beq
 V_{\rm NR} = \abs{\frac{\del W}{\del \phi} }^2
 = \lambda^2 \frac{\abs{\phi}^{2n-2}}{\M^{2n-6}}, 
 \label{potential for W ne 0}
\eeq
where $n \ge 4$. 

In the next subsection, we briefly explain the dynamics of the flat direction 
in the early Universe. 
We numerically follow its dynamics 
and calculate its GW signal, 
and the results are shown in Sec.~\ref{results1}. 
Then we give physical interpretation of the results in Sec.~\ref{interpretation1}. 
Finally, we discuss the detectability of GW signals in Sec.~\ref{prediction1}.

\subsection{\label{dynamics1}Dynamics of the flat direction}

The equation of motion for the flat direction is given by
\beq
 \ddot{\phi} + 3 H \dot{\phi} - \frac{1}{a^2} \nabla^2 \phi + \frac{\del V}{\del \phi^*} = 0,
 \label{EOM}
\eeq
with dots denoting derivatives with respect to $t$.
The potential for the flat direction $V (\phi)$ is written as 
\beq
V (\phi) = m_\phi^2 \abs{\phi}^2 + c_H H^2 (t) \abs{\phi}^2 + a_{H} H^2(t) \frac{\abs{\phi}^{2m-2}}{\M^{2m-4}}. 
\eeq
Note that we collectively denote $c_{H_{\rm inf}}$ and $c_{H_{\rm osc}}$ as $c_H$. 
As explained in the previous section, 
we consider a flat direction which has a positive Hubble induced mass term ($c_{H_{\rm inf}} >0$) during inflation 
and a negative one ($c_{H_{\rm osc}} < 0$) after inflation. 
During inflation, the flat direction stays at the origin, $\phi = 0$, due to the positive Hubble induced mass term. 
Since the coefficient $c_H$ becomes negative after the end of inflation, 
the flat direction rolls down to the stable point 
and becomes to obtain a large VEV.%
\footnote{
One might wonder if the flat direction is affected by thermal effects 
and is forced to stay at the origin by the effective thermal mass for the flat direction~\cite{DRT, 
Allahverdi:2000zd, Anisimov:2000wx, Asaka:2000nb}. 
However, we can neglect thermal effects on the flat direction by the following discussion. 
During the inflaton oscillation dominated era, 
the temperature of the plasma is given by $T \sim (H(t) \Gamma_I \Mpl^2)^{1/4}$, 
where $\Gamma_I$ is the decay rate of the inflaton. 
The flat direction obtains a thermal mass of the order of the temperature. 
The ratio of the thermal mass to the Hubble induced mass 
is thus roughly given by $T / H(t) \sim (\Gamma_I \Mpl^2 / H(t)^3)^{1/4}$. 
This is smaller than unity just after inflation 
if we consider a Planck suppressed decay of inflaton. 
Thus, the flat direction obtains a large VEV soon after the end of inflation 
due to the negative Hubble induced mass term. 
Once the flat direction obtains a large VEV, 
thermal effects become suppressed 
and can be neglected 
because particles in thermal bass obtain large effective masses~\cite{Asaka:2000nb, Anisimov:2000wx}. 
Therefore, we can neglect thermal effects on the flat direction. 
\label{footnote2}
}
The VEV is given by 
\beq
 \la \abs{\phi} \ra = 
 \lmk \frac{\abs{c_{H_{\rm osc}}}}{\lambda^2 (n-1)} \frac{H^2 (t)}{\M^2} \rmk^{1/(2n-4)} \M, 
 \label{VEV1}
\eeq
as long as $\abs{c_{H_{\rm osc}}} H^2(t) \gg \mphi^2$.

The potential of Eq.~(\ref{potential for W ne 0}) 
has a global $U(1)$ symmetry, which corresponds to the 
baryon charge (or the lepton charge 
or a combination of them) in the context of the Affleck-Dine baryogenesis. 
After the end of inflation, the non-zero VEV of the flat direction breaks the global $U(1)$ symmetry.%
\footnote{
It also breaks SM gauge symmetries because flat directions have 
nonzero SM gauge charges. 
In the unitary gauge, only gauge fields may obtain a nontrivial configuration. 
Since the broken gauge field obtains the effective mass of $g \la \phi \ra$, 
the gauge field configuration begins to oscillate and 
becomes trivial ($A^\mu = 0$) soon after the phase transition. 
Therefore, we focus only on the global symmetry of the flat direction. 
See also the discussion in Appendix~\ref{flavor symmetry}. 
}
Since the flat direction stays at the origin during inflation, 
cosmic string network 
forms 
after this phase transition. 
The energy of cosmic strings per unit length $\mu$ is roughly given by
\beq
 \mu \sim \la \abs{\phi} \ra^2, 
 \label{mu}
\eeq
depending on dimensionless parameters only logarithmically~\cite{Hindmarsh:1994re}. 
By using numerical simulations (see the next subsection), 
we confirm that 
cosmic string network reaches a scaling regime well before the string disappears. 
In this regime, 
the number of cosmic strings in the Hubble volume is $\mathcal{O}(1)$ 
and the energy density of cosmic strings $\rho_{\rm cs}$ is given by 
\beq
 \rho_{\rm cs} \sim \mu \times H^{-1} \times H^{3} = \mu H^2. 
 \label{rho_cs}
\eeq
A typical width of cosmic strings is roughly given by the curvature of the potential, 
which is of the order of the Hubble length in the present case.%
\footnote{
Since a typical width of cosmic strings is of the order of the Hubble radius, 
the Nambu-Goto approximation, which was used in Ref.~\cite{Albrecht:1984xv} for example, 
is inappropriate to describe these cosmic strings. 
}
This means that the width of cosmic strings increases with time. 
This is one of the outstanding characteristics of cosmic strings in this scenario 
compared with other cosmic strings considered in the literature.

Since the Hubble induced mass squared 
decreases with time as $\propto a^{-3} (t)$ after the inflation ends, 
the soft mass term $\mphi^2 \abs{\phi}^2$ eventually dominates 
the potential for the flat direction 
and the flat direction starts to oscillate around $\phi = 0$. 
This implies that 
cosmic strings disappear at the time of $\abs{c_{H_{\rm osc}}} H^2 (t) \simeq m_\phi^2$, 
which can be rewritten as 
\beq
 t \simeq t_{\rm decay} \equiv \frac{2 \sqrt{\abs{c_{H_{\rm osc}}}}}{3 m_\phi}. 
\label{t_decay}
\eeq
With $\mphi = \mathcal{O}(1)$ TeV, 
cosmic strings disappear much earlier than the bing bang nucleosynthesis (BBN) epoch, 
so that observational constraints on long-lived topological defects are not applicable to our scenario. 
The flat direction decays into radiation 
through gauge interactions 
soon after cosmic strings disappear.

We define the reheating temperature $T_{\rm RH}$ as the temperature when
the energy density of the Universe is dominated by that of radiation. 
The Hubble parameter at that time is given by 
\beq
 H_{\rm RH} = \sqrt{\frac{g_{*} \pi^2 T_{\rm RH}^4}{90 \Mpl^2}},
 \label{T_RH}
\eeq
where $g_*$ is the effective relativistic degrees of freedom for the energy density.
Note that 
in low-scale SUSY models 
we should require 
$\TR \lesssim 10^9 \GeV$ 
to avoid the gravitino problem~\cite{Kawasaki:2004qu, Kawasaki:2008qe}. 
Even for high-scale SUSY models such as pure gravity mediation, 
$\TR \lesssim 10^{10} \GeV$ is required 
to avoid an overproduction of LSP ($= \mathcal{O}(100) \GeV$)~\cite{Ibe:2004tg, Ibe:2004gh, Ibe:2011aa}. 
In such well-motivated cases, 
the cosmic strings disappear $(H_{\rm decay} \equiv (2/3) t_{\rm decay}^{-1} = \mathcal{O}(10^{2{\mathchar`-}3}) \GeV)$ before reheating completes $(H_{\rm RH} \lesssim\mathcal{O}(1) \GeV)$. 
Hereafter we consider such a case.

After they form at the end of inflation 
and before they disappear at $t \simeq t_{\rm decay}$, 
cosmic strings emit GWs. 
We numerically follow the evolution of the flat direction 
and calculated its GW signal. 
The results are shown in the next subsection.

\subsection{\label{results1}Results of numerical simulations}

In this subsection, we show results of our numerical simulations for 
the dynamics of the flat direction and its GW signal. 
Since we consider the evolution of cosmic strings during the oscillation dominated era ($H \ll m_\chi$), 
the Hubble parameter decreases with time as $H \propto a^{-3/2}$. 

By redefining the variables, 
we rewrite the equation of motion such as 
\beq
 \cphi'' - \frac{\del^2 \cphi}{\del \tilde{ {\bm x}}^2} + \lmk c_{H_{\rm osc}} - \half \rmk \lmk \frac{\widetilde{\tau}_i}{\widetilde{\tau}} \rmk^2 \cphi 
 + \lmk \frac{\widetilde{\tau}}{\widetilde{\tau}_i} \rmk^4 \widetilde{m}_\phi^2 \cphi + (n-1) \lmk \frac{\widetilde{\tau}_i}{\widetilde{\tau}} \rmk^{4n-12} 
 \abs{\cphi}^{2n-4} \cphi=0, 
 \label{EOM2}
\eeq
where 
\beq 
 \cphi &\equiv& (\lambda^{-1} \M^{n-3} H_i)^{-1/(n-2)} a(t) \phi \nonumber \\
 &=& \lmk \abs{c_{H_{\rm osc}}}/(n-1) \rmk^{1/(2n-4)} a(t) \phi /  \la \abs{\phi_i} \ra, \\
 \dd \widetilde{\tau} &\equiv& H_i \dd t / a(t), \\
 \widetilde{m}_\phi &\equiv& m_\phi / H_i \quad (\ll 1), \\
 \widetilde{\bm x} &\equiv& H_i {\bm x}. 
\eeq
Let us emphasize that 
the parameter $\lambda$ can be absorbed by the redefinition of $\phi$, 
so that our numerical simulation can be applied to any cases with an arbitrary value of $\lambda$. 
The variable $\widetilde{\tau}$ is a dimensionless conformal time and is related with comoving horizon size $\tau$ through $\widetilde{\tau} = \tau a_i H_i$.
Primes denote derivatives with respect to $\widetilde{\tau}$.
We rewrite important parameters by the conformal time 
and obtain
\beq
 \widetilde{\tau}_i &=& 2, \\
 H (\widetilde{\tau}) &=& \lmk \frac{\widetilde{\tau}_i}{\widetilde{\tau}} \rmk^3 H_i, \\
 \widetilde{\tau}_{\rm decay} &=& \lmk \frac{\abs{c_{H_{\rm osc}}} H_i^2}{\mphi^2} \rmk^{1/6} \widetilde{\tau}_i, \label{tau_decay}\\
 \widetilde{\tau}_{\rm RH} &=& \lmk \frac{90 \Mpl^2 H_i^2}{g_* \pi^2 T_{\rm RH}^4} \rmk^{1/6} \widetilde{\tau}_i, 
\eeq
where we implicitly assume $a_i \equiv a(\widetilde{\tau}_i) = 1$. 
The subscript $i$ represents the values at the time of $\widetilde{\tau}_i$ 
at which we set initial conditions for numerical simulations.

We have performed three-dimensional lattice simulations 
to solve the classical equation of motion Eq.~(\ref{EOM2}) numerically. 
We use a numerical method similar to the one used in Ref.~\cite{Hiramatsu:2013qaa}. 
They use the 4-th order symplectic integration method 
to evolve the equation of motion for the flat direction 
in the expanding background, while we have used the 6-th order symplectic integration method. 
We ignore the backreaction of metric perturbations on the field evolution, 
which is negligible for $\rho_{\rm cs} \ll H^2(t) \Mpl^2$, that is, for $\la \phi \ra \ll \M$ 
(see Eqs.~(\ref{mu}) and (\ref{rho_cs})). 
In the set of simulations, 
we take a unit of $H_i \equiv H(\widetilde{\tau}_i) = 1$ and set $m_\phi/H_i = 5 \times 10^{-4}$ and $c_{H_{\rm osc}}=-15$. 
The final time $\widetilde{\tau}_{f}$ and the time step $\Delta \widetilde{\tau}$ is set to be 
$50$ ($100$) and $0.2$ ($0.1$), 
for $N=128^3$ ($256^3$) grid points, respectively. 
The comoving box size is $L = 100$ ($200$) 
with $N=128^3$ ($256^3$) grid points, 
which implies that the comoving grid size is $\Delta x = 100/128 (= 200/256) \simeq 0.75$. 
Since a typical width of cosmic strings is of the order of the Hubble radius 
$H^{-1} (\widetilde{\tau}) = (\widetilde{\tau}/\widetilde{\tau}_i)^3$, 
it is correctly resolved in our numerical simulation. 
We use the periodicity boundary condition for spatial directions. 
We require that the horizon scale is smaller than the simulation size (i.e., $L/2 \geq \widetilde{\tau}$) 
to avoid the boundary effect on the cosmic string network. 
The initial condition for simulations 
is seeded by quantum fluctuations around the origin of the configuration space.
To be concrete, we set initial conditions in the following way.
First, we draw two samples $f_{{\widetilde k},1}, f_{{\widetilde k},2}$ from a normal distribution with variance of
\beq
\la \abs{f_{\widetilde k}}^2 \ra = 
\frac{\lmk \abs{c_{H_{\rm osc}}}/(n-1) \rmk^{1/(n-2)} N^6}{2 k L^3  \la \abs{\phi_i} \ra^2},
\eeq
for each discretized wavenumber vector ${\widetilde {\bf k}} = \lmk 2\pi / L \rmk {\bf n}$,
where ${\bf n}$ is a triplet of integers from $-N/2$ to $N/2-1$.
In the above expression, we set $\lambda$ such that $\la \abs{\phi_i} \ra = 10^{2}$ in units of $H_i$.
Let us stress that once the cosmic string network follows the scaling evolution, the initial condition (i.e., $\lambda$) becomes irrelevant.
We generate initial conditions $\cphi_i, \cphi'_i$ by taking a superposition of (discrete) Fourier components given by 
\beq
\cphi_i({\widetilde {\bf k}}) &=& \frac{1}{\sqrt{2}} \lmk f_{{\widetilde k},1} + f_{{\widetilde k},2} \rmk, \\
\cphi'_i({\widetilde {\bf k}}) &=& \frac{1}{\sqrt{2}} i {\widetilde k} \lmk f_{{\widetilde k},1} - f_{{\widetilde k},2} \rmk.
\eeq

\begin{figure}[t]
\centering 
\begin{tabular}{l l}
\includegraphics[width=.40\textwidth, bb=0 0 869 943]{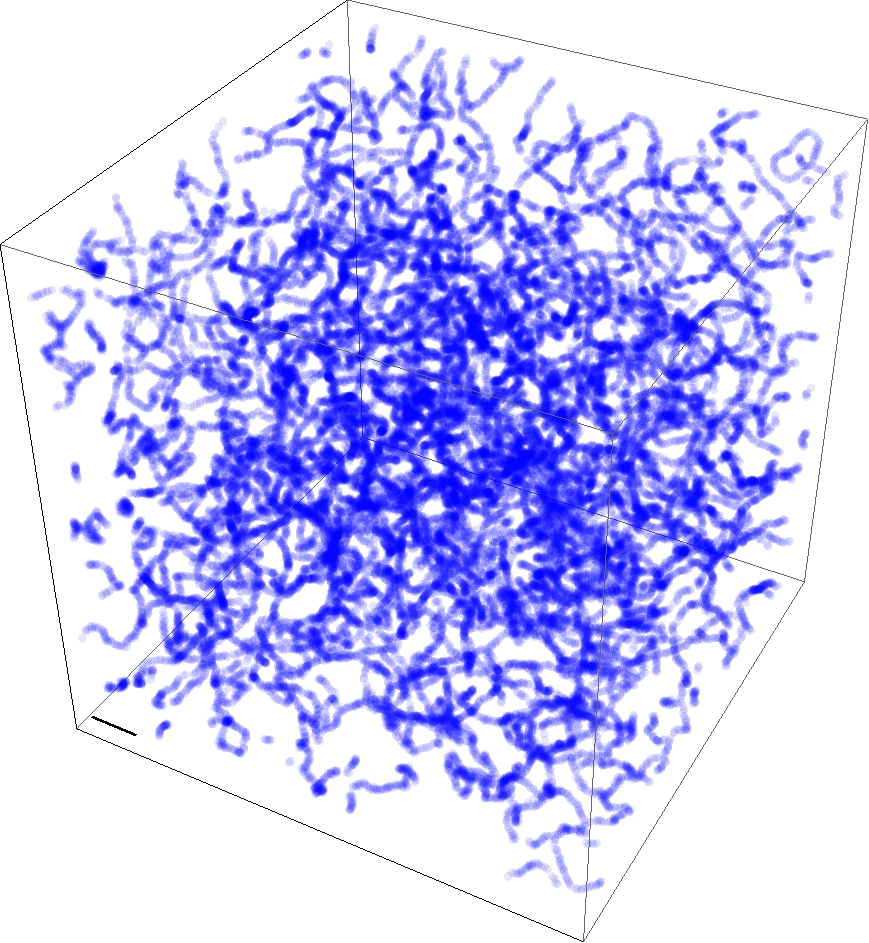} & \quad 
\includegraphics[width=.40\textwidth, bb=0 0 872 946]{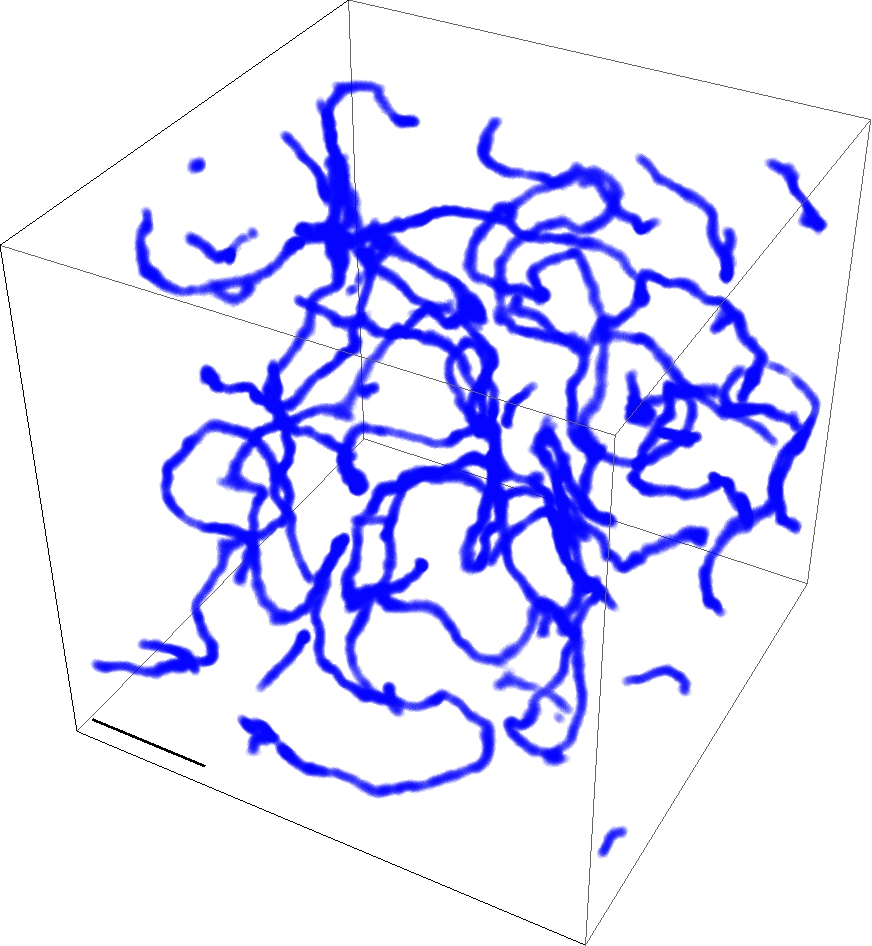} \vspace{0.5cm}\\
\includegraphics[width=.40\textwidth, bb=0 0 868 942]{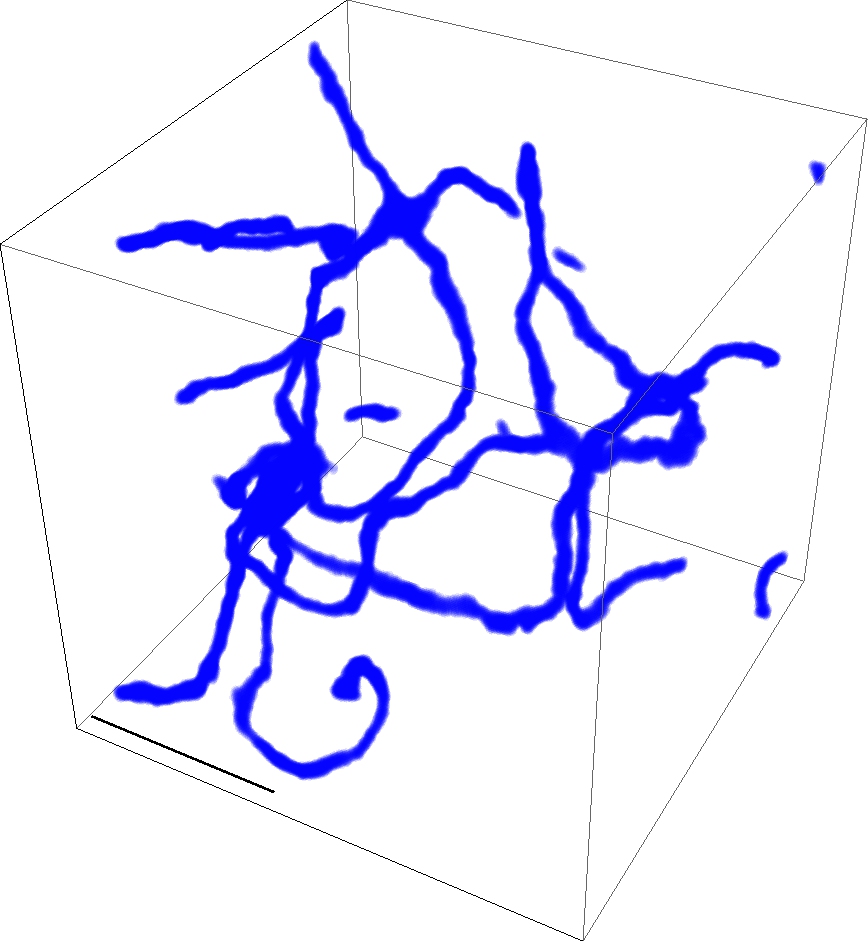} & \quad 
\includegraphics[width=.40\textwidth, bb=0 0 872 947]{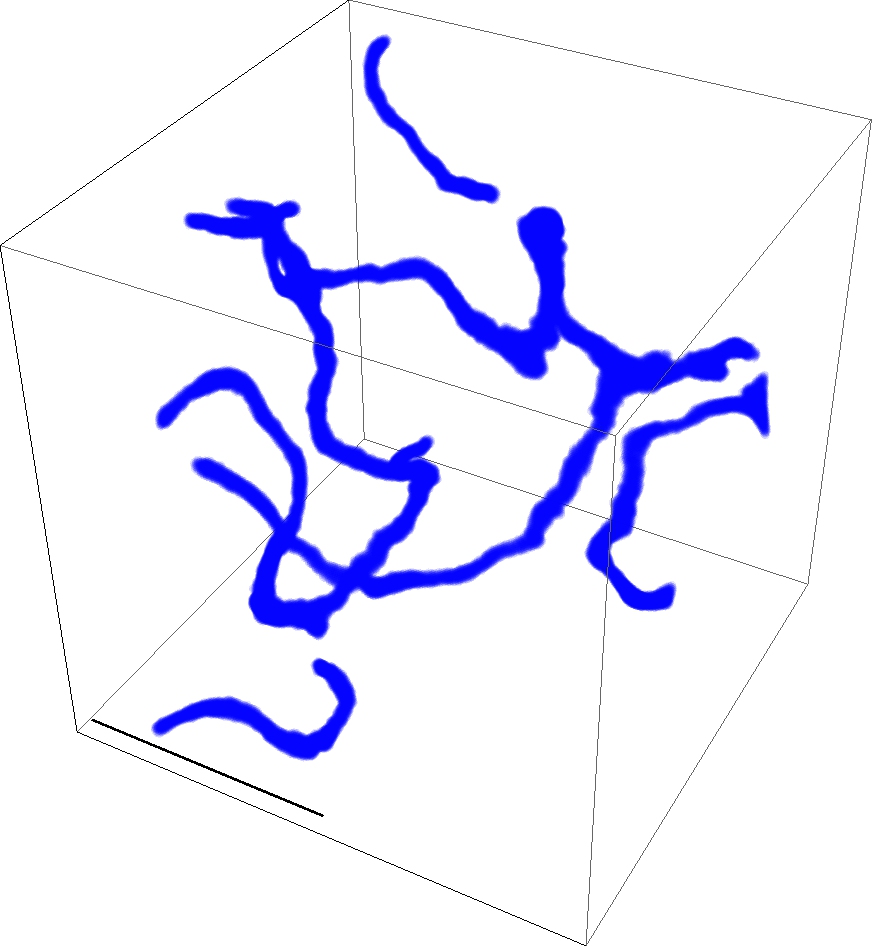} 
\end{tabular}
\caption{
Cosmic string networks obtained by our numerical simulation 
at $\tau a_i H_i = 20$ (upper left panel), $50$ (upper right panel), 
$80$ (lower left panel), $100$ (lower right panel). 
We assume $c_{H_{\rm osc}} = -15$ and $n=6$. 
In the blue regions, the magnitude of the flat direction satisfies $\abs{\phi ({\bf x})} < \la \abs{\phi} \ra / 5$, 
where $\la \abs{\phi} \ra$ is the VEV given by Eq.~(\ref{VEV1}). 
The black line at the bottom-left corner 
represents the length of the comoving horizon size $\tau$ at each time. 
}
  \label{CS network}
\end{figure}

Figure~\ref{CS network} shows one example of cosmic string network evolution 
obtained by our numerical simulation. 
The blue region represents the inside of cosmic strings, 
which we define by $\abs{\phi ({\bf x})} < \la \abs{\phi} \ra / 5$. 
One can find that a typical width of cosmic strings 
increases with time 
and is roughly proportional to the Hubble length. 
The black line at the bottom-left corner in Fig.~\ref{CS network} 
describes a unit of the comoving horizon size (given by $\tau$) at each time. 
One can see that each Hubble volume contains 
$\mathcal{O}(1)$ cosmic strings, 
which means that cosmic string network is in a scaling regime. 
From our numerical simulation, we find that 
the spatially averaged magnitude of $\phi$ grows with time and 
eventually reaches 
of the order of the VEV (Eq.~(\ref{VEV1}))
around $\widetilde{\tau}_{\rm form} \simeq 10$ for the case of $c_{H_{\rm osc}} = -15$. 

\begin{figure}[t]
\centering 
\begin{tabular}{l l}
\includegraphics[width=.40\textwidth, bb=0 0 360 353]{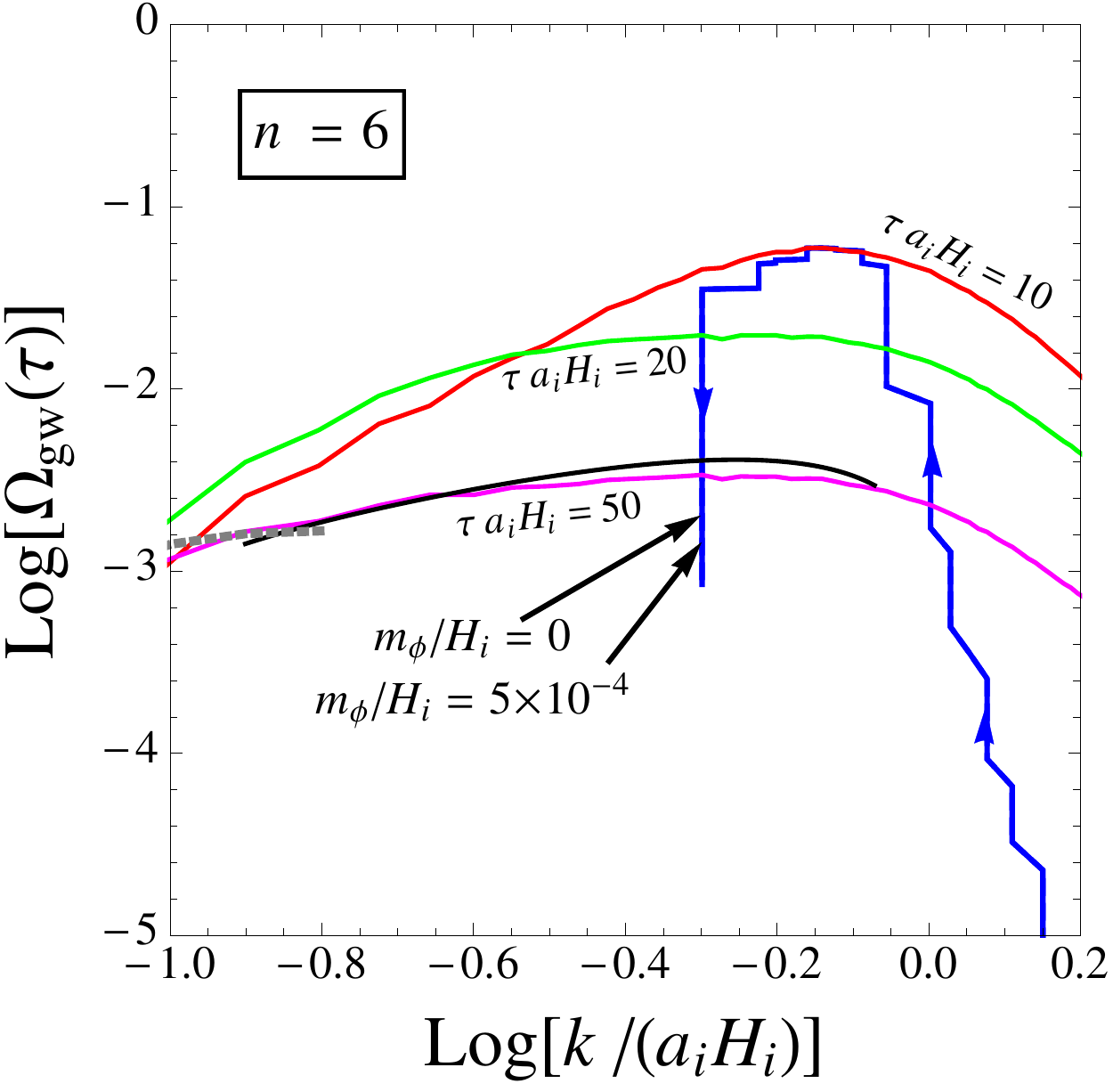} & \quad 
\includegraphics[width=.40\textwidth, bb=0 0 360 353]{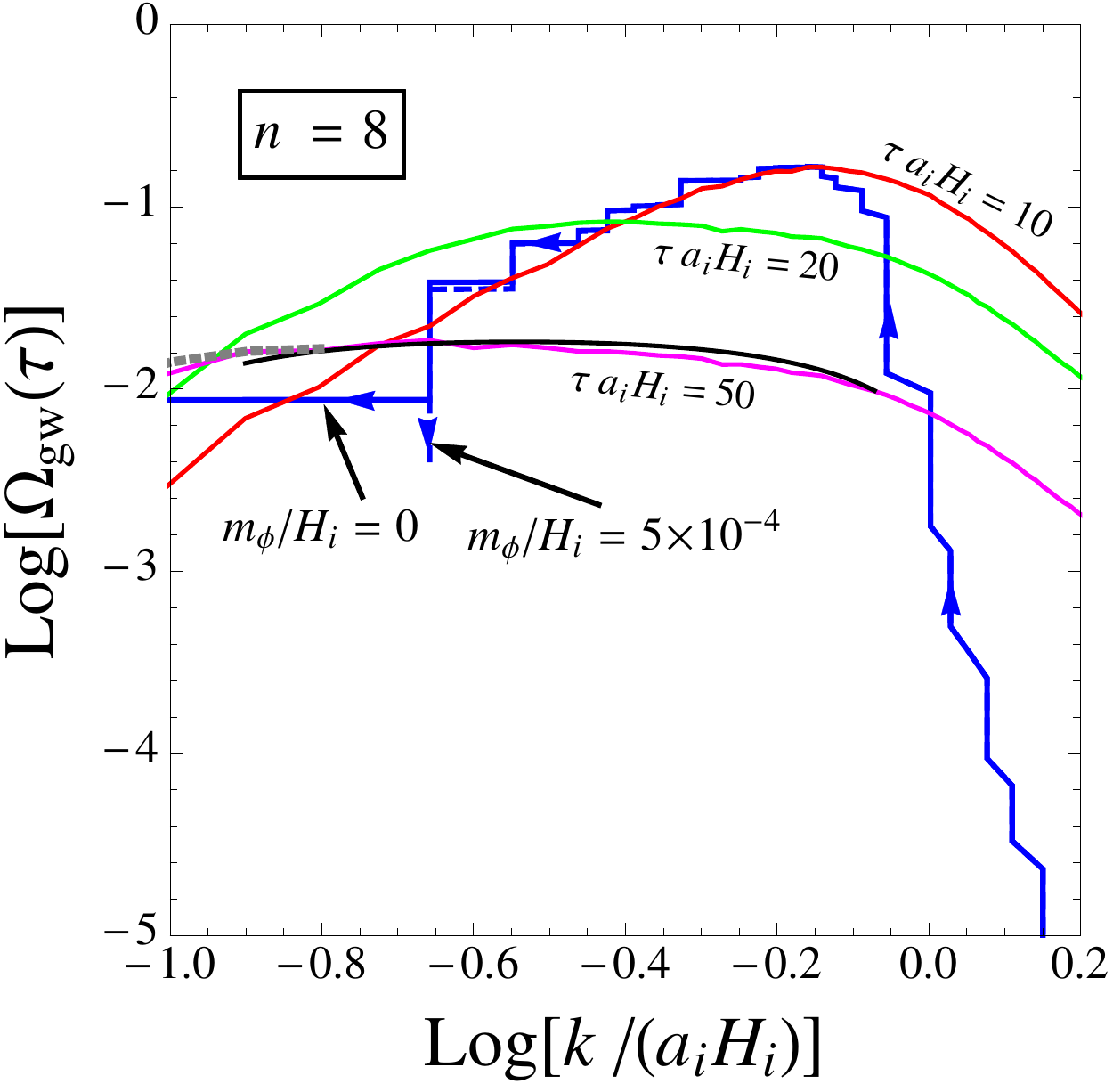} \vspace{0.5cm}\\
\includegraphics[width=.40\textwidth, bb=0 0 360 353]{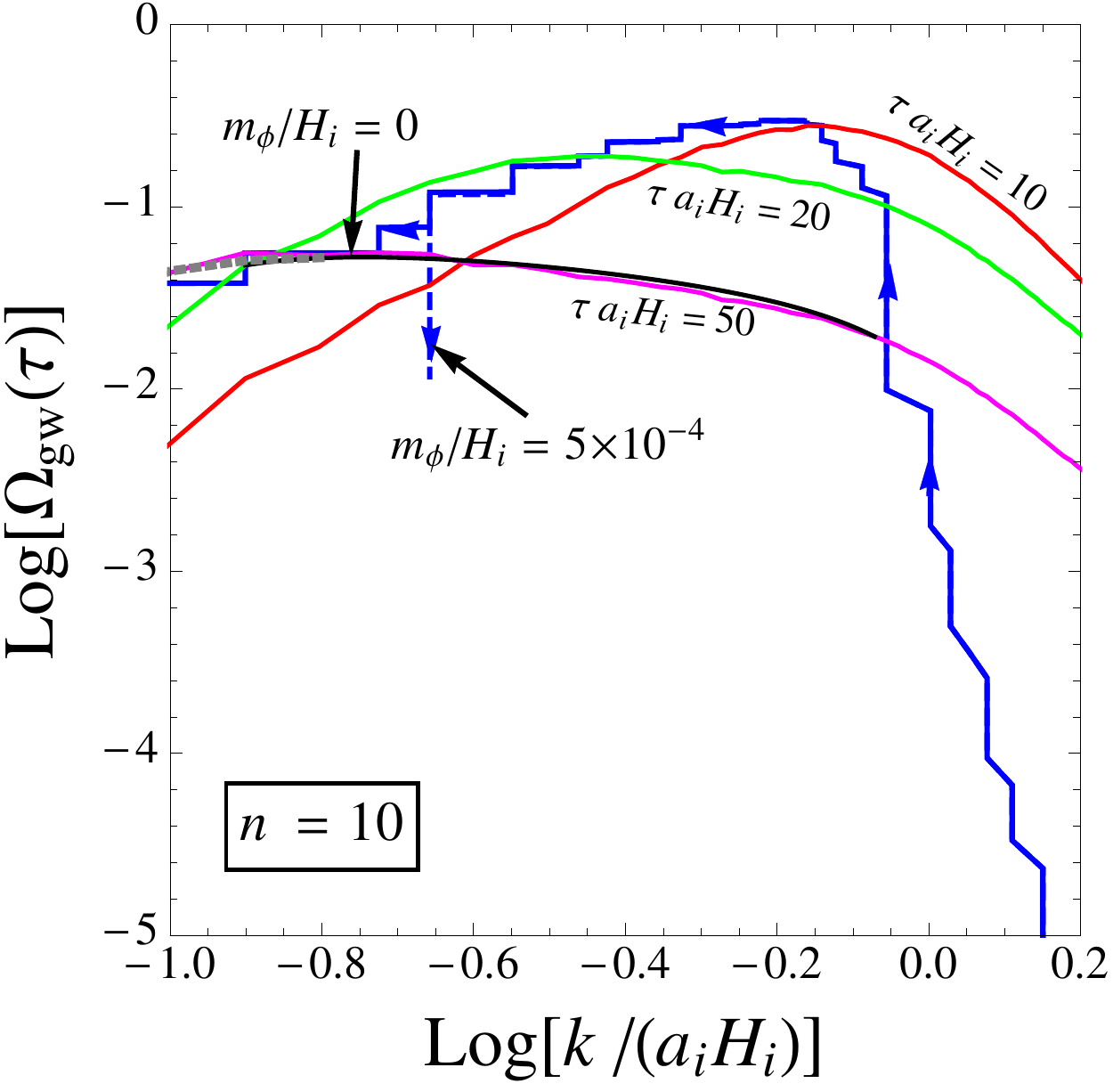} & \quad 
\includegraphics[width=.40\textwidth, bb=0 0 360 353]{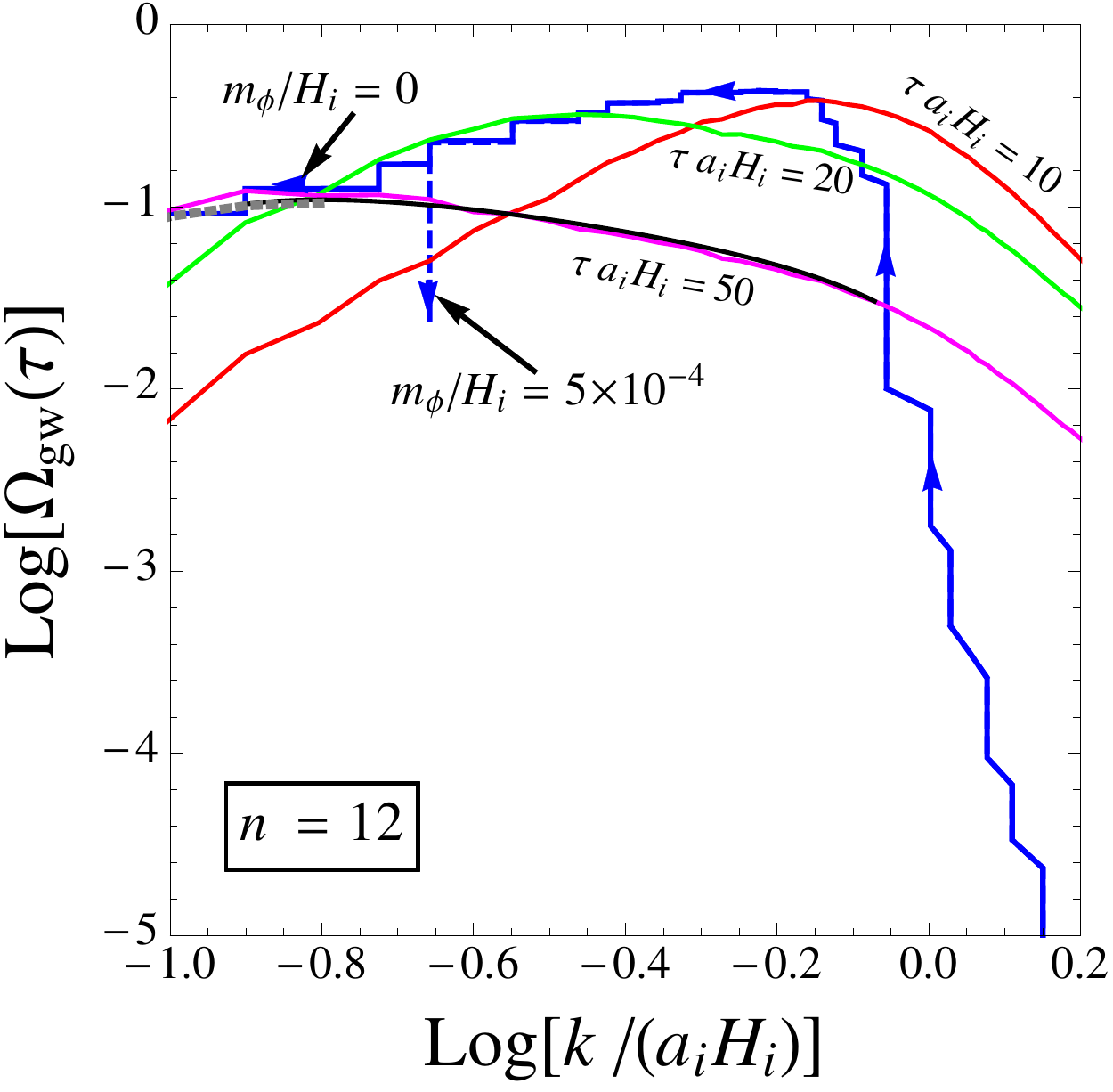} 
\end{tabular}
\caption{
GW spectra obtained from our numerical simulations. 
The energy density of GWs are rescaled by $(H_i/\lambda \Mpl )^{-4/(n-2)}$. 
The red, green, and magenta curves are GW spectra at the conformal time of 
$\tau a_i H_i = 10, 20,$ and $50$, respectively. 
We also show the time evolution of the GW peak location as the blue solid (dashed) line 
for the case of $m_\phi / H_i = 0$ 
($5 \times 10^{-4}$). 
We take $n=6$ (upper left panel), $8$ (upper right panel), $10$ (lower left panel), 
and $12$ (lower right panel), fixing $c_{H_{\rm osc}} = -15$. 
The black curves are plotted by Eq.~(\ref{large scale}) at $\tau a_i H_i = 50$
and well describes large wavenumber modes. 
The gray dotted curves are estimated 
from the discussion of causality at $\tau a_i H_i = 50$. 
}
  \label{GWspectrum}
\end{figure}

From the numerical simulation for cosmic string dynamics, 
we can obtain the transverse-traceless part of the anisotropic stress by the relation of 
\beq
 T_{ij}^{\text{TT}} ( \tau, {\bf x} ) 
 = \Lambda_{ij,lm} ({\bf x}) \lmk \del_l \phi^* \del_m \phi + {\rm c.c.} \rmk,
\eeq
where the projection operator in Fourier space is given by
\beq
\Lambda_{ij,lm} ({\hat k}) &=& P_{il} ({\hat k}) P_{jm} ({\hat k}) - \frac{1}{2} P_{ij} ({\hat k}) P_{lm} ({\hat k}), \\
P_{ij} ({\hat k}) &=& \delta_{ij} - {\hat k}_i {\hat k}_j,
\eeq
with ${\bf k} = \abs{k} {\hat k}$ and Kronecker's delta $\delta_{ij}$. 
Given $\tau_{\rm RH}$, 
we can calculate the energy density of GWs from Eqs.~(\ref{A}), (\ref{B}), and (\ref{omega_gw}) 
in Appendix~\ref{calculation}. 
We define the energy spectrum of GWs $\rho_{\rm gw}$ such as 
\beq
\Omega_{\rm gw} (\tau) 
 &\equiv& 
 \frac{1}{\rho_{\rm tot} \lmk \tau \rmk} \frac{ \dd \rho_{\rm gw} ( \tau) }{\dd \log k}, 
\eeq
where $\rho_{\rm tot} (\tau )$ ($= 3 \M^2 H^2 (\tau)$) is the total energy density of the Universe. 
Figure~\ref{GWspectrum} shows evolution of GW spectra 
obtained from our numerical simulations. 
The GW energy density grows 
as a typical value of the flat direction increases until 
it reaches the potential minimum at $\widetilde{\tau} = \widetilde{\tau}_{\rm form} \simeq 10$. 
Then the GW peak energy density decreases with time as $\Omega_{\rm gw} \propto \tau^{-12/(n-2)}$ 
and the peak wavenumber $k_{\rm peak}$ decreases with time as $k_{\rm peak} \propto \tau^{-1}$ 
for the case of $n \ge 8$. 
On the other hand, 
the GW peak energy density decreases with time as $\Omega_{\rm gw} \propto \tau^{-2}$ 
and its peak wavenumber is constant in time 
for the case of $n = 6$. 
We give physical interpretation of these results 
in the next subsection.

Taking a non-zero value of $m_\phi$ into account, 
we find that the flat direction starts to oscillate around the origin of the potential 
and cosmic strings disappear eventually
around $\widetilde{\tau} = \widetilde{\tau}_{\rm decay} \simeq 40$, 
which is given by Eq.~(\ref{tau_decay}) with $m_\phi/H_i = 5 \times 10^{-4}$ and $c_{H_{\rm osc}} = -15$. 
After that, 
the GW spectrum is freezed out, 
which means that 
the GW peak wavenumber becomes constant 
and the GW spectrum 
decrease adiabatically 
as $\propto \tau^{-2}$ ($\propto a^{-1}$). 
Note that we have taken $\widetilde{m}_\phi = 5 \times 10^{-4}$ above just for computational time saving, 
though it does not change the dynamics of flat direction until $\widetilde{\tau} \simeq \widetilde{\tau}_{\rm decay}$. 
Since $H_i$ is roughly given by $m_\chi \chi_i/\M$ (see Eq.~(\ref{chi})), $\widetilde{m}_\phi$ can be much smaller in general.
When we take $m_{\chi} = 10^{13}\,{\rm GeV}$ and $\chi_i = \M$ motivated by chaotic inflation model and $\mphi / \abs{c_{H_{\rm osc}}}^{1/2}= 10^3\,{\rm GeV}$, $\widetilde{\tau}_{\rm decay}\simeq4000$ is larger than $\widetilde{\tau}_{\rm form} \simeq 10$ by three orders of magnitude.

\subsection{\label{interpretation1}Physical interpretations}

In this subsection, we discuss how we can understand the results of our numerical simulations 
shown in Fig.~\ref{GWspectrum}. 

During the scaling regime, the typical length scale of cosmic string dynamics is the Hubble length. 
As a result, the emission peak wavenumber of GWs emitted from these cosmic strings 
is given by the Hubble scale: 
\beq
 \frac{k^{\Delta}_{\rm peak}}{a(t)} \sim H(t). 
 \label{peak0}
\eeq
We can estimate 
the energy density of GWs in the following way. 
Its quadrupole moment $Q$ can be roughly given by~\cite{Maggiore:1999vm} 
\beq
 Q \sim H^{-2} \times \mu H^{-1}, 
\eeq
where $\mu H^{-1}$ is the total energy of cosmic strings within a Hubble volume. 
Such cosmic strings emit GWs with the luminosity of 
\beq
 \mathcal{L} \sim \Mpl^{-2} \dddot{Q}^2 \sim \Mpl^{-2} (H^3 Q)^2, 
 \label{luminosity}
\eeq
where we replace the time derivative with $H$ in the last line. 
Thus, the produced energy density of GWs can be estimated as 
\beq
 \lkk \frac{\Delta \Omega_{\rm gw} }{\Delta \log \tau} \rkk_{\rm peak} 
 &\sim& 
 \frac{H^3 \times H^{-1} \mathcal{L}}{H^2 \Mpl^2} 
 \sim 
 \lmk \frac{\la \abs{\phi} \ra}{\M} \rmk^4 
 \label{omega_gw2}\\
 &\sim& 
 \lmk \frac{\abs{c_{H_{\rm osc}}}}{\lambda^2 (n-1)} \frac{H^2(t)}{\M^2} \rmk^{2/(n-2)},
 \label{omega_gw3}
\eeq
where 
we use Eqs.~(\ref{mu}) and (\ref{VEV1}) in the second and last equality, respectively. 
During each Hubble time, 
cosmic strings emit GWs 
with the emission peak wavenumber of Eq.~(\ref{peak0}) 
and 
with the produced energy density of Eq.~(\ref{omega_gw3}).

\begin{figure}[t]
\centering 
\begin{tabular}{l l}
\includegraphics[width=.40\textwidth, bb=0 0 360 351]{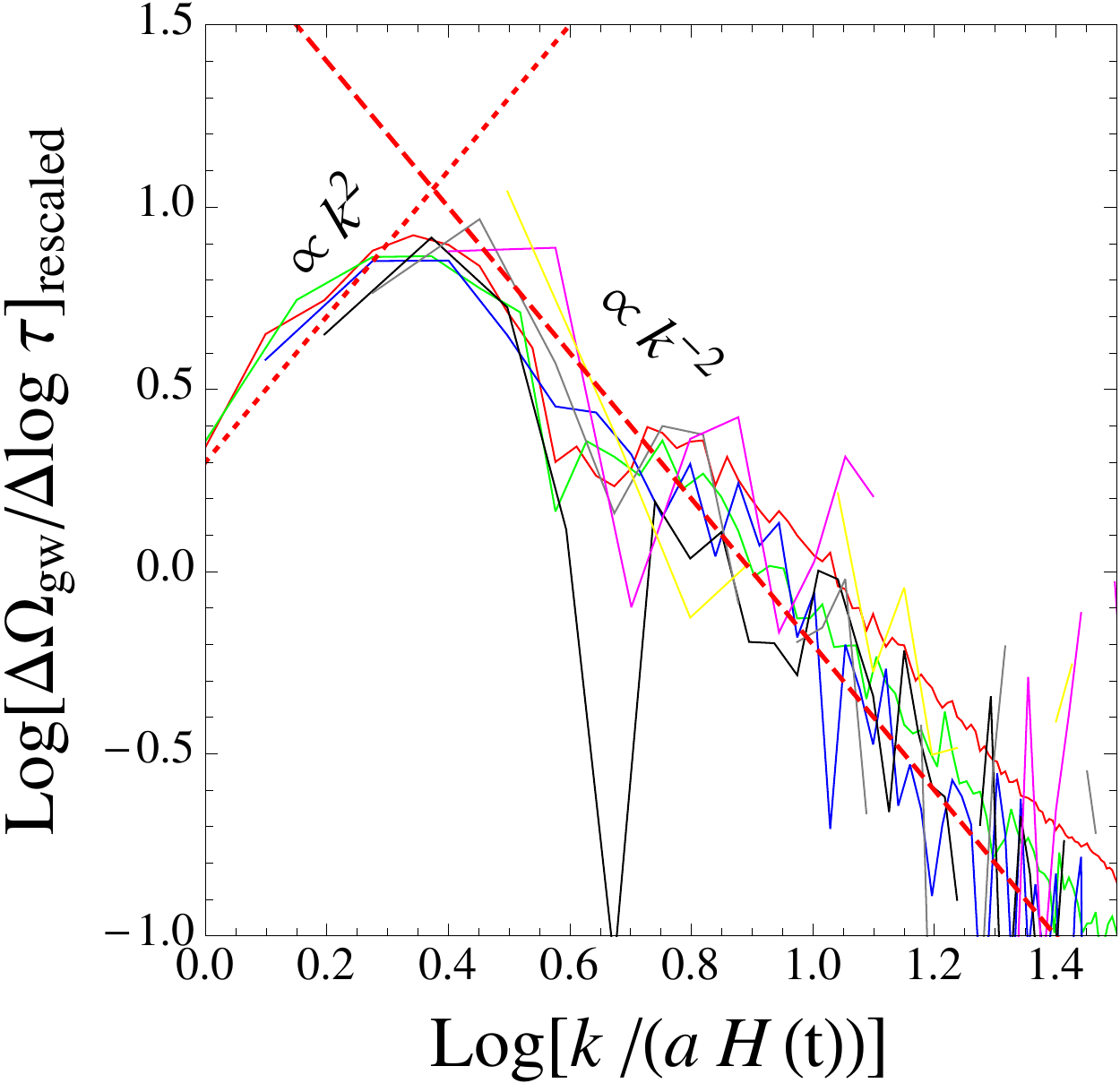} & \quad 
\includegraphics[width=.40\textwidth, bb=0 0 360 351]{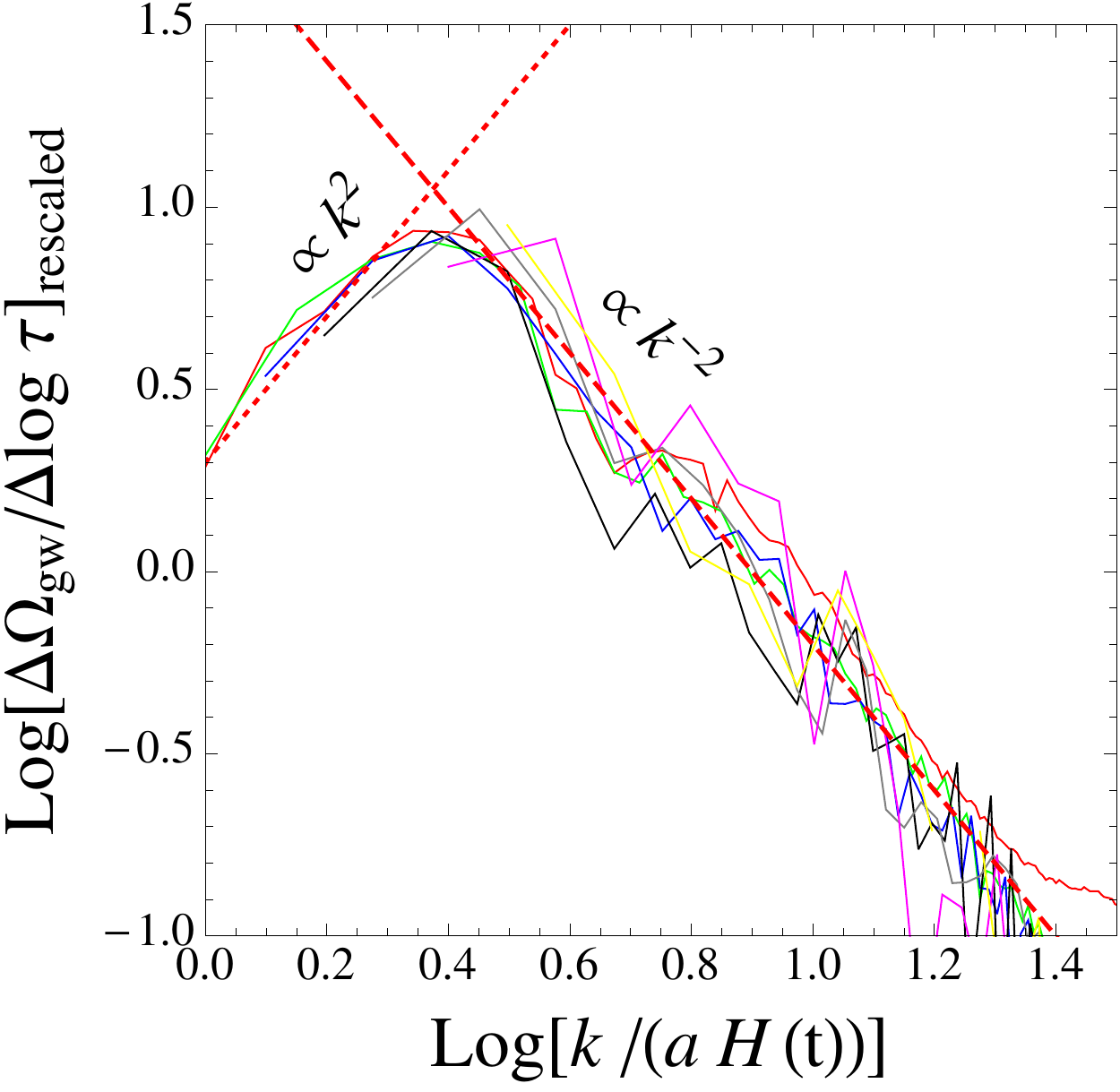} \vspace{0.5cm}\\
\includegraphics[width=.40\textwidth, bb=0 0 360 351]{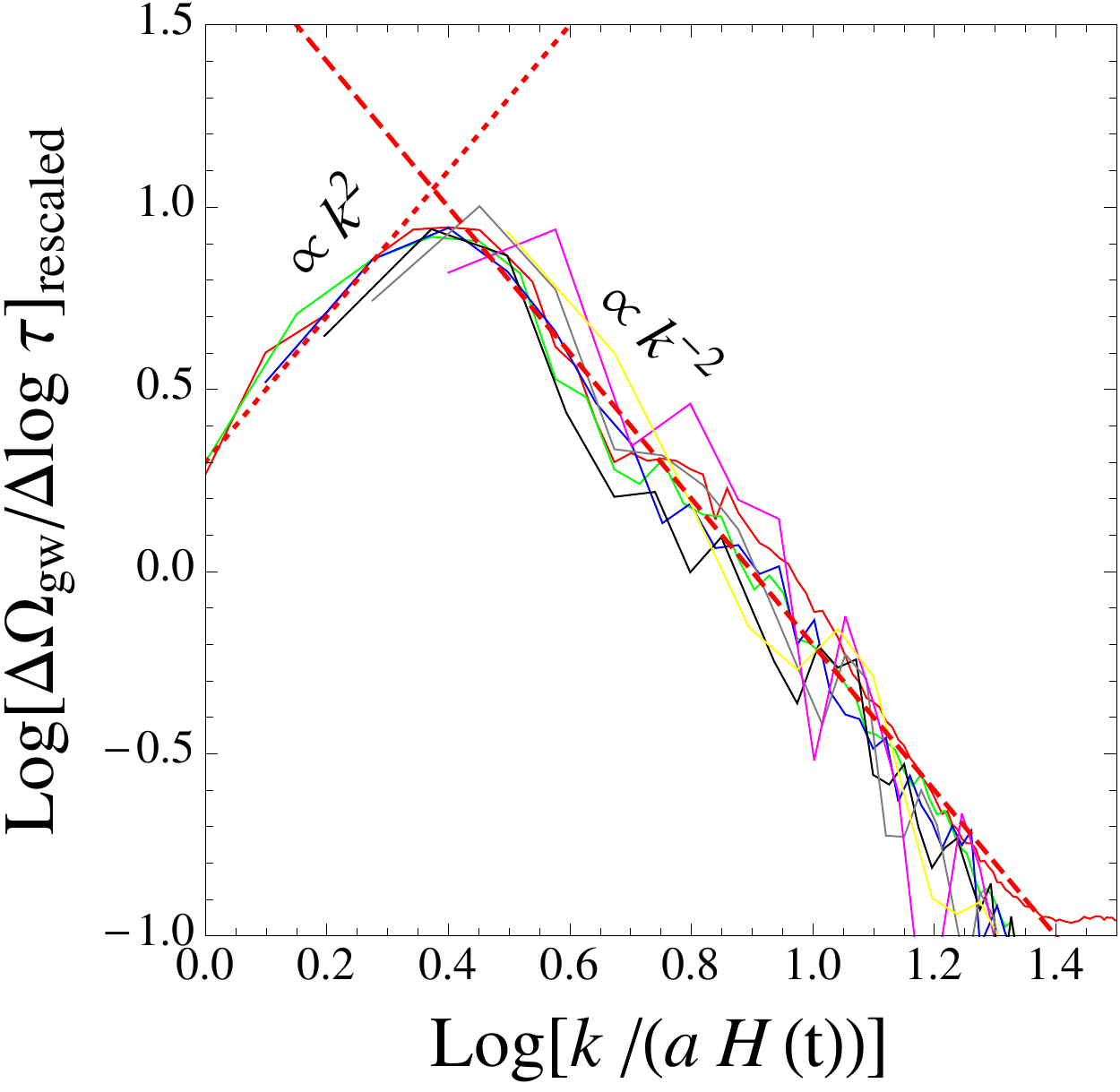} & \quad 
\includegraphics[width=.40\textwidth, bb=0 0 360 351]{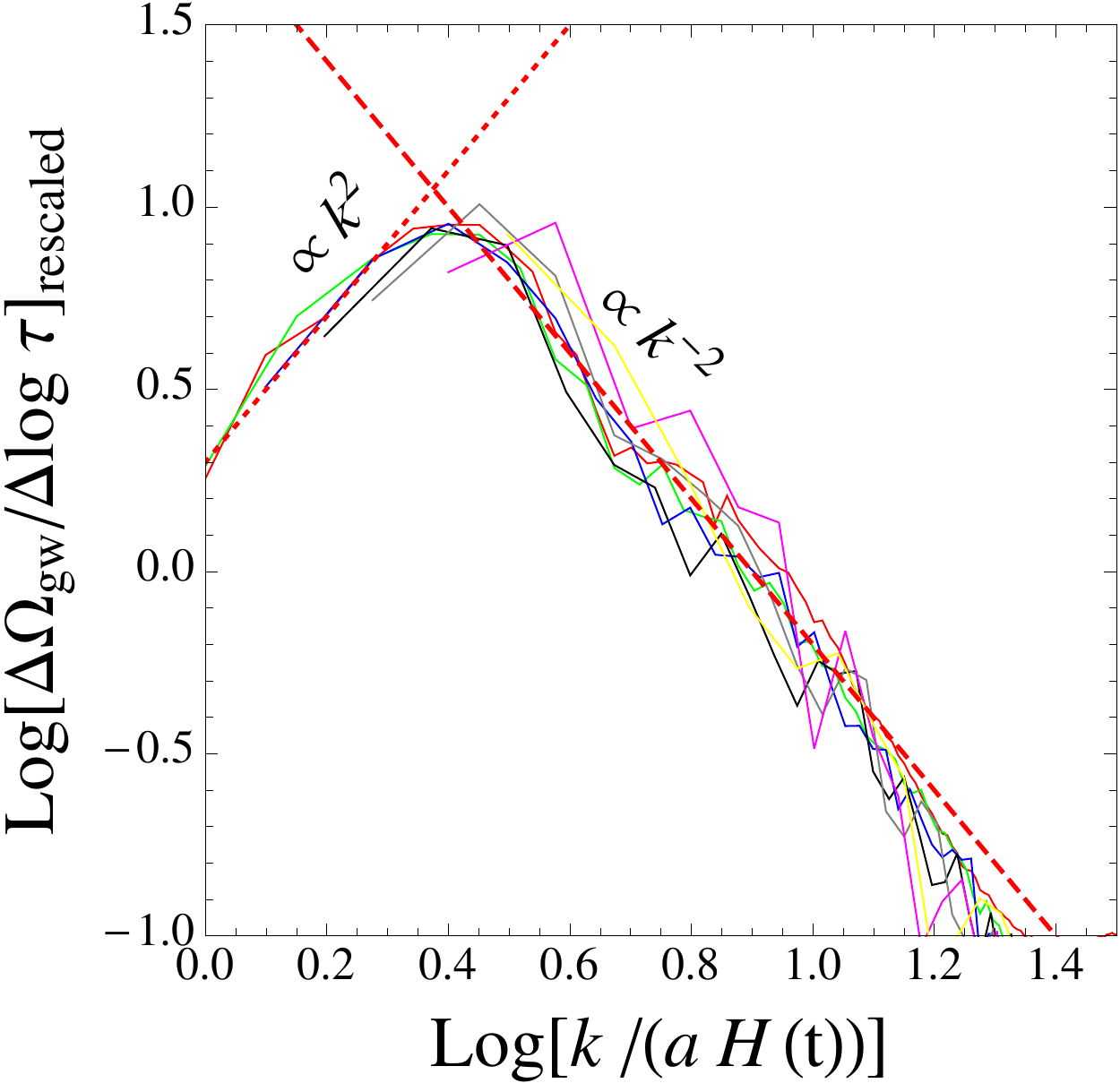} 
\end{tabular}
\caption{
Derivatives for GW spectra in terms of conformal time. 
The energy density of GWs are rescaled by the time dependent factor given by Eq.~(\ref{omega_gw3}). 
We plot GW spectra at the conformal time of $\tau a_i H_i = 20, 30, 40, 50, 60, 80,$ and $100$ for 
$n=6$ (upper left panel), $8$ (upper right panel), $10$ (lower left panel), 
and $12$ (lower right panel), fixing $c_{H_{\rm osc}} = -15$.
We fit 
small and large wavenumber modes by 
the dot and dashed red lines, which are proportional to $k^2$ and $k^{-2}$, respectively. 
}
  \label{del omega1}
\end{figure}

We have measured
$\Delta \Omega_{\rm gw} / \Delta \log \tau$ 
from our lattice simulation as shown in Fig.~\ref{del omega1}. 
We determine the numerical prefactors for 
the emission peak wavenumber and produced energy density of GWs 
from the numerical results such as 
\beq
 \frac{k^{\Delta}_{\rm peak}}{a(t)} &\simeq& 2.5 H(t), 
 \label{peak2} 
 \\
 \lkk \frac{\Delta \Omega_{\rm gw} }{\Delta \log \tau} \rkk_{\rm peak} 
 &\simeq& 10 \lmk \frac{\abs{c_H}}{\lambda^2 (n-1)} \frac{H^2(t)}{\M^2} \rmk^{2/(n-2)}. 
 \label{GW amplitude2}
\eeq 
We find that the numerical prefactors are independent of $n$ and $\tau$ (see Fig.~\ref{del omega1}), 
so that Eqs.~(\ref{peak2}) and (\ref{GW amplitude2}) can be used for any $n$ and $\tau$.

From Fig.~\ref{del omega1}, 
we can find that 
$\Delta \Omega_{\rm gw} / \Delta \log \tau$ 
is proportional to a certain power of $k$ 
for large and small wavenumber modes: 
\beq
 \frac{\Delta \Omega_{\rm gw} }{\Delta \log \tau} 
 \simeq  
 \left\{
 \bea{ll}
 \lkk \frac{\Delta \Omega_{\rm gw} }{\Delta \log \tau} \rkk_{\rm peak} 
 \lmk \frac{k}{k^{\Delta}_{\rm peak}} \rmk^{2} \qquad \text{for } k \lesssim k^{\Delta}_{\rm peak} \\
 \lkk \frac{\Delta \Omega_{\rm gw} }{\Delta \log \tau} \rkk_{\rm peak} 
 \lmk \frac{k}{k^{\Delta}_{\rm peak}} \rmk^{-\alpha} \qquad \text{for } k \gtrsim k^{\Delta}_{\rm peak} 
 \eea
 \right., 
 \label{large wavenumber}
\eeq
where $\alpha \simeq 2$. 
We also find that these values are 
almost independent of $n$ and $\tau$ from Fig.~\ref{del omega1}. 
The GW spectra for large wavenumber modes 
in Fig.~\ref{GWspectrum} 
is expected to be given by integrating the GW peak energy density from $\tau_{\rm form}$ to $\tau$, such as 
\beq
 && \Omega_{\rm gw} \lmk \tau \rmk  \nonumber \\
 &=&\int^{\tau}_{\tau_{\rm form}} \dd \log \tau' \frac{a(\tau')}{a(\tau)} 
 \frac{\Delta \Omega_{\rm gw} (\tau')}{ \Delta \log \tau' } \nonumber \\
 &\simeq& 
 \int^{\tau_k}_{\tau_{\rm form}} \dd \log \tau' \frac{a(\tau')}{a(\tau)} 
 \lkk \frac{\Delta \Omega_{\rm gw} (\tau')}{ \Delta \log \tau' } \rkk_{\rm peak} 
 \lmk \frac{k}{k^{\Delta}_{\rm peak}(\tau')} \rmk^{2} \nonumber \\
 &&+ 
 \int^{\tau}_{\tau_k} \dd \log \tau' \frac{a(\tau')}{a(\tau)} 
 \lkk \frac{\Delta \Omega_{\rm gw} (\tau')}{ \Delta \log \tau' } \rkk_{\rm peak} 
 \lmk \frac{k}{k^{\Delta}_{\rm peak}(\tau')} \rmk^{-\alpha} \nonumber \\
 &=& 
 \lmk 2 + \frac{2n-16}{n-2} \rmk^{-1} 
 \lkk \frac{\Delta \Omega_{\rm gw} (\tau)}{ \Delta \log \tau } \rkk_{\rm peak} 
 \lmk \frac{k^{\Delta}_{\rm peak} (\tau)}{k} \rmk^{(2n-16)/(n-2)} 
 \lkk 1 - \lmk \frac{k}{k^{\Delta}_{\rm peak} (\tau_{\rm form})} \rmk^{ 2 + (2n-16)/(n-2)} \rkk \nonumber \\ 
 &&+ 
 \lmk \alpha - \frac{2n-16}{n-2} \rmk^{-1} 
 \lkk \frac{\Delta \Omega_{\rm gw} (\tau)}{ \Delta \log \tau } \rkk_{\rm peak} 
 \lmk \frac{k^{\Delta}_{\rm peak} (\tau)}{k} \rmk^{(2n-16)/(n-2)} 
 \lkk 1 - \lmk \frac{k^{\Delta}_{\rm peak} (\tau)}{k} \rmk^{\alpha - (2n-16)/(n-2)} \rkk. \nonumber \\
 \label{large scale}
\eeq
The scale factor is included in the integrant because 
the GW energy density decreases with time as $\propto a^{-1} (\tau) \propto \tau^{-2}$ 
in the oscillation dominated era. 
We define $\tau_k$ by the time 
the GW emission peak wavenumber reaches $k$, that is, $k^{\Delta}_{\rm peak} (\tau_k) \equiv k$. 
We implicitly assume $\tau > \tau_k$ in this calculation, 
which means that Eq.~(\ref{large scale}) is valid for $k > k^{\Delta}_{\rm peak} (\tau)$. 

The result of Eq.~(\ref{large scale}) at $\widetilde{\tau} = 50$ is plotted as the black curves in Fig.~\ref{GWspectrum} 
and well describes the GW spectra. 
Note that 
the GW energy density is roughly proportional to 
$\propto (k^{\Delta}_{\rm peak} / k)^{(2n-16)/(n-2)}$ 
for $k^{\Delta}_{\rm peak} (\tau_{\rm form})$ $(= k^{\Delta}_{\rm peak} (\tau) \tau / \tau_{\rm form})$ $\gg k \gg k^{\Delta}_{\rm peak} (\tau)$ and $n \geq 6$. 
Thus, in principle, we can obtain 
the value of $n$ from the observation of GWs with high frequencies. 
Note that for the case of $n = 6$ and $7$, the power-law index of GW spectra is positive ($16-2n > 0$). 
This implies that the GW spectrum has a peak at $\tau = \tau_{\rm form}$. 
This is because 
the emitted GW energy density of Eq.~(\ref{GW amplitude2}) decreases with time faster than the redshift $\propto a^{-1}$ 
for $n = 6$ and $7$. 
This implies that
the peak energy density emitted at $\tau < \tau_{\rm form}$ is smaller than 
the redshifted one emitted at $\tau =\tau_{\rm form}$
when
the spatially averaged magnitude of the flat direction reaches the minimum of the potential. 
Thus, the GW peak wavenumber is determined by $k_{\rm peak} \simeq k^{\Delta}_{\rm peak} (\tau_{\rm form}) \sim \tau_{\rm form}^{-1}$. 
Then the GW peak energy density decreases with $\propto a^{-1}$ due to the redshift until reheating completes. 
The redshift factor is given by $a_{\rm form}/a_{\rm RH} = (H_{\rm RH} / H_{\rm form})^{2/3}$. 
This is the reason that 
the GW peak wavenumber is constant in time for the case of $n=6$ 
in Fig.~\ref{GWspectrum}. 
The GW spectrum for $n=8$ is too flat to determine its peak wavenumber precisely, 
though the GW spectra are well fitted by the above estimation.

Next, let us consider the GW spectrum 
for very small wavenumber modes $k \lesssim k_{\rm peak}$. 
For such super-horizon modes, 
we expect 
that 
the leading contribution for 
the Fourier transformed transverse-traceless part of the anisotropic stress 
$\TT$ is independent of $k$ 
due to the loss of causality at the large scale~\cite{Dufaux:2007pt, Kawasaki:2011vv}. 
As a result, 
the GW spectrum is proportional to $k^2$ 
for the modes smaller than but around $k_{\rm peak}$,
which is in agreement with our numerical simulations in Fig.~\ref{del omega1}.
It is proportional to $k$ 
for the modes much smaller than $k_{\rm peak}$, 
which are out of reach of our numerical simulations. 
In addition, we have to take into account the effect of 
the reheating, at which the expansion law of the Universe changes. 
These effects on super-horizon modes 
leave their imprints on GW signals at present, 
which we discuss in the next subsection. 
We can calculate the GW spectrum for super-horizon modes 
by substituting a constant $\TT$ into Eqs.~(\ref{A}) and (\ref{B}) in Appendix~\ref{calculation} 
and plot the results at $\widetilde{\tau}=50$ as the gray dotted curves in Fig.~\ref{GWspectrum}.

From Fig.~\ref{GWspectrum}, we can see 
that GW spectra drop off 
for modes $k \gtrsim \tau_{\rm form}^{-1}$ ($\gg k_{\rm peak}$), which enter 
the horizon before the spatially averaged magnitude of the flat direction
reaches VEV $\la \abs{\phi} \ra$. 
This is because 
the produced energy density of GWs is proportional to the fourth power of the field value of the flat direction 
as shown in Eq.~(\ref{omega_gw2}). 
The result depends nontrivially on $c_{H_{\rm osc}}$ 
because 
the dynamics of the flat direction 
around the origin of the potential 
is mainly determined by the Hubble induced mass term. 
Let us note that we are less interested in such high frequency modes 
in light of the detection of GW signals (see Fig.~\ref{detectability1}).

The GW emission terminates 
around $\tau_{\rm decay}$, 
which is defined by $\abs{c_{H_{\rm osc}}} H^2(t) = m_\phi^2$ (see Eq.~(\ref{tau_decay})). 
This is because the flat direction starts to oscillate around the origin of the potential 
and cosmic strings disappear at that time. 
This implies that the GW spectrum has a peak (or at least bends) at the wavenumber 
corresponding to the soft mass of the flat direction. 
In fact, 
the GW peak energy density and frequency at that time 
is given by 
\beq
 \lkk \Omega_{\rm gw} (\tau_{\rm decay}) \rkk_{\rm peak} &\simeq& 
 14 \lmk 2+\frac{2n-16}{n-2} \rmk^{-1} \lmk \frac{1}{\lambda^2 (n-1)} \frac{m_\phi^2}{\M^2} \rmk^{2/(n-2)}, 
  \label{omega_gw4}
  \\
  \frac{k_{\rm peak}}{a(\tau_{\rm decay})} &\simeq& 3 \frac{\mphi}{\sqrt{\abs{c_{H_{\rm osc}}}}}, 
  \label{peak1}
\eeq
for the case of $n \ge 8$. 
Here, we determine the numerical prefactors from the results of our numerical simulations. 
Although GW spectrum has a peak at $k \simeq \tau_{\rm form}^{-1}$ for the case of $n = 6$ and $7$, 
it bends at the wavenumber of Eq.~(\ref{peak1}) and has the energy density of Eq.~(\ref{omega_gw4}). 
In either case, 
the GW spectrum contains information of the soft mass of the flat direction 
at that wavenumber because of the relation of Eq.~(\ref{peak1}). 
Hereafter, we denote the wavenumber defined by Eq.~(\ref{peak1}) 
as the peak wavenumber $k_{\rm peak}$. 
The shape of GW spectrum is in fact freezed out
around $\widetilde{\tau} = \widetilde{\tau}_{\rm decay} \simeq40$, 
which is given by Eq.~(\ref{tau_decay}) with $m_\phi/H_i = 5 \times 10^{-4}$ and $c_{H_{\rm osc}} = -15$.
After that, the peak wavenumber does not change and 
the GW energy density decreases as $\propto a^{-1}$ 
until reheating completes. 

\begin{figure}[t]
\centering 
\includegraphics[width=.40\textwidth, bb=0 0 360 354]{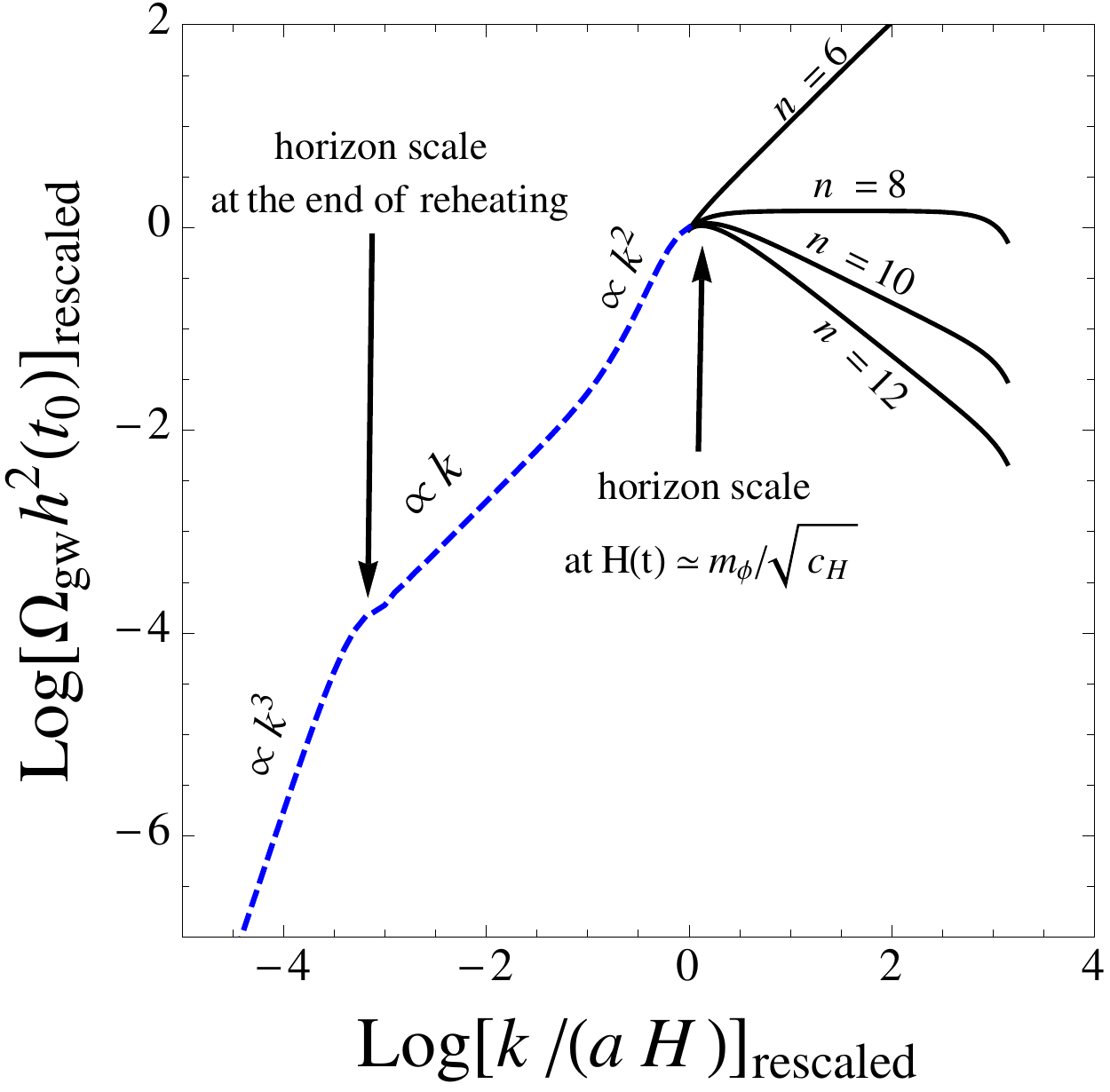}
\caption{
Schematic diagram for the GW spectrum emitted from cosmic strings. 
The horizontal and vertical axes are rescaled such that 
the GW energy density and the GW peak wavenumber are one. 
The black curves are plotted by using Eq.~(\ref{large scale}) 
for the cases of $n=6, 8, 10,$ and $12$. 
The blue dashed curve is calculated by using Eqs.~(\ref{A}) and (\ref{B}) 
with a constant $\TT$. 
We take into account a nonzero value of reheating temperature, 
which modifies the GW spectrum for wavenumber smaller than 
the inverse of the horizon scale at the end of reheating. 
}
  \label{schematic diagram1}
\end{figure}

Figure~\ref{schematic diagram1} is a schematic diagram for the resulting GW spectrum 
emitted from the cosmic strings. 
We calculate super-horizon modes, i.e., small wavenumber modes 
by using Eqs.~(\ref{A}) and (\ref{B}) 
with a constant $\TT$. 
The result is plotted as the blue dashed curve. 
The GW spectrum for small wavenumber modes near the peak wavenumber is proportional to $k^2$, 
while that for the smaller wavenumber modes is proportional to $k$. 
We also take into account the effect of reheating, 
so that the GW spectrum bends at the wavenumber corresponding to the inverse of the horizon size
at the end of reheating~\cite{Seto:2003kc},
\beq
  \frac{k_{\rm bend}}{a(t_{\rm RH})} &\simeq& H(t_{\rm RH}).
  \label{bend1}
\eeq 
We explain this phenomena in the next subsection in detail. 
We calculate large wavenumber modes by using Eq.~(\ref{large scale})
for the cases of $n=6, 8, 10$, and $12$, 
and the results are plotted as the black curves. 
One can see that the GW spectrum for high frequency modes 
depends on the value of $n$. 
For $n \ge 9$, 
the GW spectrum has a peak at the wavenumber corresponding to 
the inverse of the horizon scale when cosmic strings disappear, 
that is, at the time of $H(t) \simeq m_\phi / \sqrt{\abs{c_{H_{\rm osc}}}}$. 
For $n = 6, 7,$ and $8$, 
the GW spectrum does not have a peak, but bends at that wavenumber. 
Note that 
the magnitude of GW energy density depends on the value of the coupling in the superpotential $\lambda$, 
though we use it to rescale the vertical axis in Fig.~\ref{schematic diagram1}.

\subsection{\label{prediction1}GW signals at present}

The above calculations and discussions 
focus on the GW spectrum during the oscillation dominated era. 
In this subsection, we derive 
the GW spectrum at the present epoch, 
taking into account the redshift.

The energy density of GWs at $\tau_{\rm decay}$,
$\Omega_{\rm gw} (\tau_{\rm decay})$ 
is given by our numerical simulations 
as shown and discussed in the previous two sections. 
Note that 
we consider the case that 
cosmic strings disappear before 
reheating completes, 
as we explained in Sec.~\ref{dynamics1}. 
Since the GW energy density $\Omega_{\rm gw}$ decreases with time as $a^{-1}$ during 
the oscillation dominated era (i.e., the matter dominated era), 
its present value is given by 
\beq
 && \Omega_{\rm gw} h^2 (t_0) \nonumber\\ 
 &=& 
 \Omega_r h^2 
 \lmk \frac{g_s (t_0)}{g_s(t_{\rm RH})} \rmk^{4/3} 
 \lmk \frac{g_*(t_{\rm RH})}{g_* (t_0)} \rmk 
 \Omega_{\rm gw} ( t_{\rm RH}) \nonumber\\ 
 &\simeq& 
 \Omega_r h^2 
 \lmk \frac{g_s (t_0)}{g_s(t_{\rm RH})} \rmk^{4/3} 
 \lmk \frac{g_*(t_{\rm RH})}{g_* (t_0)} \rmk 
 \lmk \frac{H_{\rm RH}}{H_{\rm decay}} \rmk^{2/3}
 \Omega_{\rm gw} ( \tau_{\rm decay}), 
\label{present energy density}
\eeq
where $t_0$ is the present time, 
$\Omega_r h^2$ ($\simeq 4.15 \times 10^{-5}$) is the present energy density of radiation, 
and $g_*$ ($g_s$) is the effective relativistic degrees of freedom for the energy (entropy) density. 
In the following calculations, we set $g_s (t_0)=43/11$, $g_* (t_0)=2+21/11(4/11)^{1/3}$, and $g_s(t_{\rm RH}) = g_*(t_{\rm RH}) = 915/4$ in the following calculations.
Substituting $\lkk \Omega_{\rm gw} (\tau_{\rm decay}) \rkk_{\rm peak}$ into Eq.~(\ref{present energy density}), 
we obtain 
\beq
 && \lkk \Omega_{\rm gw} h^2 (t_0) \rkk_{\rm peak} \nonumber \\ 
  &\simeq& 
 3 \times 10^{-6} 
 \lmk \frac{\abs{c_{H_{\rm osc}}}^{-1/2} \mphi}{10^3 \GeV} \rmk^{-2/3}
 \lmk \frac{T_{\rm RH}}{10^9 \GeV} \rmk^{4/3}
 \lmk 2+\frac{2n-16}{n-2} \rmk^{-1} \lmk \frac{1}{\lambda^2 (n-1)} \frac{m_\phi^2}{\M^2} \rmk^{2/(n-2)}, 
 \nonumber \\
 \label{result omega}
\eeq
where we use 
Eqs.~(\ref{t_decay}), (\ref{T_RH}), and (\ref{omega_gw4}). 
The GW peak frequency at present $f_0$ is given by 
\beq
 f_0 
 &=& 
 \lmk \frac{g_s (t_0)}{g_s(t_{\rm RH})} \rmk^{1/3} 
 \lmk \frac{T_0}{T_{\rm RH}} \rmk 
 \frac{k_{\rm peak}}{2 \pi a(t_{\rm RH})} \nonumber\\ 
 &\simeq& 
 \lmk \frac{g_s (t_0)}{g_s(t_{\rm RH})} \rmk^{1/3} 
 \lmk \frac{T_0}{T_{\rm RH}} \rmk 
 \lmk \frac{H_{\rm RH}}{H_{\rm decay}} \rmk^{2/3}
 \frac{k_{\rm peak}}{2 \pi a(t_{\rm decay})}. 
\label{present peak frequency}
\eeq
Substituting Eq.~(\ref{peak1}) into this equation, 
we obtain 
\beq
 f_0 
 \simeq 700 \text{ Hz} \lmk \frac{\abs{c_{H_{\rm osc}}}^{-1/2} \mphi}{10^3 \GeV} \rmk^{1/3} \lmk \frac{\TR}{10^9 \GeV} \rmk^{1/3}. 
 \label{result f}
\eeq
We should emphasize that this peak frequency 
depends only on $\abs{c_{H_{\rm osc}}}^{-1/2} \mphi$ ($\simeq H (\tau_{\rm decay})$) and $\TR$ and 
is independent of 
$n$ and $\lambda$.

Here, let us 
consider the effect of reheating on the GW spectrum, 
which we mentioned in the previous section. 
In the absence of GW source, 
the metric perturbation is constant for super-horizon modes 
and decreases with $a^{-1}$ for sub-horizon modes.
Let us define $a_k$ as the scale factor at which GW with wavenumber $k$ enters the horizon: 
$k /a_k = H(a_k)$. 
For $H(a) \propto a^{-\nu}$, $a_k$ scales like $a_k \propto k^{-1/(\nu-1)}$.
Noting that $\Omega_{\rm gw}$ is proportional to the metric perturbation squared, 
we obtain $\Omega_{\rm gw} \propto k^{-2/(\nu-1)} \Omega_{\rm gw}^{p}$ for sub-horizon modes, 
which have primordial GW spectrum of $\Omega_{\rm gw}^{p}$ before entering the horizon.
If the primordial GW source is scale invariant, i.e., $T^{TT}_{ij} = {\rm const.}$ with respect to $k$, 
the primordial GW spectrum scales like $\Omega_{\rm gw}^{p} \propto k^5$.
Since the expansion law of the Universe changes from $H \propto a^{-3/2}$ to $H \propto a^{-2}$,
we expect $\Omega_{\rm gw} \propto k$ and $\propto k^3$ for GWs which enter the horizon before and after reheating, respectively, if the primordial GW source is scale invariant. 
As a result, 
the GW spectrum is expected to bend at the 
wavenumber corresponding to the horizon scale at the end of reheating~\cite{Seto:2003kc, 
Nakayama:2008ip, previous work}. 
Since we consider the case that 
cosmic strings disappear before 
reheating completes, 
the relevant GW with wavenumber corresponding to the horizon scale at the end of reheating 
is super-horizon mode when the GW emission terminates ($\tau \simeq \tau_{\rm decay}$). 
Thus, we can use $\TT = \text{const.}$ in Eqs.~(\ref{A}) and (\ref{B}) to estimate the GW spectrum 
due to the loss of causality for super-horizon scales~\cite{Dufaux:2007pt, Kawasaki:2011vv}. 
Then, 
using $j_l(x) \to  2^l l!  x^l/(2l+1)!$ and $n_l(x) \to -(2l)!/ (2^l l! x^{l+1})$ for $x \ll 1$ 
in Eq.~(\ref{functions}), 
we obtain 
$\Omega_{\rm gw} \propto k$ for $\tau_{\rm RH}^{-1} \ll k \ll \tau_{\rm decay}^{-1}$ 
and 
$\Omega_{\rm gw} \propto k^3$ for $k \ll \tau_{\rm RH}^{-1}$ as expected. 
This means 
that the GW spectrum actually bends at the wavenumber around $k \simeq \tau_{\rm RH}^{-1}$. 
We denote the frequency corresponding to that wavenumber at present as bend frequency $f_{\rm bend}$ 
(see Eq.~(\ref{bend1}). 
It is calculated as 
\beq
 f_{\rm bend} 
 &=& 
 \lmk \frac{g_s (t_0)}{g_s(t_{\rm RH})} \rmk^{1/3} 
 \lmk \frac{T_0}{T_{\rm RH}} \rmk 
 \frac{k_{\rm bend}}{2 \pi a(t_{\rm RH})} \nonumber\\ 
 &\simeq& 30 \text{ Hz} 
 \lmk \frac{\TR}{10^9 \GeV} \rmk. 
 \label{f_bend}
\eeq
Note that this bend frequency is different from the one 
corresponding to $k_{\rm peak}$ for $n=6$ and $7$. 
In these cases, GW spectrum bends at two frequencies: 
at $f=f_0$ ($k = k_{\rm peak}$) corresponding to the horizon scale at $\tau = \tau_{\rm decay}$ 
and at $f = f_{\rm bend}$ ($k= k_{\rm bend}$) corresponding to the horizon scale at the end of reheating.

Here, let us explain what kind of information we can obtain 
through the observation of GW spectrum emitted from cosmic strings. 
First, we can obtain the value of reheating temperature through 
detection of the bend frequency of Eq.~(\ref{f_bend}). 
Then we can obtain the soft mass scale of the flat direction 
through Eq.~(\ref{result f}). 
We can also obtain the value of the power $n$ 
because the slope of the GW spectrum for wavenumber larger than Eq.~(\ref{result f}) 
is dependent on $n$ (see Eq.~(\ref{large scale}) or Fig.~\ref{schematic diagram1}). 
Finally, we can obtain the value of $\lambda$ through 
the measurement of the GW energy density Eq.~(\ref{result omega}). 
Thus, the observation of the whole GW spectrum 
gives us the values of $\abs{c_{H_{\rm osc}}}^{-1/2} \mphi$, $T_{\rm RH}$, $n$, and $\abs{c_{H_{\rm osc}}}^{-1/4} \lambda$ .

\begin{figure}[t]
\centering 
\includegraphics[width=.45\textwidth, bb=0 0 360 339]{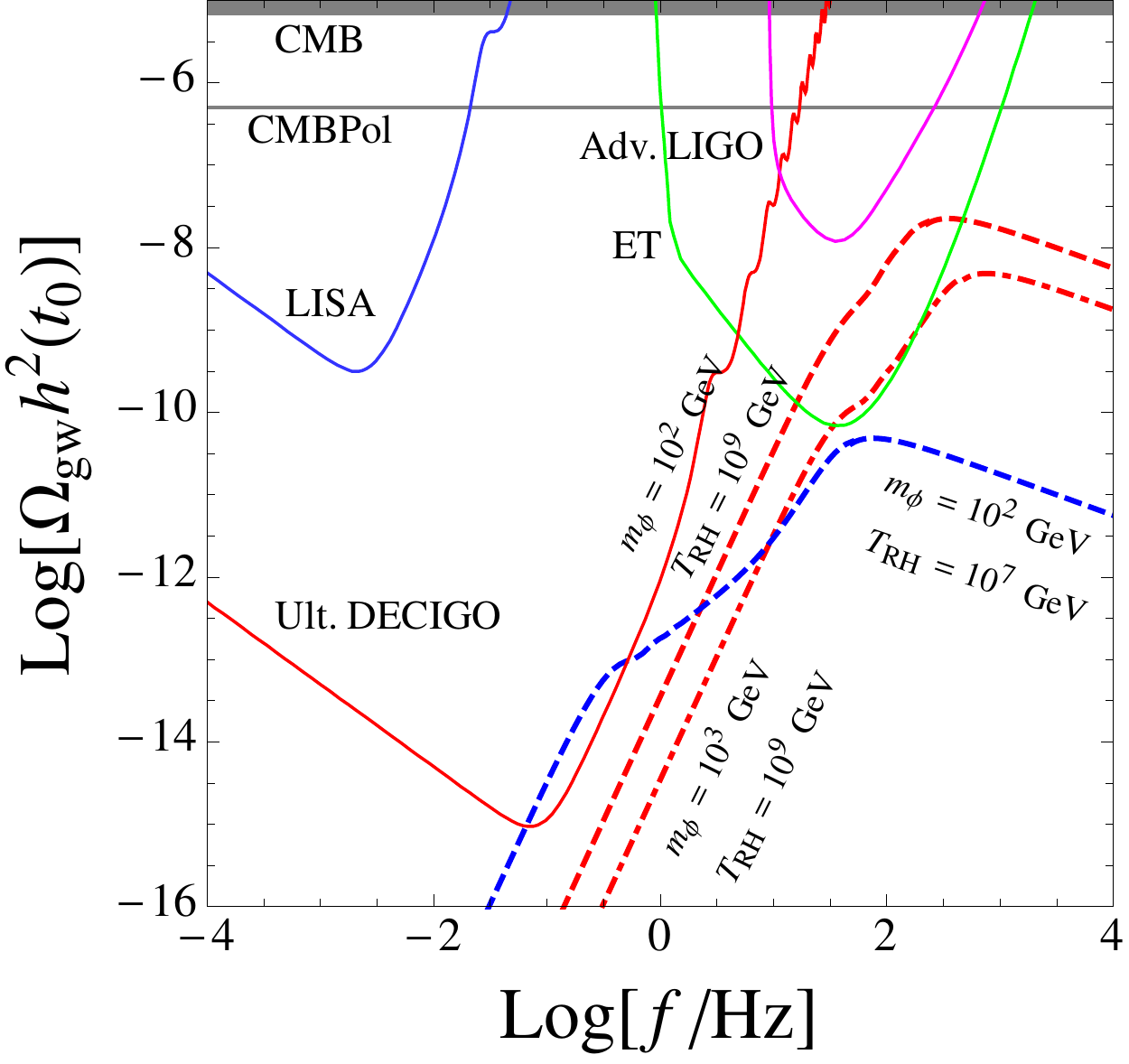} 
\caption{
GW spectra generated by cosmic strings 
(dash and dot-dashed curves) 
and sensitivities of planned interferometric detectors (solid curves). 
We assume 
$\mphi = 10^2 \GeV$ 
with $\TR = 10^7 \GeV$ (blue dashed curve), 
$\mphi = 10^2 \GeV$ 
with $\TR = 10^9 \GeV$ (red dashed curve), 
and 
$\mphi = 10^3 \GeV$ 
with $\TR = 10^9 \GeV$ (red dot-dashed curve). 
We also assume $c_{H_{\rm osc}} = 10$ and $n=10$. 
We set the value of $\lambda$ so that 
the GW peak energy density is given by $0.01$ at $\tau = \tau_{\rm decay}$. 
}
  \label{detectability1}
\end{figure}

Figure~\ref{detectability1} shows examples of GW spectrum 
predicted by the present mechanism. 
The GW peak energy density and peak frequency are 
given by Eqs.~(\ref{result omega}) and (\ref{result f}), which are obtained from our numerical results. 
While the shape of the GW spectrum for low frequency modes are calculated 
from Eqs.~(\ref{A}) and (\ref{B}) with a constant $\TT$ and $\tau_f = \tau_{\rm decay}$, 
that for high frequncy modes are calculated from Eq.~(\ref{large scale}). 
Note that the high frecuency modes corresponding to 
the horizon scale at the time of cosmic string formation $\tau_{\rm form}$ 
are out of the range of Fig.~\ref{detectability1}, 
because the energy scale of inflation is much larger than the mass scale of the flat direction. 
We take 
$\mphi = 10^2 \GeV$ 
with $\TR = 10^7 \GeV$ (blue dashed curve), 
$\mphi = 10^2 \GeV$ 
with $\TR = 10^9 \GeV$ (red dashed curve), 
and 
$\mphi = 10^3 \GeV$ 
with $\TR = 10^9 \GeV$ (red dot-dashed curve).

To discuss the detectability of GW signals, 
we plot single detector sensitivities for LISA~\cite{lisa} 
and Ultimate DECIGO~\cite{decigo} 
by using the online sensitivity curve generator in~\cite{sens_curves} 
with the parameters in Table $7$ of Ref.~\cite{Alabidi:2012ex}. 
Note that in some parameter space we can obtain $\TR$ by using only one single detector for Ultimate DECIGO. 
One may wonder why our sensitivity curve is different from that in Ref.~\cite{Nakayama:2008ip}, 
where they claim that Ultimate DECIGO can reach well below $\Omega_{\rm gw} \sim 10^{-16}$ 
for frequencies of $0.1-1$\,Hz. 
This is because they assume taking two detectors correlation for a decade 
to obtain $T_{\rm RH}$ by observation of the inflationary GW background, 
while we assume a single detector. 
A single detector is sufficient for our purpose because GW signals from cosmic strings 
are much stronger than inflationary ones. 
We also plot cross-correlation sensitivities for Advanced LIGO~\cite{adv_ligo} 
and ET (ET-B configuration)~\cite{et}, 
assuming two detectors are co-aligned and coincident. 
In Fig.~\ref{detectability1}, we take the signal to noise ratio ${\rm SNR} = 5$, 
the angular efficiency factor $F=2/5$, 
the total observation time $T=1 \,{\rm yr}$, and the frequency resolution $\Delta f / f =0.1$. 
CMB constraints (horizontal lines) are put on the integrated energy density of GWs 
$\int \dd \log f \Omega_{\rm gw} h^2 (t_0)$~\cite{Smith:2006nka}. 
We find that 
GW signals would be observed by ET or Ultimate DECIGO.

Note that we assume a very small $\lambda$ in Fig.~\ref{detectability1} 
so that the future GW detector can observe GW signals. 
However, such a small $\lambda$ may result in a VEV (Eq.~(\ref{VEV1})) larger than the cutoff scale just after inflation, 
even though the VEV is smaller than the cutoff scale around the time of $H(t) \simeq m_\phi$. 
In this case, the potential for the flat direction should be dominated by 
higher dimensional superpotential or \Kahler potentials 
to stabilize its VEV at a smaller scale than the cutoff scale for a while. 
As the Hubble parameter decreases, 
only the lowest dimensional nonrenormalizable superpotential 
becomes relevant.
Then the GW signals are determined by the discussion in this section. 
Finally, the potential is dominated by the soft mass term 
and the GW emission terminates. 
We should emphasize that 
the power-law index of GW spectrum for high frequency modes 
around the peak frequency depends on the value of $n$ 
as shown in Fig.~\ref{schematic diagram1}. 
Thus, for larger wavenumber modes, 
the GW spectrum bends at the wavenumber corresponding to the 
horizon scale when the potential becomes dominated by the lowest dimensional nonrenormalizable superpotential. 
However, such a high frequency modes are beyond the reach of future GW detecters, 
so that we neglect such effects in Fig.~\ref{detectability1}.

\section{\label{W=0}Case B: $W=0$ with $U(1)$ symmetry}

In this section, we consider the case that 
the potential for the flat direction 
is given by a 
higher dimensional term of Eq.~(\ref{potential2}), 
which comes from 
a nonrenormalizable K\"{a}hler potential. 
It is given by 
\beq
 V_{\rm NR} (\phi) = 
 a_{H} H^2(t) \frac{\abs{\phi}^{2m-2}}{\M^{2m-4}}, 
 \label{potential2-2}
\eeq
where 
$m \ge 3$. 
The coefficient $a_{H}$ is given by 
\beq
 a_H = a_H' \lmk \frac{\M}{M_*} \rmk^{2m-2}, 
 \label{a_H}
\eeq
where we assume $a_H' = \mathcal{O}(1)$. 

The following discussion and calculation in this section are essentially the same as the ones 
given in the previous section. 
In the next subsection, we briefly explain the dynamics of the flat direction 
in the early Universe. 
We numerically follow its dynamics 
and calculate its GW signal, 
and the results are shown in Sec.~\ref{results2}. 
Then we give physical interpretation of the results in Sec.~\ref{interpretation2}. 
Finally, we discuss the detectability of GW signals in Sec.~\ref{prediction2}.

\subsection{Dynamics of the flat direction}

The whole potential for the flat direction $V(\phi)$ is given by 
\beq
 V(\phi) = m_\phi^2 \abs{\phi}^2 + c_H H^2(t) \abs{\phi}^2 + 
 a_H H^2(t) \frac{\abs{\phi}^{2m-2}}{\M^{2m-4}}, 
 \label{EOM without W}
\eeq
where we assume that $c_H$ is positive during inflation ($c_{H_{\rm inf}} >0$) and is negative ($c_{H_{\rm osc}} < 0$) after inflation. 
The positive $c_{H_{\rm inf}} $ makes the flat direction stay at the origin during inflation. 
Then, after inflation ends, the negative $c_{H_{\rm osc}}$ makes it obtain a large VEV. 
The potential takes the minimum at
\beq
 \la \abs{\phi} \ra = 
 \lmk \frac{\abs{c_{H_{\rm osc}}}}{a_{H} (m-1)} \rmk^{1/(2m-4)} \M \quad \lmk \sim M_* \rmk, 
 \label{VEV2}
\eeq
as long as $\abs{c_{H_{\rm osc}}} H^2(t) \gg \mphi^2$.
Here we have used $\abs{c_{H_{\rm osc}}} \sim \lmk \Mpl / M_* \rmk^2$ (see Eq.~(\ref{c_H})) 
and $a_H \sim \lmk \Mpl / M_* \rmk^{2m-2}$ (Eq.~(\ref{a_H0})).
The VEV is as large as the cutoff scale $M_*$. 
Note that 
the VEV of the flat direction
is independent of the Hubble parameter and is constant in time. 
This is an important difference from the previous section (see Eq.~(\ref{VEV1})).

Since the above potential possesses a global $U(1)$ symmetry, 
cosmic strings form after the end of inflation. 
By using numerical simulations (see the next subsection), 
we confirm that 
cosmic string network reaches a scaling regime well before cosmic strings disappear. 
In this regime, 
the number of cosmic strings in the Hubble volume is $\mathcal{O}(1)$ 
and the energy density of cosmic strings $\rho_{\rm cs}$ is given 
by Eq.~(\ref{rho_cs}). 
The energy density of cosmic strings per unit length $\mu$ is given by Eq.~(\ref{mu}). 
A typical width of cosmic strings is roughly given by the curvature of the potential, 
which is of the order of the Hubble length in the present case. 
This means that the width of cosmic strings increases with time. 
This is one of the outstanding characteristics of cosmic strings in this scenario 
compared with other cosmic strings considered in the literature. 

When the Hubble induced mass decreases down to the mass of the flat direction $m_\phi$, 
the flat direction starts to oscillate around the origin of the potential 
and cosmic strings disappear. 
The time when cosmic strings disappear is given by Eq.~(\ref{t_decay}) as in the previous section. 
For the case around $\mphi = \mathcal{O}(1)$ TeV, 
cosmic strings disappear much earlier than the BBN epoch, 
so that observational constraints on long-lived topological defects is not applicable to our scenario. 
As we explained in the previous section, 
we consider the well motivated case that 
the cosmic strings disappear before reheating completes.

After they form at the end of inflation 
and before they disappear at $t \simeq t_{\rm decay}$, 
cosmic strings emit GWs. 
We numerically follow the evolution of the flat direction 
and calculated its GW signal. 
The results are shown in the next subsection.

\subsection{\label{results2}Results of numerical simulations}

In this subsection, we show our results of numerical simulations for 
the dynamics of the flat direction and its GW signal.

By redefining the variables, 
we rewrite the equation of motion such as 
\beq
 \cphi'' - \frac{\del^2 \cphi}{\del \widetilde{ {\bm x}}^2} + \lmk c_{H_{\rm osc}} - \half \rmk \lmk \frac{\widetilde{\tau}_0}{\widetilde{\tau}} \rmk^2 \cphi 
 + \lmk \frac{\widetilde{\tau}}{\widetilde{\tau}_0} \rmk^4 \widetilde{m}_\phi^2 \cphi + (m-1) \lmk \frac{\widetilde{\tau}_0}{\widetilde{\tau}} \rmk^{4m-6} 
 \abs{\cphi}^{2m-4} \cphi =0, 
 \label{EOM rewritten}
\eeq
where 
\beq
 \label{cphi rewritten}
 \cphi &\equiv& a_H^{1/(2m-4)} \M^{-1} a(t) \phi \nonumber \\
 &=& \lmk \abs{c_{H_{\rm osc}}} / (m-1) \rmk^{1/(2m-4)} a(t) \phi /  \la \abs{\phi} \ra, \\
 \label{tau rewritten}
 \dd \widetilde{\tau} &\equiv& H_i \dd t / a(t),  \\
 \label{mphi rewritten}
 \widetilde{m}_\phi &\equiv& m_\phi / H_i \quad (\ll 1), \\
 \label{x rewritten}
 \widetilde{\bm x} &\equiv& H_i {\bm x}. 
\eeq 
The parameter $a_H$ can be absorbed by the redefinition of $\phi$, 
so that our numerical simulation can be applied to any case with an arbitrary value of $a_H$.

We follow the time evolution of the flat direction by solving the equation 
of motion Eq.~(\ref{EOM rewritten}) numerically. 
Details of the numerical method are explained in the previous section. 
We confirm that each Hubble volume contains 
$\mathcal{O}(1)$ cosmic strings  
and a typical width of cosmic strings is roughly proportional to the Hubble radius. 
From our numerical simulation, we find that 
the spatially averaged magnitude of $\phi$ grows with time and 
eventually reaches of the order of the VEV (Eq.~(\ref{VEV2}))
around $\tau H_i = \widetilde{\tau}_{\rm form}  \simeq 10$ for the case of $c_{H_{\rm osc}} = -15$.

\begin{figure}[t]
\centering 
\begin{tabular}{l l}
\includegraphics[width=.40\textwidth, bb=0 0 360 349]{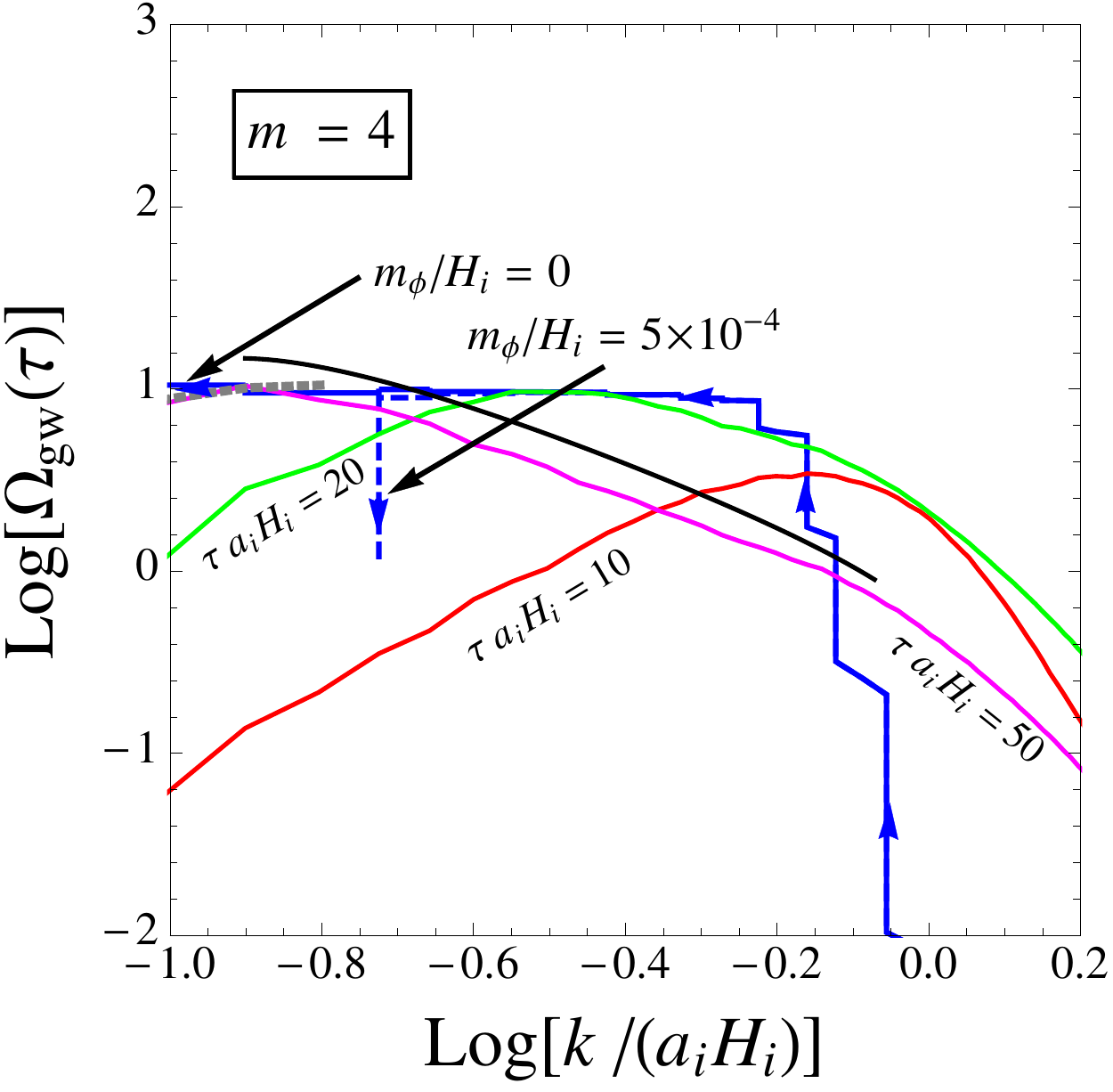} & \quad 
\includegraphics[width=.40\textwidth, bb=0 0 360 349]{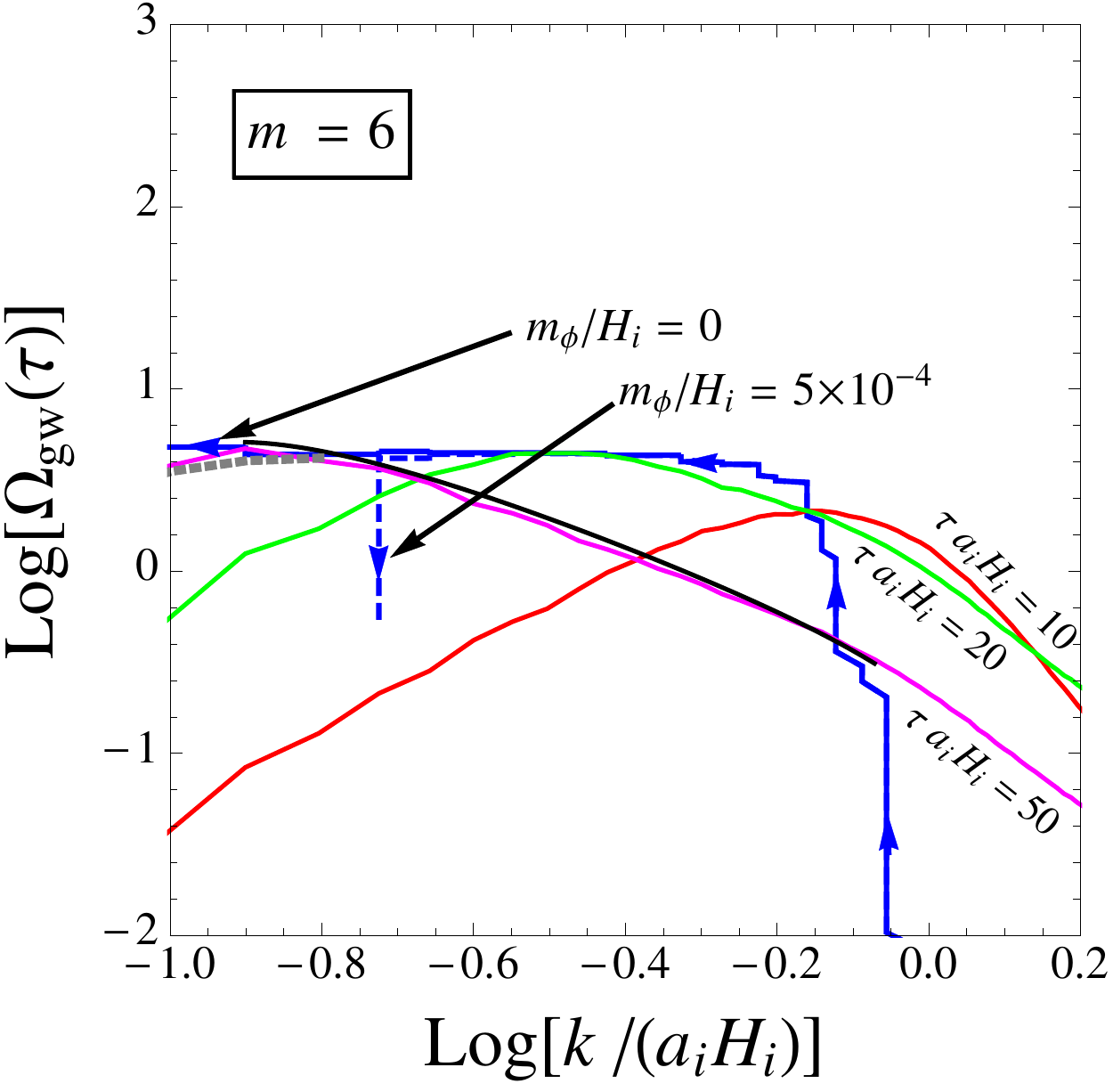} \vspace{0.5cm}\\
\includegraphics[width=.40\textwidth, bb=0 0 360 349]{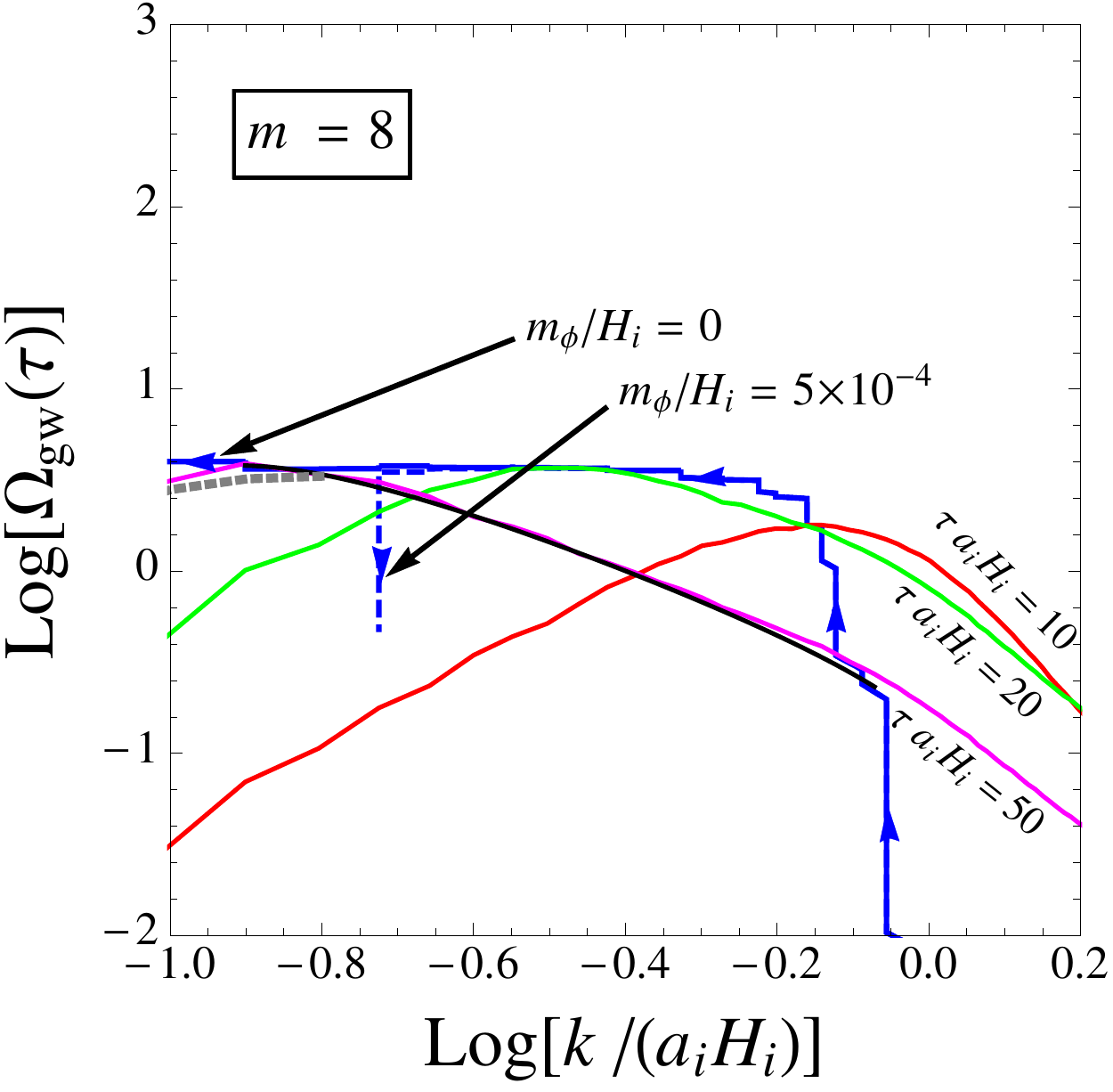} & \quad 
\includegraphics[width=.40\textwidth, bb=0 0 360 349]{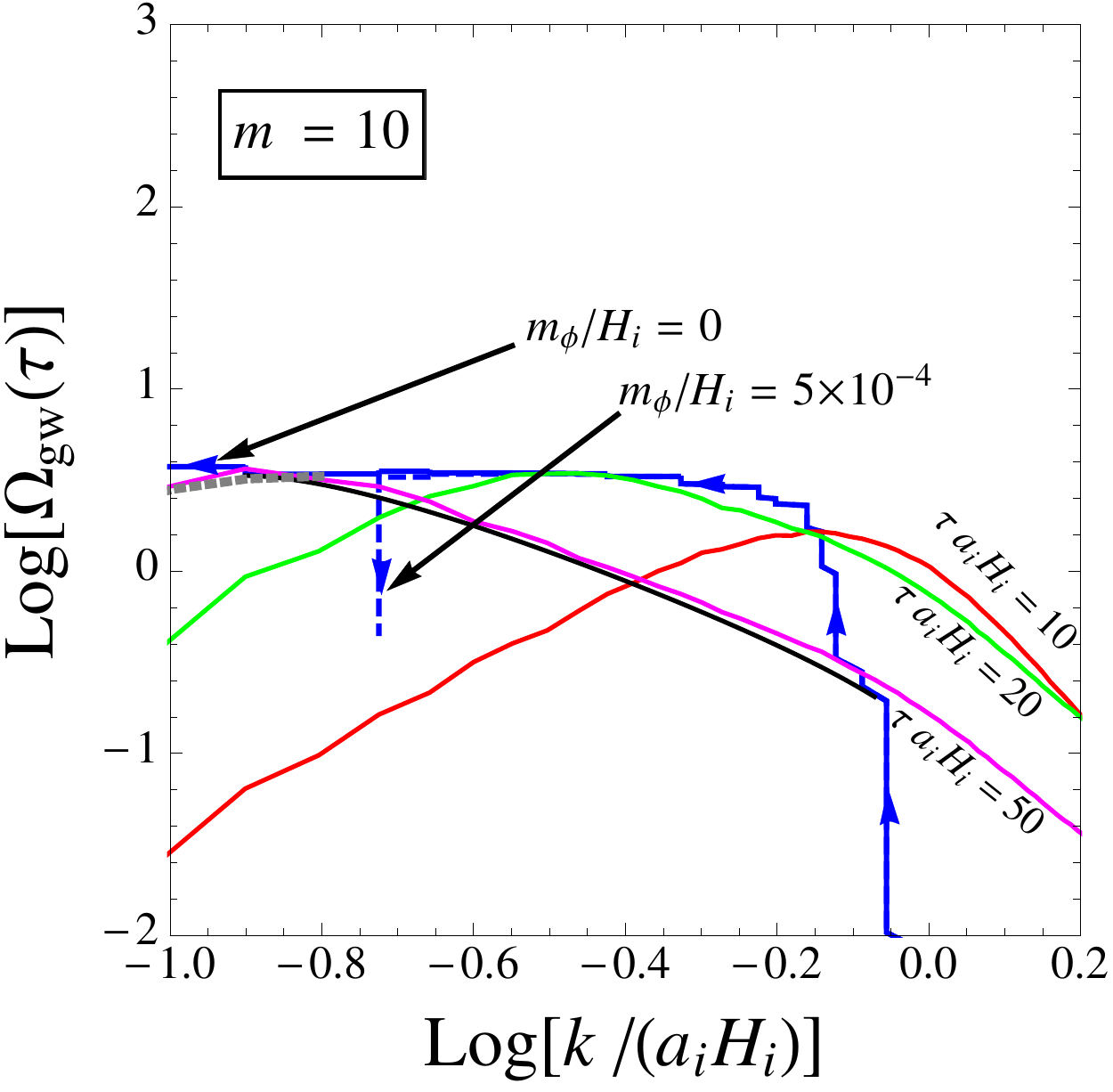} 
\end{tabular}
\caption{
GW spectra obtained from our numerical simulations. 
The energy density of GWs are rescaled by $a_H^{-2/(m-2)}$. 
The red, green, and magenta curves are GW spectra at the conformal time of 
$\tau a_i H_i = 10, 20,$ and $50$, respectively. 
We also show the time evolution of the GW peak location as the blue solid (dashed) line 
for the case of $m_\phi / H_i = 0$ 
($5 \times 10^{-4}$). 
We take $m=4$ (upper left panel), $6$ (upper right panel), $8$ (lower left panel), 
and $10$ (lower right panel), fixing $c_{H_{\rm osc}} = -15$. 
The black curves are plotted by Eq.~(\ref{large scale2}) at $\tau a_i H_i = 50$
and well describes large wavenumber modes. 
The gray dotted curves are estimated 
from the discussion of causality at $\tau a_i H_i = 50$
and well describes small wavenumber modes. 
}
  \label{GWspectrum wo W}
\end{figure}

We calculate the transverse-traceless part of the anisotropic stress from the numerical simulation 
and obtain the energy density of GWs using the equations in Appendix~\ref{calculation}. 
The results are plotted in Fig.~\ref{GWspectrum wo W}. 
One can find that 
the GW peak energy density is almost constant in time and is almost independent of $m$. 
We give physical interpretation of these results in the next subsection.

Taking a non-zero value of $m_\phi$ into account, 
we find that the flat direction starts to oscillate around the origin of the potential 
and cosmic strings disappear 
around $\widetilde{\tau} = \widetilde{\tau}_{\rm decay} \simeq 40$, 
which is given by Eq.~(\ref{tau_decay}) with $m_\phi/H_i = 5 \times 10^{-4}$ and $c_{H_{\rm osc}} = -15$. 
After that, 
the GW peak wavenumber becomes constant (blue dashed line in Fig.~\ref{GWspectrum wo W}) 
and the GW spectrum 
decrease adiabatically 
as $\propto \tau^{-2}$ ($\propto a^{-1}$). 
As we discussed in Sec.\,\ref{results1}, the above choice of $m_\phi/H_i$ is convenient for numerical simulations, 
though it does not change the dynamics of flat direction until $\widetilde{\tau} \simeq \widetilde{\tau}_{\rm decay}$. 
In reality, it can be many orders of magnitude smaller depending on inflation models.

\subsection{\label{interpretation2}Physical interpretations}

In this subsection, we discuss how we can understand the results of our numerical simulations 
shown in Fig.~\ref{GWspectrum wo W}.

Since during the scaling regime the typical length scale of cosmic string dynamics is the Hubble length, 
the GW emission peak wavenumber is typically given by the Hubble scale like Eq.~(\ref{peak0}). 
The produced energy density of GWs is estimated as Eq.~(\ref{omega_gw2}) 
but with the VEV given by Eq.~(\ref{VEV2}) in this case: 
\beq
 \lkk \frac{\Delta \Omega_{\rm gw} }{\Delta \log \tau} \rkk_{\rm peak}
 \sim 
 \lmk \frac{\la \phi \ra}{\M} \rmk^4 
\simeq 
 \lmk \frac{\abs{c_{H_{\rm osc}}}}{a_H (m-1)} \rmk^{2/(m-2)} 
 \sim \lmk \frac{M_*}{\Mpl} \rmk^4. 
 \label{omega_gw5}
\eeq 
Cosmic strings emit GWs with a constant energy density expressed above.

\begin{figure}[t]
\centering 
\begin{tabular}{l l}
\includegraphics[width=.40\textwidth, bb=0 0 360 351]{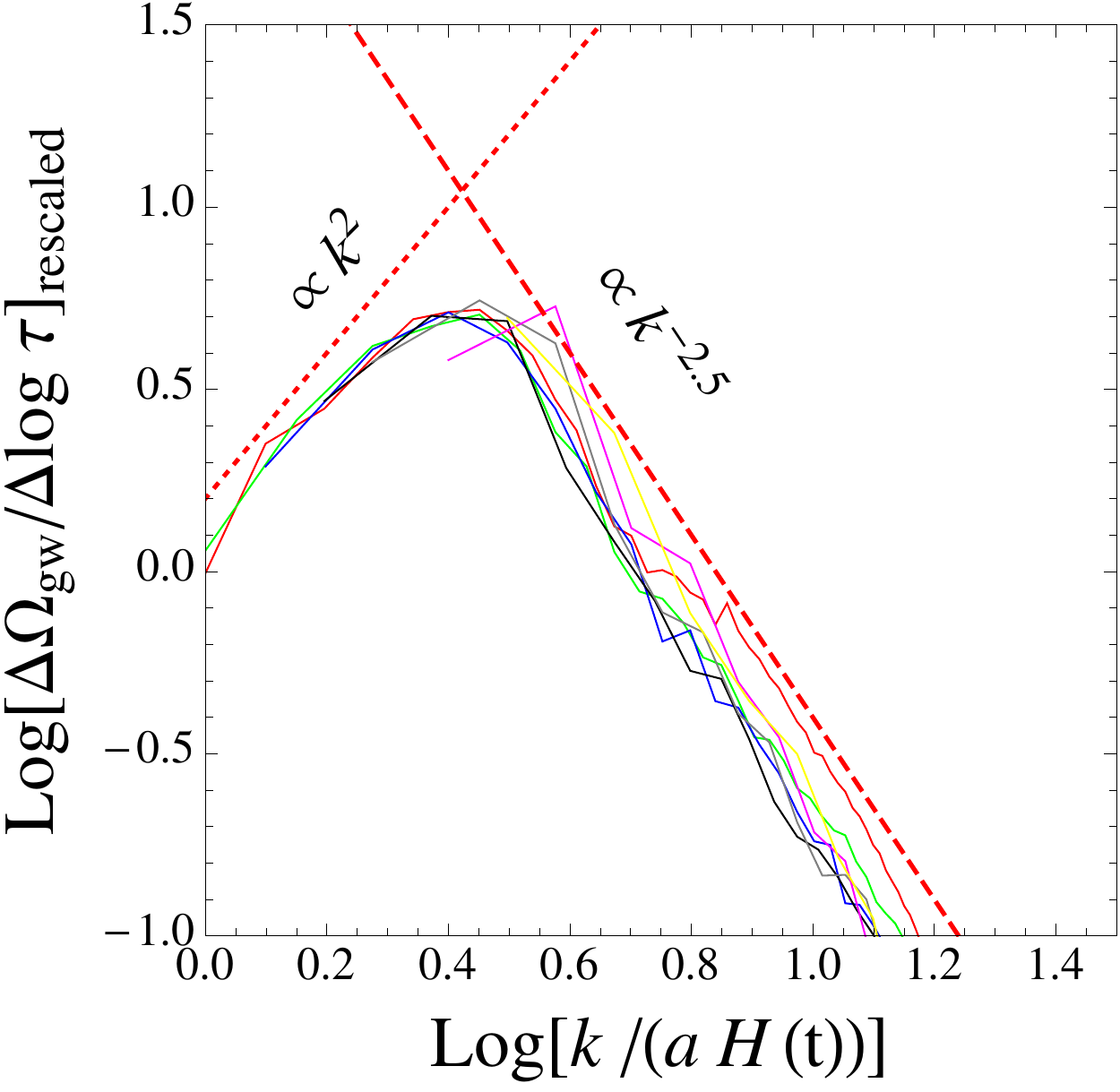} & \quad 
\includegraphics[width=.40\textwidth, bb=0 0 360 351]{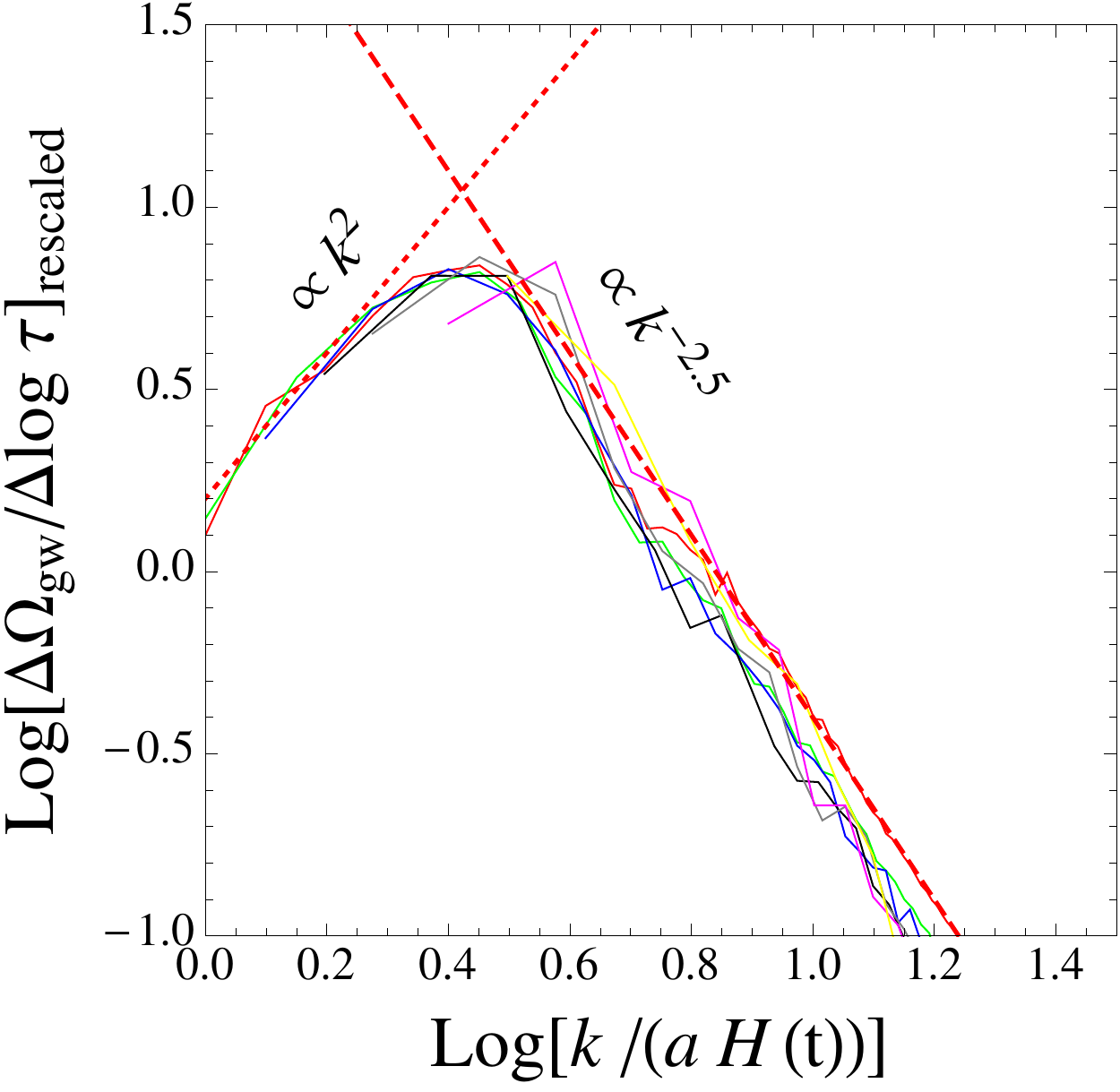} \vspace{0.5cm}\\
\includegraphics[width=.40\textwidth, bb=0 0 360 351]{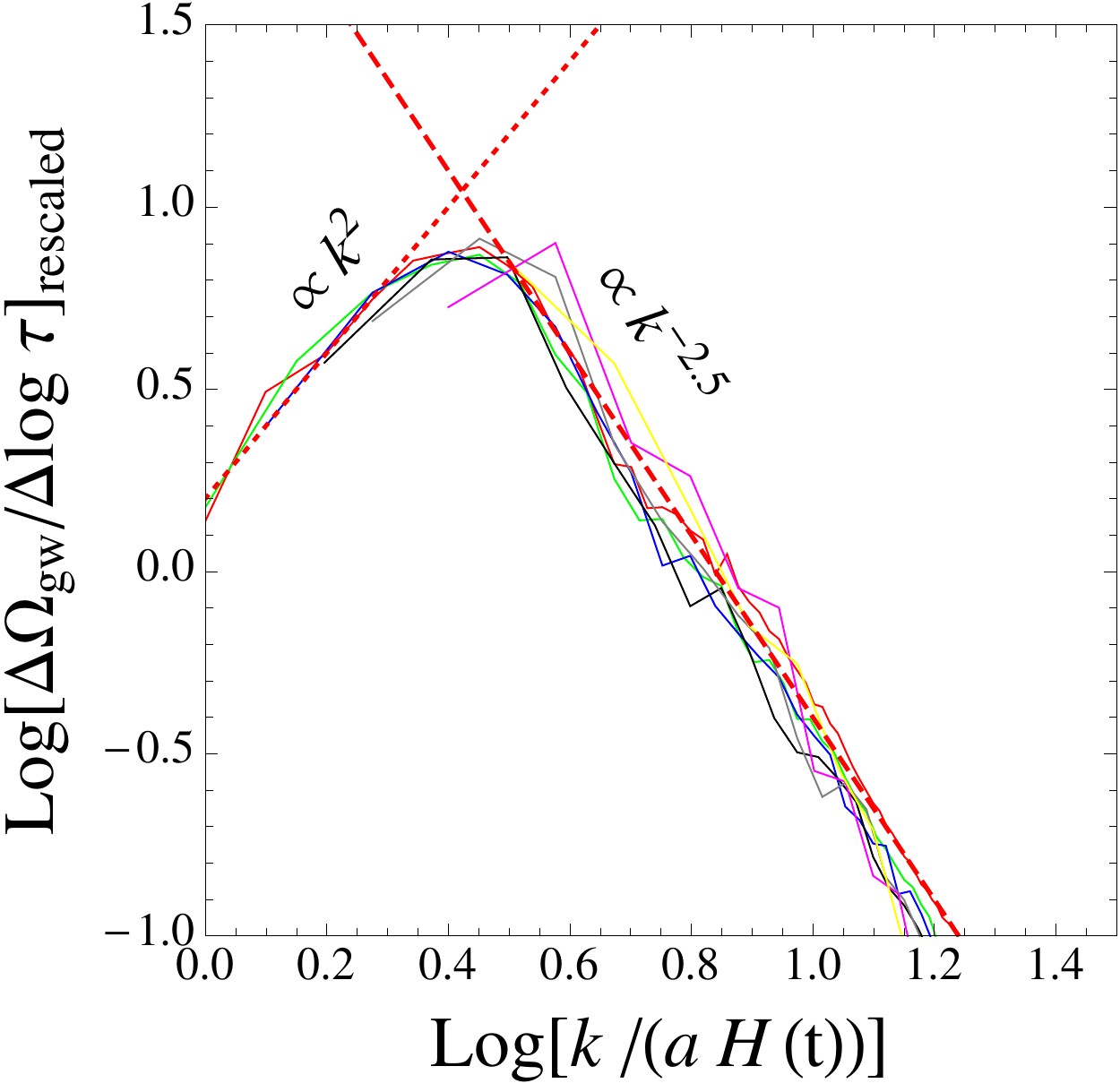} & \quad 
\includegraphics[width=.40\textwidth, bb=0 0 360 351]{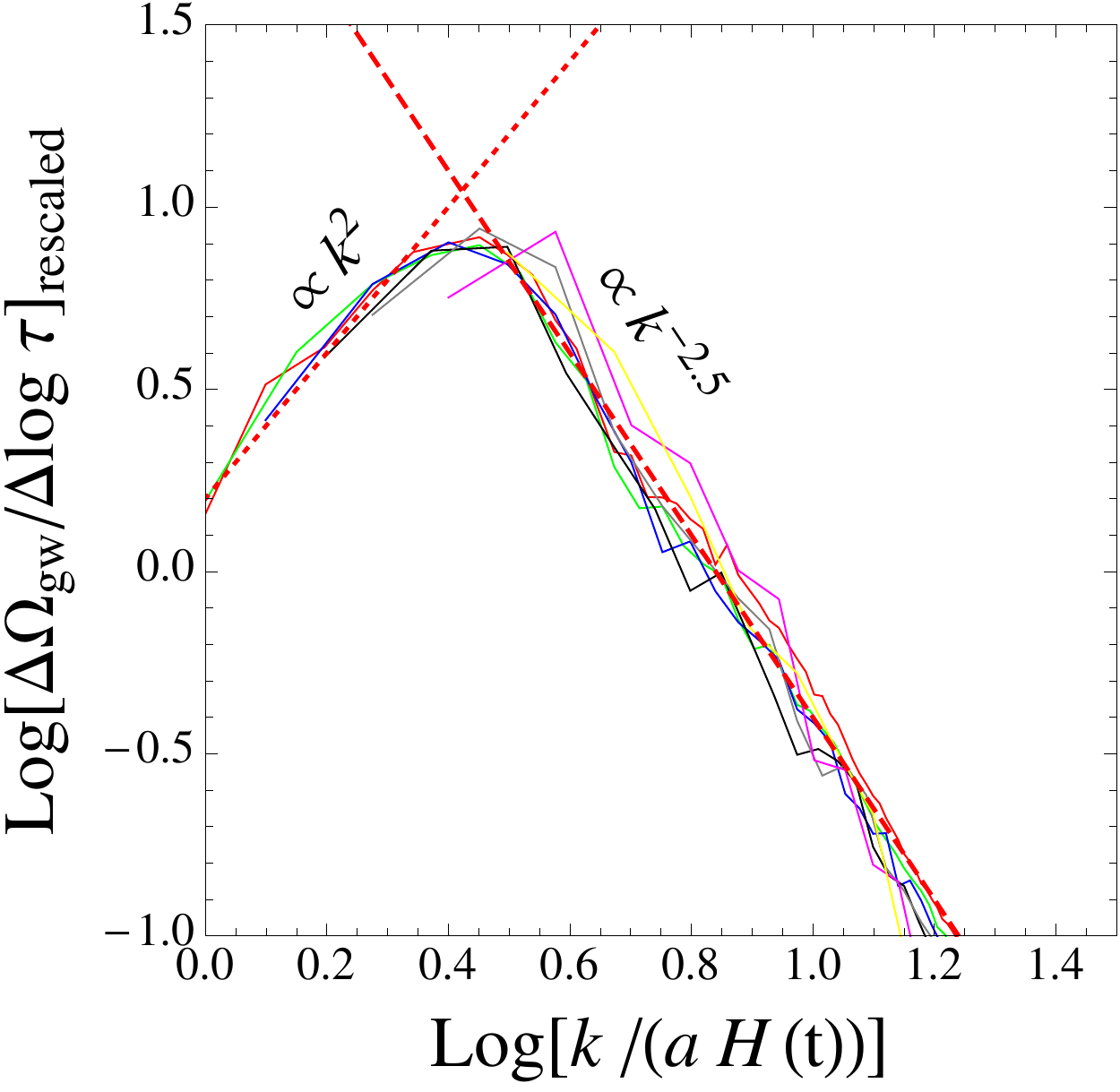} 
\end{tabular}
\caption{
Derivatives for GW spectra in terms of the conformal time. 
The energy density of GWs are rescaled by the factor given by Eq.~(\ref{omega_gw5}). 
We plot GW spectra for $\tau a_i H_i = 20, 30, 40, 50, 60, 80,$ and $100$. 
We take $m=4$ (upper left panel), $6$ (upper right panel), $8$ (lower left panel), 
and $10$ (lower right panel), fixing $c_{H_{\rm osc}} = -15$. 
}
  \label{del omega2}
\end{figure}

We plot our numerical results of $\Delta \Omega_{\rm gw} / \Delta \log \tau$ 
in Fig.~\ref{del omega2}. 
It shows that the produced energy density of GWs is constant in time as expected. 
We can determine the numerical prefactors for 
the emission peak wavenumber and the produced energy density of GWs 
from the numerical results 
like 
\beq
 \frac{k^{\Delta}_{\rm peak} }{a(t)} &\simeq& 2.5 H(t), 
 \label{peak3} 
 \\
 \lkk \frac{\Delta \Omega_{\rm gw} }{\Delta \log \tau} \rkk_{\rm peak} 
 &\simeq& 10 \lmk \frac{\abs{c_{H_{\rm osc}}}}{a_H (m-1)} \rmk^{2/(m-2)}. 
 \label{GW amplitude3}
\eeq 
We find that the numerical prefactors are almost independent of $m$ and $\tau$ 
up to by a factor of two.

Figure~\ref{del omega2} shows that 
$\Delta \Omega_{\rm gw} / \Delta \log \tau$ is proportional to 
a certain power of $k$ for large and small wavenumber modes like Eq.~(\ref{large wavenumber}), 
where 
$\alpha \simeq 2.5$ in this case. 
The GW spectra for large wavenumber modes ($k>k^{\Delta}_{\rm peak}(\tau)$)
in Fig.~\ref{GWspectrum wo W} 
can be understood by an integration like Eq.~(\ref{large scale}), 
where we can effectively take 
$n \to \infty$ in this case. 
Thus, we obtain 
\beq
 && \Omega_{\rm gw} \lmk \tau \rmk \\
 &=& 
 \frac{1}{4} 
 \lkk \frac{\Delta \Omega_{\rm gw} (\tau)}{ \Delta \log \tau } \rkk_{\rm peak} 
 \lmk \frac{k_{\rm peak} (\tau)}{k} \rmk^2 
 \lkk 1 - \lmk \frac{k}{k_{\rm peak} (\tau_{\rm form})} \rmk^{4} \rkk \\ 
 &&+ 
 \lmk \alpha - 2 \rmk^{-1} 
 \lkk \frac{\Delta \Omega_{\rm gw} (\tau)}{ \Delta \log \tau } \rkk_{\rm peak} 
 \lmk \frac{k_{\rm peak} (\tau)}{k} \rmk^2 
 \lkk 1 - \lmk \frac{k_{\rm peak} (\tau)}{k} \rmk^{\alpha - 2} \rkk, 
 \label{large scale2}
\eeq
where $\tau_k$ is defined by 
$k_{\rm peak} (\tau_k) = k$. 

The result of Eq.~(\ref{large scale2}) at $\widetilde{\tau} = 50$ is plotted as the black curves 
in Fig.~\ref{GWspectrum wo W}. 
It is consistent with the large-wavenumber behaviour of GW spectrum found in our numerical simulations,
though it is larger than our result by a factor of 2 in the case of $m=4$. 
Note that 
the GW spectrum is expected to be proportional to $k^{-2}$ 
for $k^{\Delta}_{\rm peak} (\tau_{\rm form}) \gg k \gg k^{\Delta}_{\rm peak} (\tau)$. 
This asymptotic behavior is different from the one obtained in the previous section 
because the VEV is constant in time in the present case. 
The gray dotted curves in Fig.~\ref{GWspectrum wo W} show that 
simulated super-horizon modes are consistent with the 
spectrum expected from the discussion of causality~\cite{Dufaux:2007pt, Kawasaki:2011vv} 
as explained in the previous section. 

The blue dashed lines in Fig.~\ref{GWspectrum wo W} 
show the evolution of GW peak wavenumber 
in the case of $m_\phi/H_i = 5 \times 10^{-4}$. 
Since the flat direction starts to oscillate around the origin 
due to its soft mass term, 
cosmic strings disappear 
at $\tau \simeq \tau_{\rm decay}$. 
The GW peak energy density and frequency at that time 
is given by 
\beq
 \lkk \Omega_{\rm gw} (\tau_{\rm decay}) \rkk_{\rm peak} &\simeq& 
 10 \frac{1}{4} \lmk \frac{\abs{c_{H_{\rm osc}}}}{a_H (m-1)} \rmk^{2/(m-2)}, 
  \label{omega_gw7}
  \\
  \frac{k_{\rm peak}}{a(\tau_{\rm decay})} &\simeq& 3 \frac{\mphi}{\sqrt{\abs{c_{H_{\rm osc}}}}}, 
\eeq
where the numerical prefactors are determined from our numerical results. 
The shape of GW spectrum is in fact freezed out
around $\widetilde{\tau} = \widetilde{\tau}_{\rm decay} \simeq40$, 
which is given by Eq.~(\ref{tau_decay}) with $m_\phi/H_i = 5 \times 10^{-4}$ and $c_{H_{\rm osc}} = -15$.
After that, the peak wavenumber does not change and 
the GW energy density decreases as $\propto a^{-1}$ 
until reheating completes. 

In summary, the schematic form of resulting GW spectrum is 
similar to the one described in Fig.~\ref{schematic diagram1}, 
except for the power-law index of large wavenumber modes. 
In the case of this section, 
the GW spectrum for large wavenumber modes is proportional to $k^{-2}$ 
and is independent of $m$. 
This is because the potential minimum for the flat direction 
is constant in time. 
Thus, in principle,
we can distinguish the cases with and without 
the higher dimensional superpotential through the observation of the GW spectrum 
with higher frequencies.

\subsection{\label{prediction2}GW signals at present}

The above calculations and discussions 
determine the GW spectrum during the oscillation dominated era. 
In this subsection, we derive 
the GW spectrum at the present epoch, 
taking into account the redshift.

The peak energy density of GWs at $\tau=\tau_{\rm decay}$,
$\lkk \Omega_{\rm gw} (\tau_{\rm decay}) \rkk_{\rm peak}$ 
is given by Eq.~(\ref{omega_gw7}) from our numerical simulations. 
Substituting $\lkk \Omega_{\rm gw} (\tau_{\rm decay}) \rkk_{\rm peak}$  into Eq.~(\ref{present energy density}), 
we obtain 
the present value of GW peak energy density such as
\beq
 \lkk \Omega_{\rm gw} h^2 (t_0) \rkk_{\rm peak}
 \simeq 
 5 \times 10^{-7} 
 \lmk \frac{\abs{c_{H_{\rm osc}}}^{-1/2} \mphi}{10^3 \GeV} \rmk^{-2/3}
 \lmk \frac{T_{\rm RH}}{10^9 \GeV} \rmk^{4/3}
 \lmk \frac{\abs{c_{H_{\rm osc}}}}{a_H (m-1)} \rmk^{2/(m-2)}, 
 \label{result omega2}
\eeq
where we use 
Eqs.~(\ref{t_decay}), (\ref{T_RH}), and (\ref{omega_gw7}). 
From Eq.~(\ref{present peak frequency}), 
the GW peak frequency at the present epoch 
is given by 
\beq
 f_0 
 \simeq 700 \text{ Hz} \lmk \frac{\abs{c_{H_{\rm osc}}}^{-1/2} \mphi}{10^3 \GeV} \rmk^{1/3} \lmk \frac{\TR}{10^9 \GeV} \rmk^{1/3}. 
 \label{result f2}
\eeq
Note that the peak frequency is independent of $m$ and $a_H$.

Since the GW spectrum is sensitive to what dominates the energy density of the Universe, 
it contains information of reheating temperature 
as explained in the previous section. 
In fact, the GW spectrum bends at 
the frequency given by Eq.~(\ref{f_bend}) 
corresponding to the wavenumber around $k \simeq \tau_{\rm RH}^{-1}$~\cite{Seto:2003kc, 
Nakayama:2008ip, previous work}.

Since there are three observables: 
Eqs.~ (\ref{f_bend}), (\ref{result omega2}) (see also Eq.~(\ref{VEV2})), and (\ref{result f2}), 
in principle, we can obtain three parameters through observation of GW spectrum: 
$\abs{c_{H_{\rm osc}}}^{-1/2} \mphi$, $\TR$, and $M_*/\Mpl$ (up to by a factor of order one).
Again, let us stress that 
we can distinguish 
the case of $W \ne 0$ in the previous section 
from the case of $W=0$ in this section.
This is 
because for high frequencies 
the GW spectrum scales as $\Omega_{\rm gw} \propto f^{(16-2n)/(n-2)}$ 
($\ne f^{-2}$) 
for the former case 
and $\Omega_{\rm gw} \propto f^{-2}$ for the latter case.

If the cutoff scale was the Planck scale ($M_* \simeq \M$), 
we could obtain the mass of the flat direction $\mphi$ and the reheating temperature $\TR$ 
from observations of GWs (see Eqs.~(\ref{f_bend}) and (\ref{result f2})). 
In this case, however, 
we can not ignore 
the effect of the backreaction of GW emission 
when we calculate the spectrum of GWs. 
In addition, the gradient energy density of the flat direction is comparable to 
the energy density of the Universe 
and thus it may spoil the isotropic evolution of the Universe. 
We can neglect these effects 
when the cutoff scale is less than the Planck scale.

\begin{figure}[t]
\centering 
\includegraphics[width=.45\textwidth, bb=0 0 360 339]{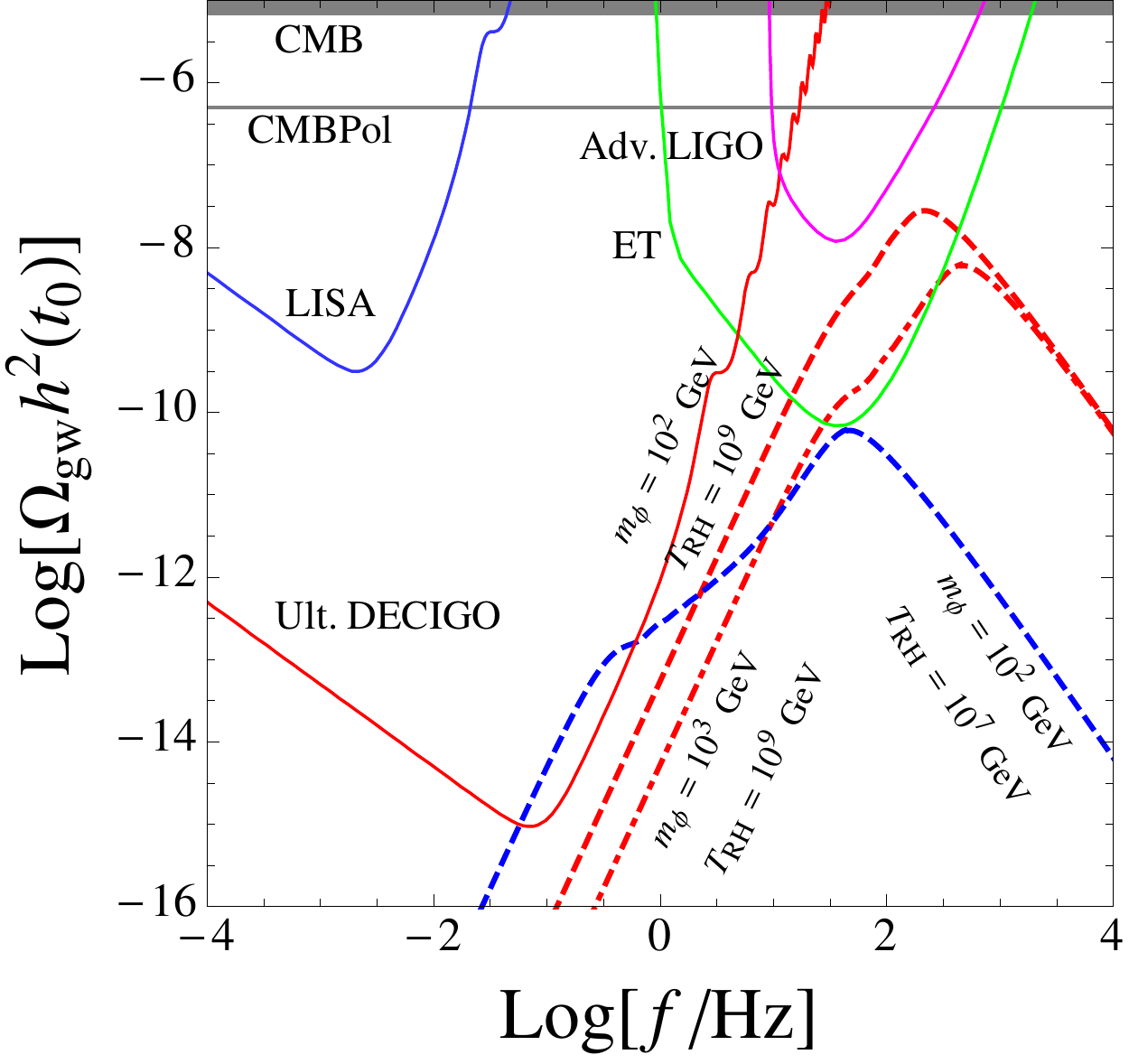} 
\caption{
GW spectra generated by cosmic strings 
(dash and dot-dashed curves) 
and sensitivities of planned interferometric detectors (solid curves). 
We assume
$\mphi = 10^2 \GeV$ 
with $\TR = 10^7 \GeV$ (blue dashed curve), 
$\mphi = 10^2 \GeV$ 
with $\TR = 10^9 \GeV$ (red dashed curve), 
and 
$\mphi = 10^3 \GeV$ 
with $\TR = 10^9 \GeV$ (red dot-dashed curve). 
We also assume $c_{H_{\rm osc}} = -15$, $a_H' = 1$, and $m=4$. 
We set the value of $M_*$ as $M_*^2 = \Mpl^2 / 10$. 
}
  \label{detectability2}
\end{figure}

Figure~\ref{detectability2} shows examples of GW spectrum 
predicted by the present mechanism. 
The GW peak energy density and peak frequency are 
given by Eqs.~(\ref{result omega2}) and (\ref{result f2}), which are obtained from our numerical results. 
While the shape of the GW spectrum for low frequency modes are calculated 
from Eqs.~(\ref{A}) and (\ref{B}) with a constant $\TT$ and $\tau_f = \tau_{\rm decay}$, 
that for large wavenumber modes are calculated from Eq.~(\ref{large scale2}). 
In the figure, we take 
$\mphi = 10^2 \GeV$ 
with $\TR = 10^7 \GeV$ (blue dashed curve), 
$\mphi = 10^2 \GeV$ 
with $\TR = 10^9 \GeV$ (red dashed curve), 
and 
$\mphi = 10^3 \GeV$ 
with $\TR = 10^9 \GeV$ (red dot-dashed curve). 
We could obtain rough scales of them from the observation of GW spectrum. 
To discuss the detectability of GW signals, 
we plot sensitivities of planned interferometric detectors.
The sensitivity curves are the same as in Fig.~\ref{detectability1},
and details are given in Sec.~\ref{prediction1}.
We find that 
GW signals would be observed by ET or Ultimate DECIGO.

\section{\label{W=0 without U(1)}Case C: $W=0$ without $U(1)$ symmetry}

In this section, we consider the same potential with the one considered in the previous section, 
Eq.~(\ref{potential2-2}), 
except for an additional $U(1)$ breaking term 
coming from higher dimensional \Kahler potentials like 
\beq
 V_{U(1) br} = -  b_H H^2(t) \lmk \frac{\phi^{m'}}{\M^{m'-2}} + {\rm c.c}. \rmk , 
 \label{potential3-2}
\eeq
where $m' \ge 2$. 
We assume $m' < 2m-2$ so that 
the potential for the flat direction is bounded from below. 
The coefficient $b_H$ is given by 
\beq
 b_H = b'_H \lmk \frac{\Mpl}{M_*} \rmk^{m'}, 
 \label{b_H}
\eeq
where we assume $b'_H = \mathcal{O}(1)$. 
Here we have redefined the phase of the field $\phi$ such that $b'_H$ is postitive.
This additional term explicitly breaks $U(1)$ symmetry, 
but there remains $Z_{m'}$ symmetry in that potential.

The following discussion and calculation in this section are essentially the same as the ones 
explained in the previous two sections, 
though it is more complicated due to the additional term breaking $U(1)$ symmetry. 
In the next subsection, we briefly explain the dynamics of the flat direction 
in the early Universe. 
We numerically follow the evolution of its dynamics 
and its GW signal, 
and the results are shown in Sec.~\ref{results3}. 
Then we give physical interpretation of the results in Sec.~\ref{interpretation3}. 
Finally, we discuss the detectability of GW signals in Sec.~\ref{prediction3}.

\subsection{Dynamics of the flat direction}

Throughout this section, we consider 
the following potential for the flat direction: 
\beq
 V(\phi) = m_\phi^2 \abs{\phi}^2 + c_H H^2(t) \abs{\phi}^2 + 
 a_H H^2(t) \frac{\abs{\phi}^{2m-2}}{\M^{2m-4}} 
 - b_H H^2(t) \lmk \frac{\phi^{m'}}{\M^{m'-2}} + {\rm c.c.} \rmk, 
 \label{EOM without W2}
\eeq
where we assume that $c_H$ $(=c_{H_{\rm inf}}) >0$ during inflation and $c_H$ $(= c_{H_{\rm osc}}) <0$ after inflation. 
The positive $c_{H_{\rm inf}}$ makes 
the flat direction stay at the origin during inflation. 
After inflation ends, 
the flat direction obtains a large VEV due to the negative $c_{H_{\rm osc}}$. 
Note that the $U(1)$ breaking term unstabilizes the potential for the flat direction around the cutoff scale. 
For a small $b_H$, the potential minimum is determined by 
the Hubble induced mass term as in the previous section (see Eq.~(\ref{VEV2})), 
while for a small $c_{H_{\rm inf}}$, 
it is determined by the $U(1)$ breaking term. 
Thus, the VEV can be estimated as 
\beq
 \la \phi \ra &\equiv& 
 C e^{2 p \pi i/ m'} \M, \\
 C &\simeq& 
 \Max \lkk 
 \lmk \frac{\abs{c_{H_{\rm osc}}}}{a_{H} (m-1)} \rmk^{1/(2m-4)}, 
 \lmk \frac{m'}{m-1} \frac{b_H}{a_H} \rmk^{1/(2m-m'-2)} 
 \rkk \\
 &\sim& \frac{M_*}{\Mpl}, 
 \label{VEV3}
\eeq
where $p = 0, 1, 2, \dots, m'-1$. 
In the last line, we use 
Eqs.~(\ref{c_H}), (\ref{a_H0}), and (\ref{b_H0}). 
One can see that the VEV is as large as the cutoff scale $M_*$.

Since the nonzero VEV of the flat direction breaks $Z_{m'}$ symmetry, 
$m'$ types of domain walls form after the end of inflation. 
In general, 
a typical width of domain walls is given by the inverse of the curvature of the potential 
at the minimum. 
In the present case, it is given by the inverse of the mass of the phase direction $m_\theta$, 
which is given by 
\beq
 m_\theta^2 \simeq (m')^2 b_H H^2(t) C^{m'-2} \ \lmk \sim \frac{\Mpl^2}{M_*^2} H^2(t) \rmk. 
\eeq
The energy of domain walls per unit area $\sigma$ can be estimated by 
\beq
 \sigma &\sim& m_\theta \lmk \frac{\abs{\la \phi \ra}}{m'} \rmk^2 \\
 &\simeq& (m')^{-1} \sqrt{b_H} H (t) C^{(m'+2)/2} \Mpl^2
 \ \lmk \sim M_* \Mpl H(t) \rmk. 
\eeq
These mean that the width of domain walls increases 
and the energy per unit area decreases with time. 
These are outstanding characteristics of domain walls in this scenario 
compared with other domain walls considered in the literature. 
By using numerical simulations (see the next subsection), 
we confirm that 
domain wall system soon reaches the scaling regime, 
in which the number of domain walls in the Hubble volume is of the order of the domain wall number $m'$. 
In this regime, 
the energy density of domain walls is roughly given by 
\beq
 \rho_{\rm DW} \sim m' \times \sigma \times H^{-2} \times H^3 
 \ \lmk \sim M_* \Mpl H^2 (t) \rmk.
 \label{rho_DW}
\eeq
We should emphasize that 
the energy density of domain walls never dominate that of the Universe. 
This is because the curvature of the potential is given by $H(t)$ 
and decreases with time as $\propto a^{-3/2}$, 
which is completely different from the ordinary domain walls~\cite{Zeldovich:1974uw}. 

When the Hubble parameter decreases down to the mass of the flat direction $m_\phi$, 
the flat direction starts to oscillate around the origin of the potential 
and domain walls disappear. 
Since domain walls disappear well before the BBN epoch, 
our scenario is free from stringent constraints on long-lived domain walls~\cite{Zeldovich:1974uw}. 
As we explained in Sec.~\ref{dynamics1}, 
we consider the well motivated case that 
the domain walls disappear before reheating completes.

After they form at the end of inflation 
and before they disappear at $t \simeq t_{\rm decay}$, 
domain walls emit GWs. 
In the subsequent subsections, we 
show our numerical results of domain wall dynamics 
and spectra of GWs emitted from these domain walls.

\subsection{\label{results3}Results of numerical simulations}

In this subsection, we show our results of numerical simulations for 
the dynamics of the flat direction and its GW signal.

We redefine the variables in the same way as in the previous subsection (see Eqs.~(\ref{cphi rewritten})-(\ref{x rewritten}))
and obtain the equation of motion equivalent to Eq.~(\ref{EOM rewritten}) 
except for the additional $U(1)$ breaking term: 
\beq
 && \cphi'' - \nabla^2 \cphi + \lmk c_{H_{\rm osc}} - \half \rmk \lmk \frac{\widetilde{\tau}_i}{\widetilde{\tau}} \rmk^2 \cphi 
 + \lmk \frac{\widetilde{\tau}}{\widetilde{\tau}_i} \rmk^4 \widetilde{m}_\phi^2 \cphi \nonumber \\
 &&\quad + (m-1) \lmk \frac{\widetilde{\tau}_i}{\widetilde{\tau}} \rmk^{4m-6} 
 \abs{\cphi}^{2m-4} \cphi 
 + \frac{m'}{2} \widetilde{b}_H \lmk \frac{\widetilde{\tau}_i}{\widetilde{\tau}} \rmk^{2m'-2} (\cphi^*)^{m'-1} = 0. 
 \label{EOM rewritten2}
\eeq
The parameter $a_H$ can be absorbed by the redefinition of $\phi$ and $b_H$, 
so that our numerical simulation depends on $a_H$ and $b_H$ 
only though the combination of $\widetilde{b}_H$ defined by
\beq
 \widetilde{b}_H \equiv b_H a_H^{-(m'-2) / (2m-4)}. 
\eeq

\begin{figure}[t]
\centering 
\begin{tabular}{l l}
\includegraphics[width=.40\textwidth, bb=0 0 947 1028]{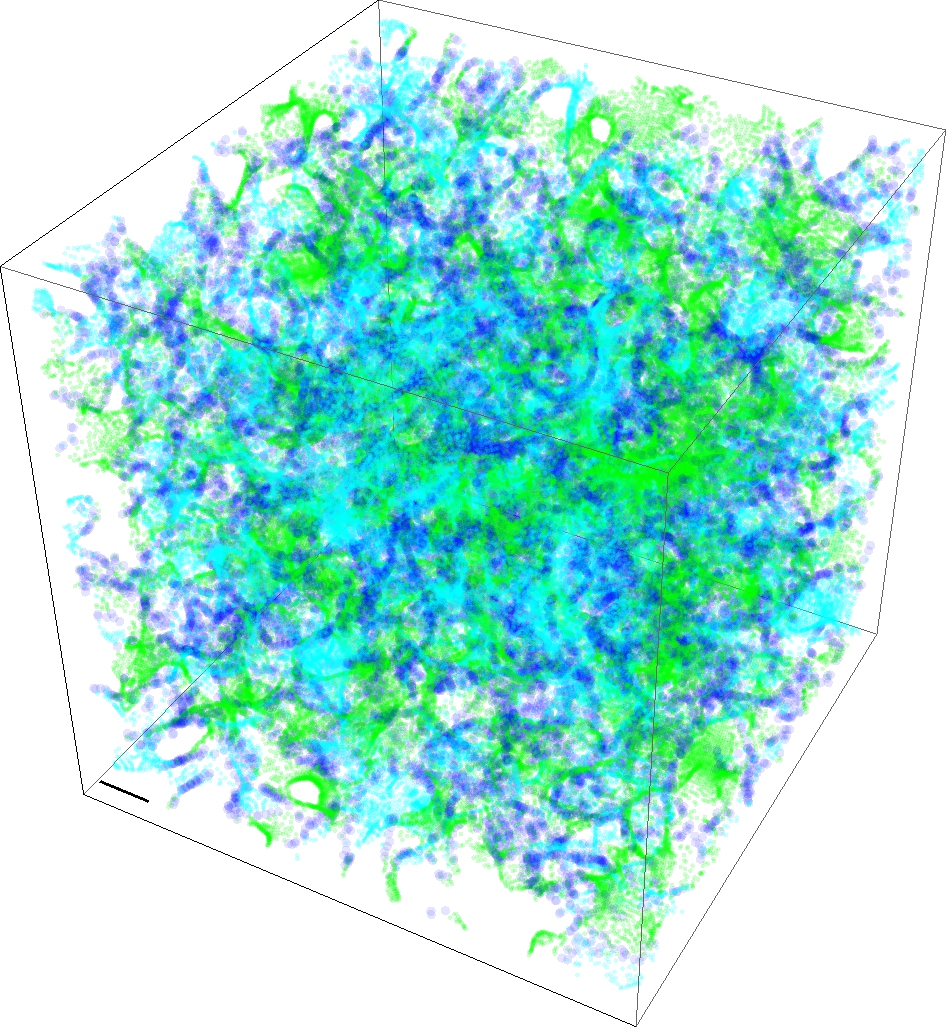} & \quad 
\includegraphics[width=.40\textwidth, bb=0 0 955 1037]{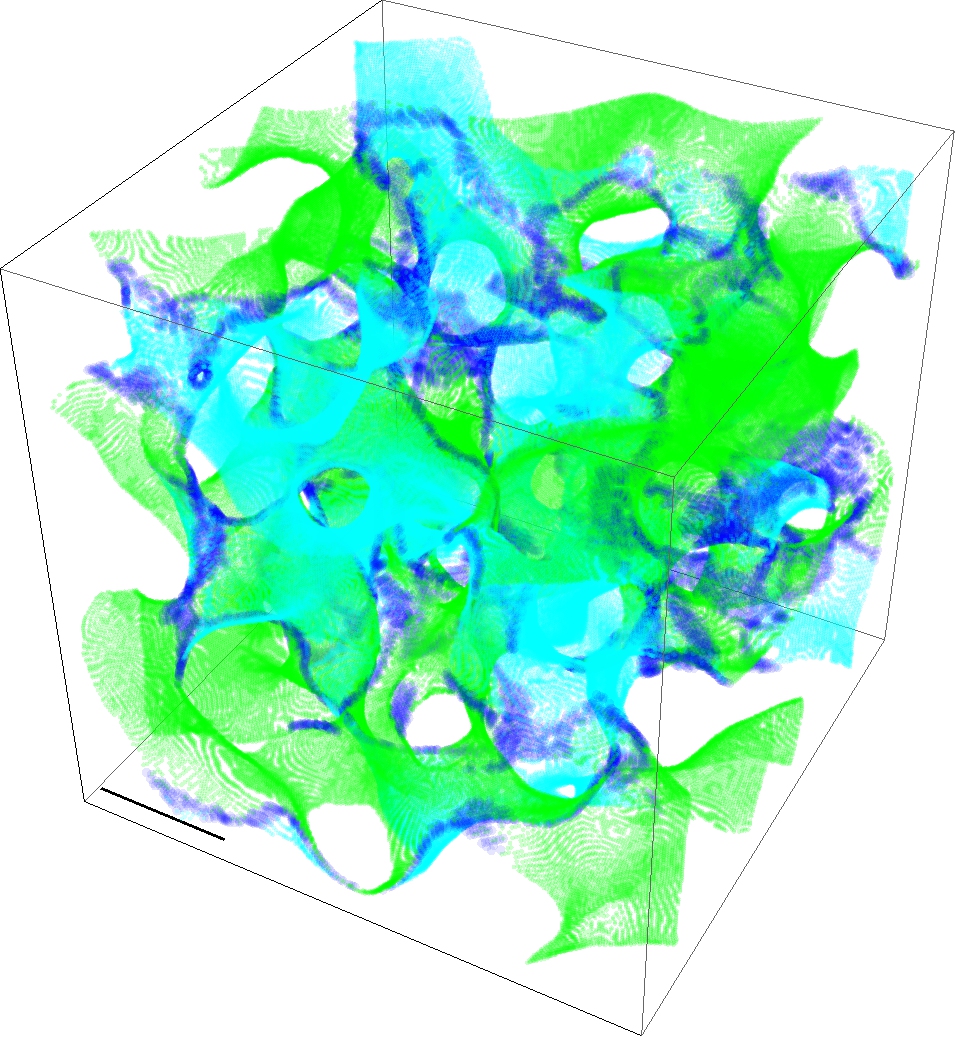} \vspace{0.5cm}\\
\includegraphics[width=.40\textwidth, bb=0 0 951 1032]{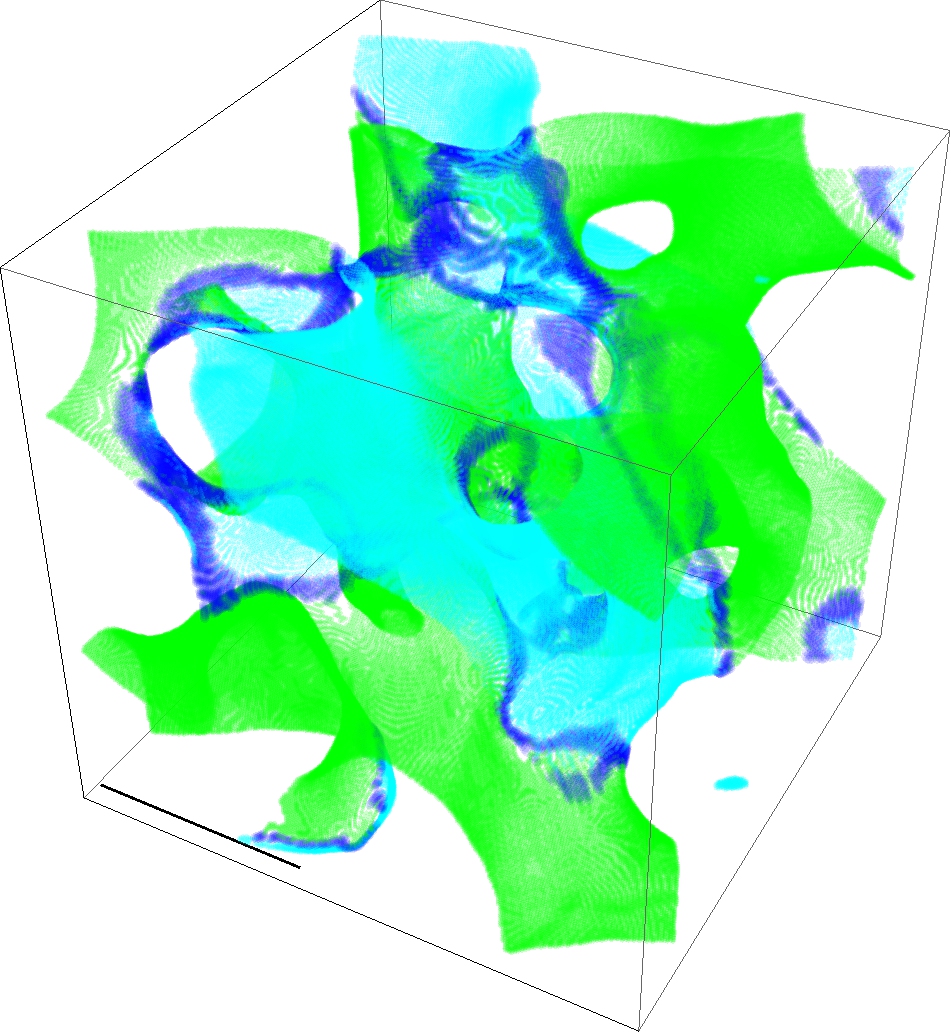} & \quad 
\includegraphics[width=.40\textwidth, bb=0 0 953 1034]{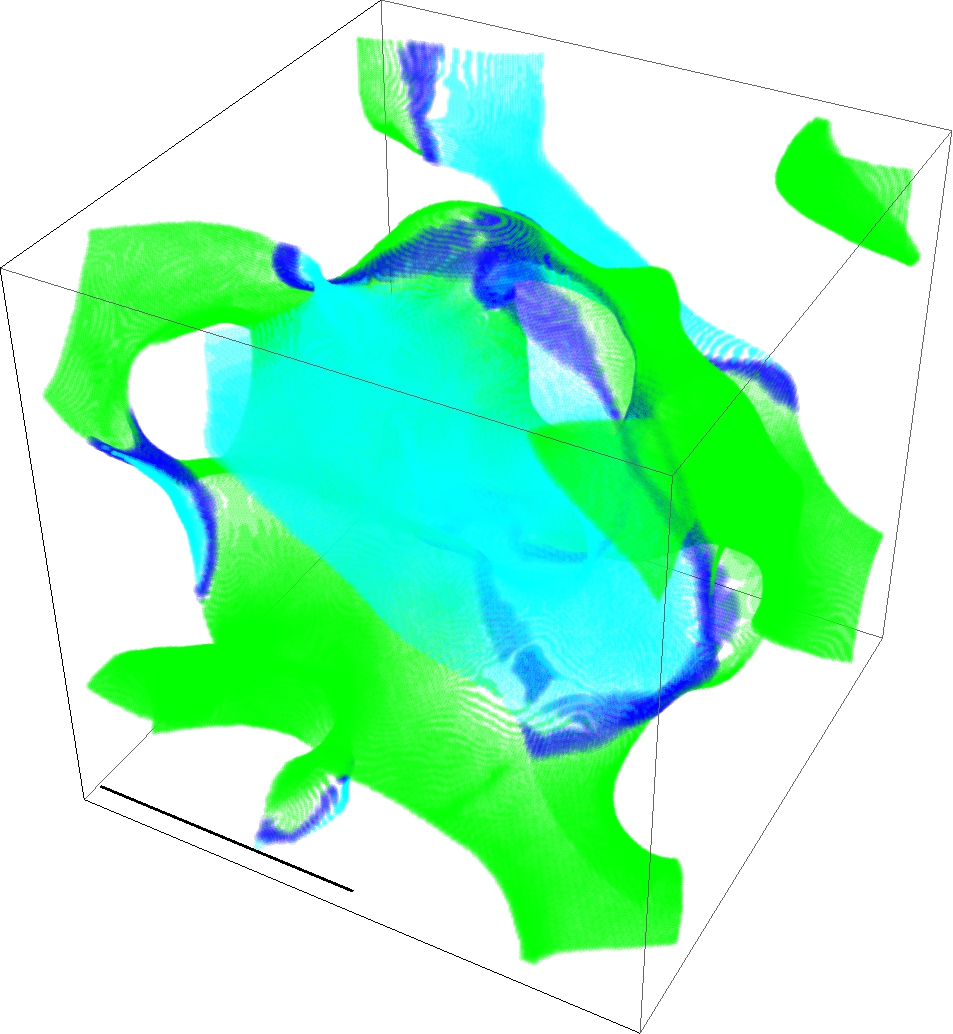} 
\end{tabular}
\caption{
Domain wall networks obtained by our numerical simulation 
at $\tau a_i H_i = 20$ (upper left panel), $50$ (upper right panel), 
$80$ (lower left panel), $100$ (lower right panel). 
We set $c_{H_{\rm osc}} = -15$, $m=4$, $b_H=5$, and $m'=2$. 
In the blue regions, the VEV of the flat direction satisfies $\abs{\phi} < \abs{\la \abs{\phi} \ra} /30$, 
where $\la \abs{\phi} \ra$ is the VEV given by Eq.~(\ref{VEV3}). 
In the green and cyan regions, the phase of the flat direction satisfies 
$- \pi / 8 < \theta < \pi / 8$ and $(7/8) \pi < \theta < (9/8) \pi$, respectively. 
The black line at the bottom-left corner 
represents the length of the comoving horizon size $\tau$ at each time. 
}
  \label{DW network}
\end{figure}

We follow the time evolution of the flat direction by solving the equation 
of motion Eq.~(\ref{EOM rewritten2}) numerically. 
Details of the numerical method are written in Sec.~\ref{results1}. 
Figure~\ref{DW network} shows one example of domain wall network evolution 
obtained by our numerical simulation. 
We set $c_{H_{\rm osc}} = -15$, $m=4$, and $m'=2$. 
The green and cyan regions 
satisfy $- \pi / 8 < \theta < \pi / 8$ and $(7/8) \pi < \theta < (9/8) \pi$, respectively, 
where $\theta$ is the phase of the flat direction. 
They represent the inside of domain walls. 
The blue region satisfies $\abs{\phi} < \la \abs{\phi} \ra / 30$ 
and describes cosmic strings in the limit of $b_H \to 0$. 
The black line at the bottom-left corner in Fig.~\ref{DW network} 
describes a unit of the comoving horizon size (given by $\tau$) at each time. 
One can see that a typical width of domain walls 
increases with time 
and is roughly proportional to the Hubble length. 
One can also see that each Hubble volume contains $\mathcal{O}(1)$ domain walls of each type ($m'=2$ in Fig.~\ref{DW network}). 
From our numerical simulation, we find that 
the spatially averaged magnitude of $\phi$ grows in time and 
eventually reaches 
of the order of the VEV (Eq.~(\ref{VEV3}))
around $\widetilde{\tau}_{\rm form} \simeq 10$ in this case.

\begin{figure}[t]
\centering 
\begin{tabular}{l l}
\includegraphics[width=.40\textwidth, bb=0 0 360 349]{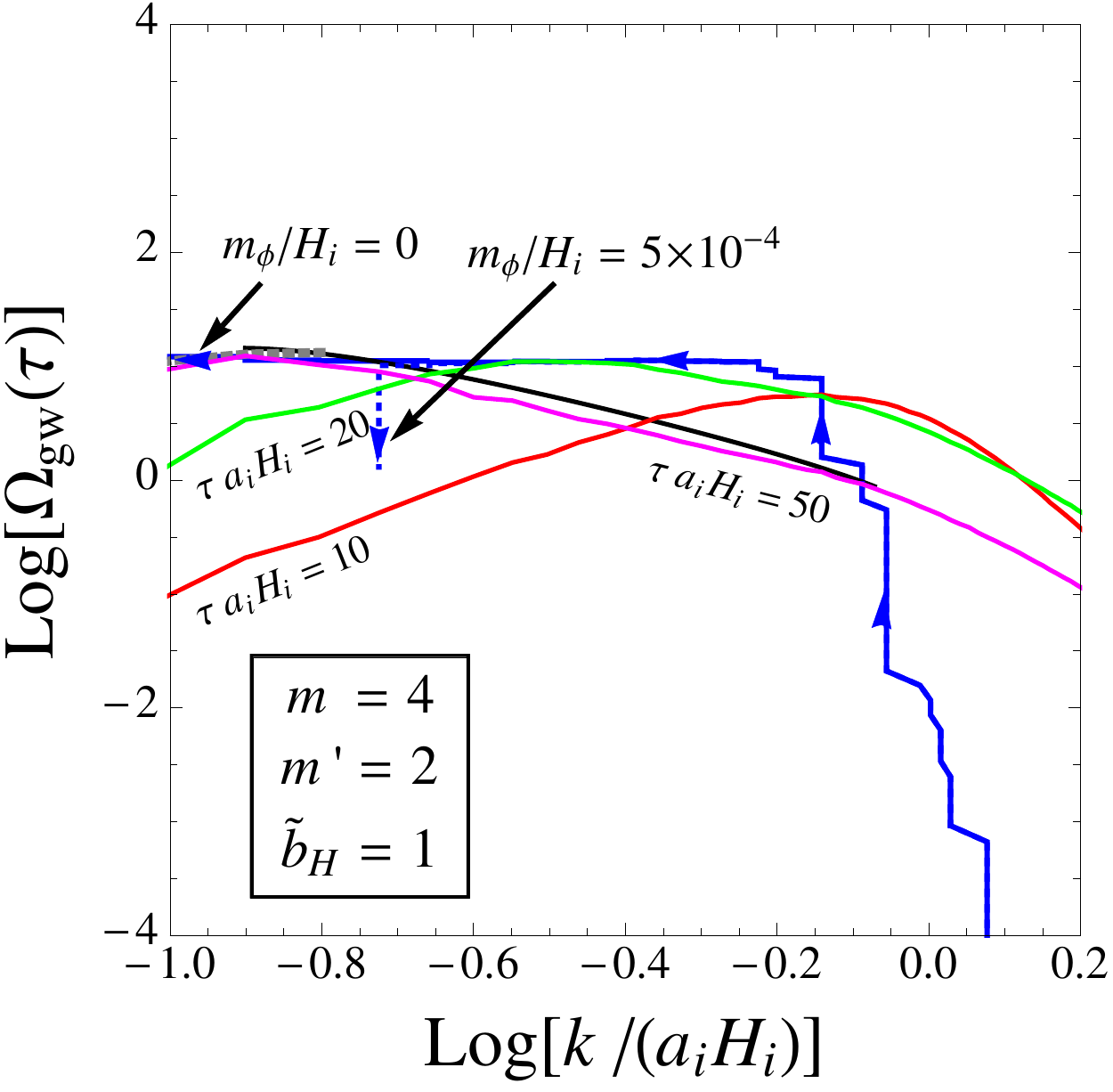} & \quad 
\includegraphics[width=.40\textwidth, bb=0 0 360 349]{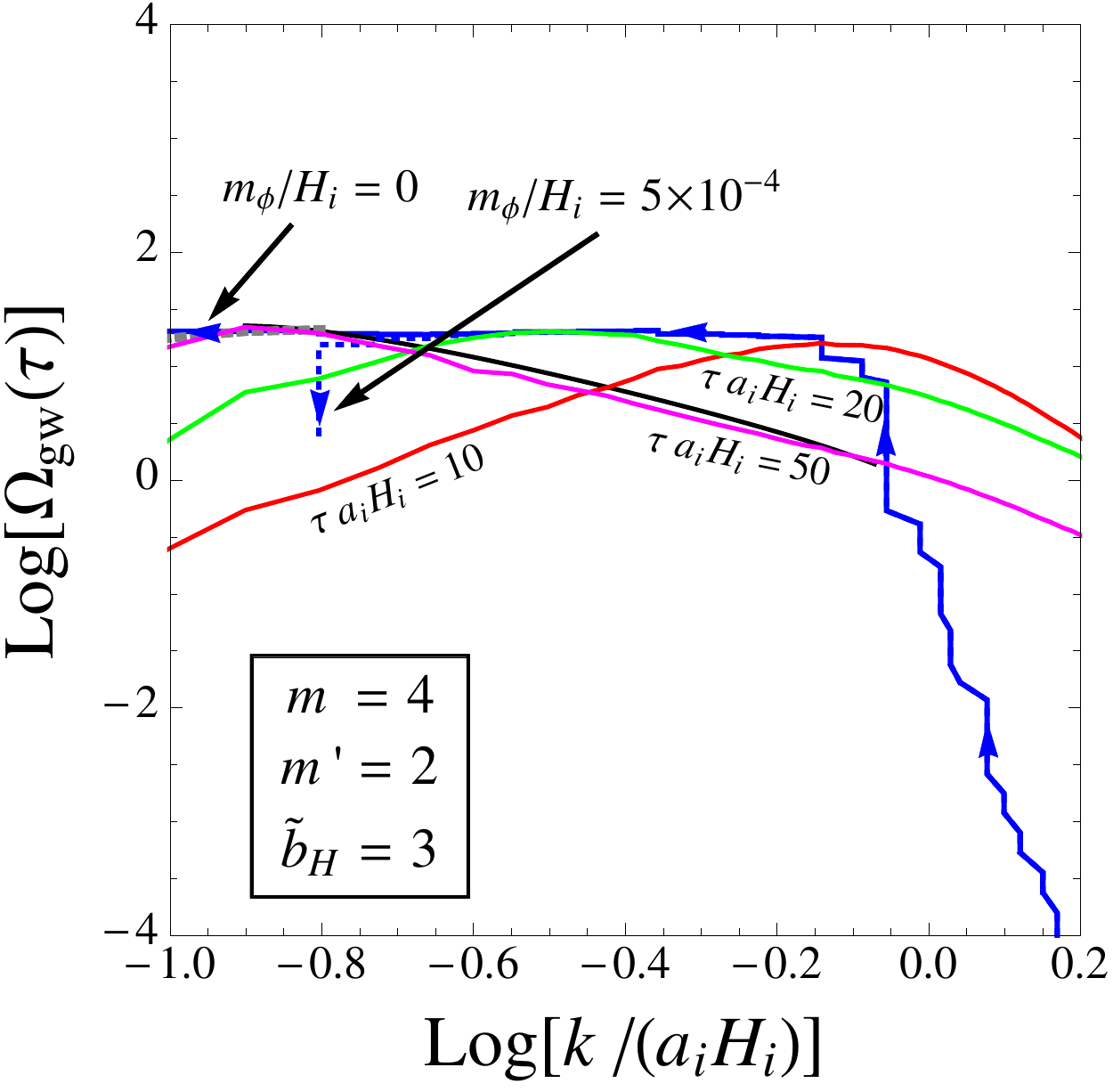} \vspace{0.5cm}\\
\includegraphics[width=.40\textwidth, bb=0 0 360 349]{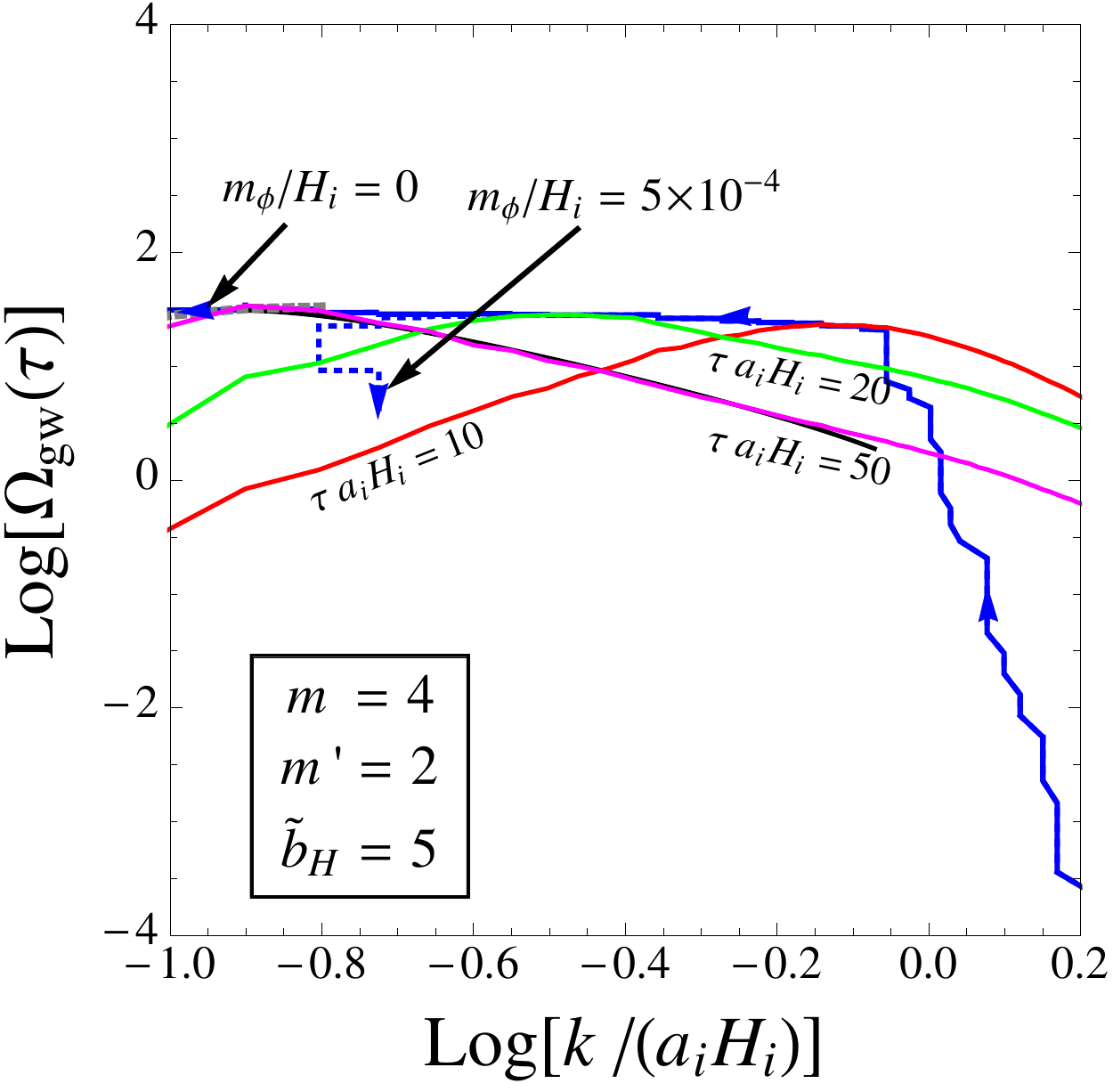} & \quad 
\includegraphics[width=.40\textwidth, bb=0 0 360 349]{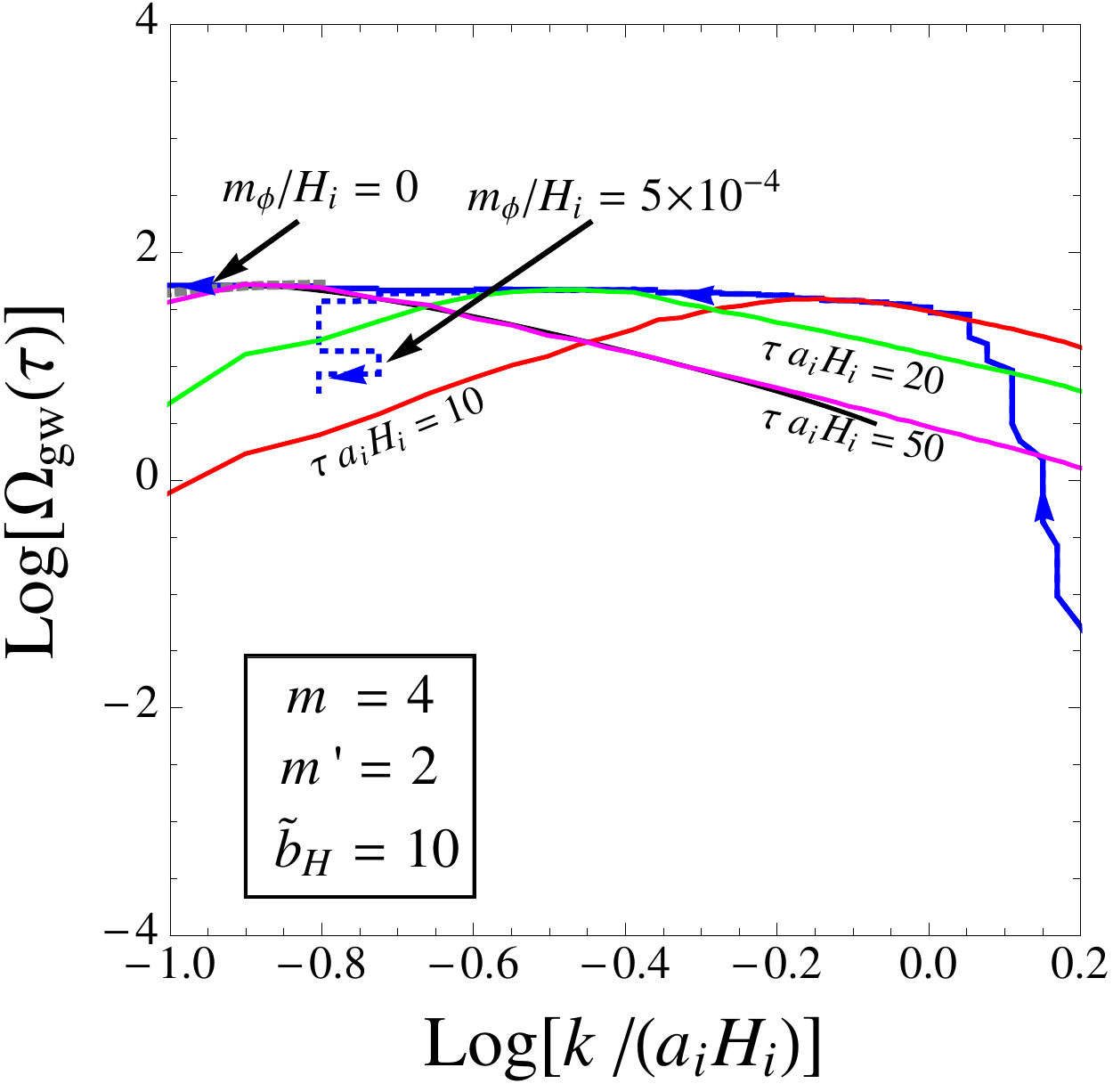} 
\end{tabular}
\caption{
GW spectra obtained from our numerical simulations. 
The energy density of GWs are rescaled by $a_H^{-2/(m-2)}$. 
The red, green, and magenta curves are GW spectra at the time of 
$\tau a_i H_i = 10, 20,$ and $50$, respectively. 
We also show the time evolution of the GW peak location as the blue solid (dashed) line 
for the case of $m_\phi / H_i = 0$ 
($5 \times 10^{-4}$). 
We take $\widetilde{b}_H = 1$ (upper left panel), $3$ (upper right panel), $5$ (lower left panel), 
and $10$ (lower right panel). 
The black curves at $\tau a_i H_i = 50$ are plotted by Eq.~(\ref{large scale2}) 
and well describes large wavenumber modes. 
The gray dotted curves are estimated 
from the discussion of causality at $\tau a_i H_i = 50$
and well describes small wavenumber modes. 
}
  \label{GWspectrum wo U(1)}
\end{figure}

\begin{figure}[t]
\centering 
\begin{tabular}{l l}
\includegraphics[width=.40\textwidth, bb=0 0 360 349]{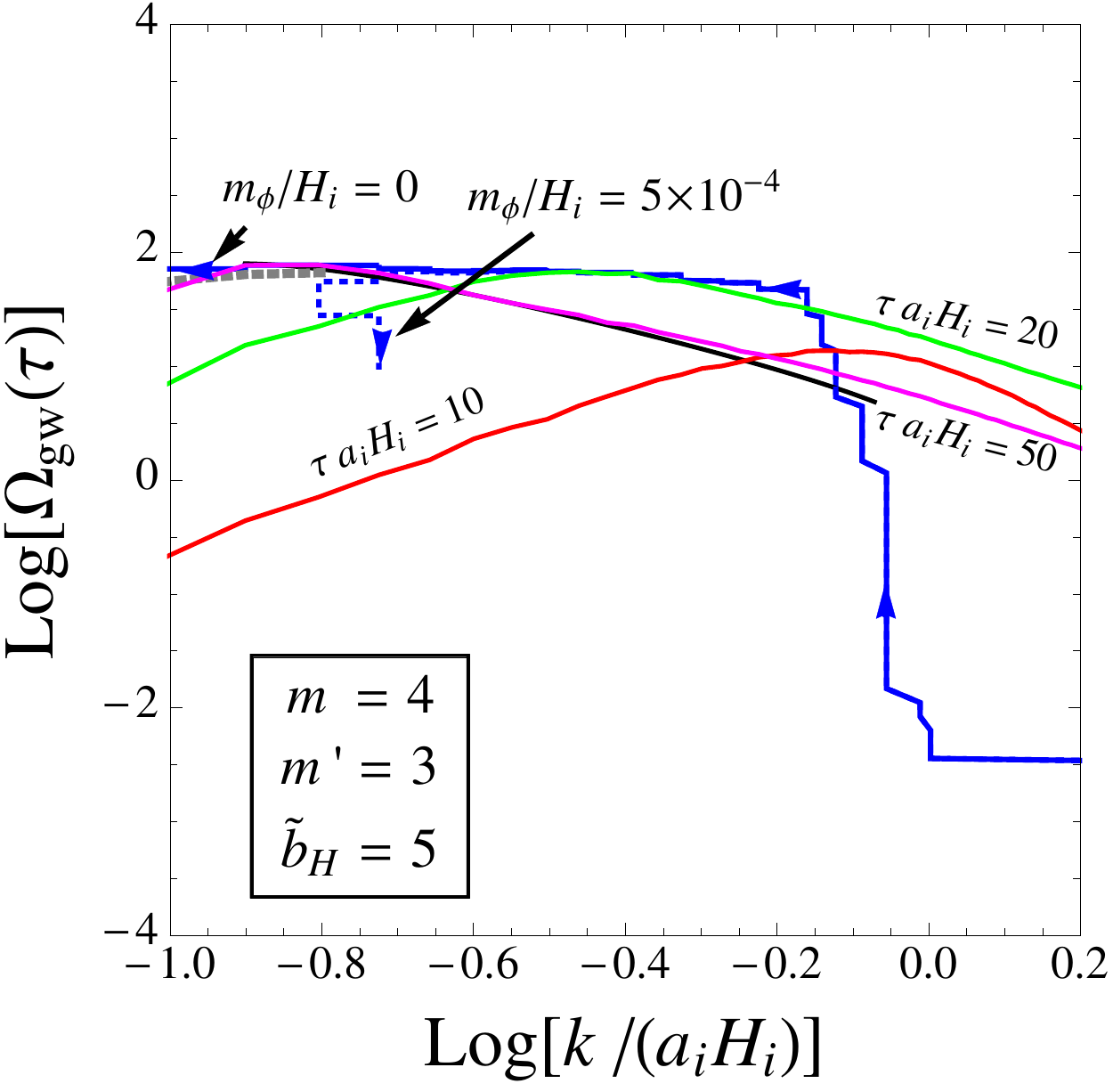} & \quad 
\includegraphics[width=.40\textwidth, bb=0 0 360 349]{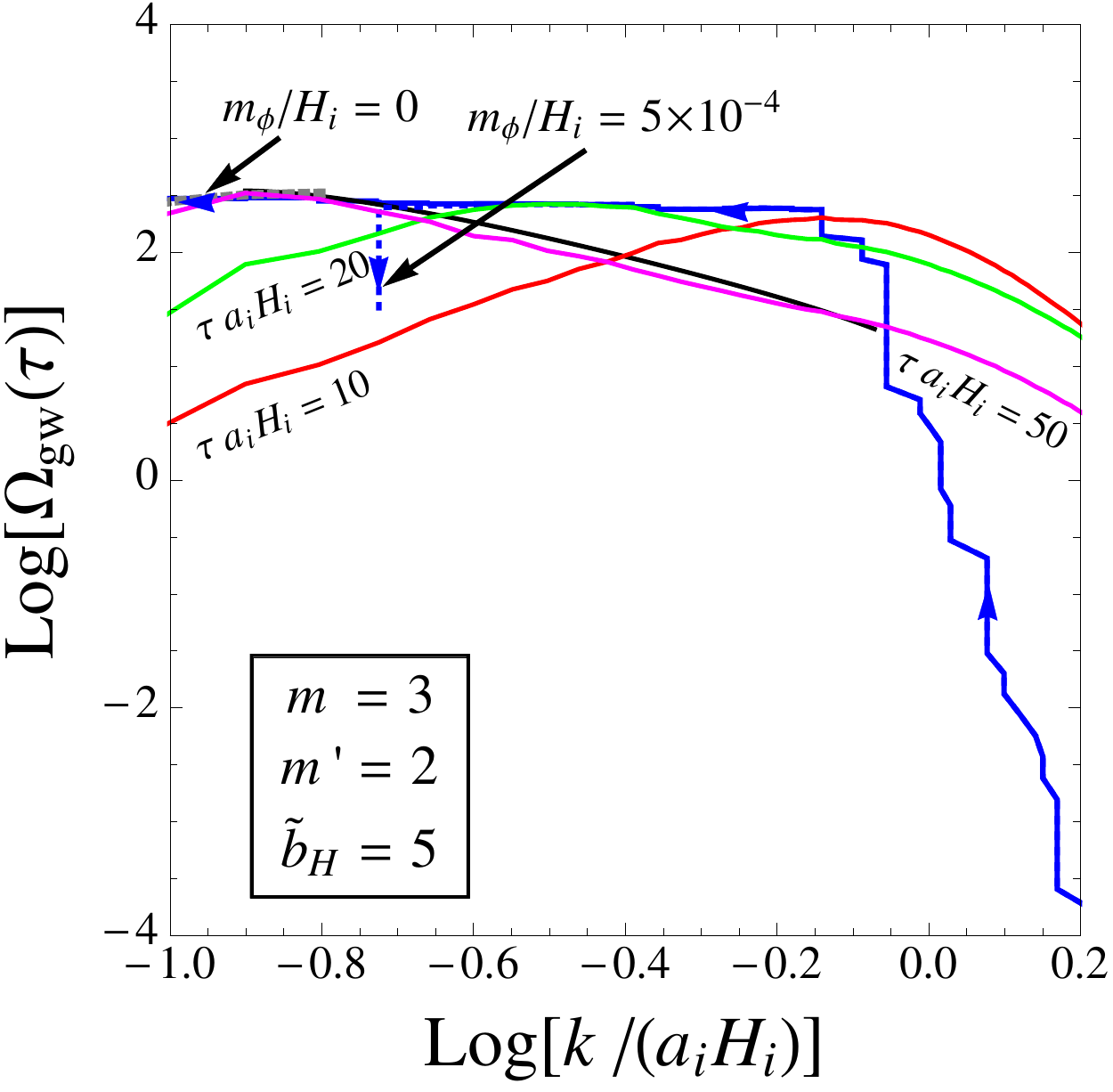} \\
\end{tabular}
\caption{
Same as Fig.~\ref{GWspectrum wo U(1)} 
but for the cases of $m=4$ and $m'=3$ (left panel), 
and $m=3$ and $m'=2$ (right panel). 
We assume $\widetilde{b}_H = 5$. 
}
  \label{GWspectrum wo U(1)_2}
\end{figure}

We calculate the transverse-traceless part of the anisotropic stress from the numerical simulation 
and obtain the energy density of GWs using the equations in Appendix~\ref{calculation}. 
The results are plotted in Figs.~\ref{GWspectrum wo U(1)} and \ref{GWspectrum wo U(1)_2}. 
One can find that 
the GW peak energy density is almost constant in time. 
We give physical interpretation of these results in the next subsection.

Taking a non-zero value of $m_\phi$ into account, 
we find that the flat direction starts to oscillate around the origin of the potential 
and domain walls disappear around $\tau \simeq \tau_{\rm decay}$, 
which is given by Eq.~(\ref{t_decay}). 
After that, 
the GW peak wavenumber becomes constant 
and the GW spectrum 
decreases adiabatically 
as $\propto \tau^{-2}$ ($\propto a^{-1}$). 
In Figs.~\ref{GWspectrum wo U(1)} and \ref{GWspectrum wo U(1)_2}, 
the GW spectrum is freezed out around 
$\widetilde{\tau} = \widetilde{\tau}_{\rm decay} \simeq40$, 
which is given by Eq.~(\ref{tau_decay}) with $m_\phi/H_i = 5 \times 10^{-4}$ and $c_{H_{\rm osc}} = -15$.
As we discussed in Sec.\,\ref{results1}, the above choice of $m_\phi/H_i$ is convenient for numerical simulations, 
though it does not change the dynamics of flat direction until $\widetilde{\tau} \simeq \widetilde{\tau}_{\rm decay}$. 
In reality, it can be many orders of magnitude smaller depending on inflation models.

\subsection{\label{interpretation3}Physical interpretations}

In this subsection, we discuss how we can understand the results of our numerical simulations 
shown in Figs.~\ref{GWspectrum wo U(1)} and \ref{GWspectrum wo U(1)_2}.

During the scaling regime, the typical length scale of domain wall dynamics is the Hubble length. 
As a result, 
the emission peak wavenumber of GWs emitted from these domain walls is 
estimated by 
the Hubble scale like Eq.~(\ref{peak0}). 
Its quadrupole moment $Q$ can be roughly given by~\cite{Maggiore:1999vm}
\beq
 Q \sim H^{-2} \times m' \sigma H^{-2}, 
\eeq
where $m' \sigma H^{-2}$ is the total energy of domain walls within a Hubble volume. 
Such domain walls emit GWs with the luminosity of 
Eq.~(\ref{luminosity}), 
which is proportional to the square of the quadrupole moment of domain walls. 
Thus, the produced energy density of GWs can be estimated as 
\beq
 \lkk \frac{\Delta \Omega_{\rm gw} }{\Delta \log \tau} \rkk_{\rm peak}
 \sim 
 b_H C^{m'+2} 
 \ \lmk \sim \frac{M_*^2}{\Mpl^2} \rmk. 
 \label{omega_gw6-0}
\eeq
Domain walls emit GWs with a constant energy density given by Eq.~(\ref{omega_gw6-0}). 
Note that cosmic strings should coexist with domain walls
because in the limit of $b_H \to 0$, the potential for the flat direction restores 
$U(1)$ symmetry. 
This implies that GWs are also emitted from cosmic strings. 
Thus, the total GW energy density is given by 
\beq
 \lkk \frac{\Delta \Omega_{\rm gw} }{\Delta \log \tau} \rkk_{\rm peak}
 \sim 
 k_{\rm br} 
 b_H C^{m'+2} 
 + 
 \lmk \frac{\abs{c_{H_{\rm osc}}}}{a_H(m-1)} \rmk^{2/(m-2)}, 
 \label{omega_gw6}
\eeq
where the second term 
comes from the contribution of cosmic strings 
and is relevant for small $b_H$. 
We include a numerical prefactor $k_{\rm br}$ to fit our numerical results. 
Note that the first term in the right-hand side of this equation 
is proportional to $M_*^2 / \Mpl^2$ (Eq.~(\ref{omega_gw6-0})), 
while the second term is proportional to $M_*^4 / \Mpl^4$ (Eq.~(\ref{omega_gw5})). 
This results from the fact that the energy density of domain walls 
is proportional to $M_*$ (see Eq.~(\ref{rho_DW})),
while that of cosmic strings is proportional to $M_*^2$ (see Eqs.~(\ref{mu}), (\ref{rho_cs}) and (\ref{VEV2})). 
Since the energy density of domain walls is much larger than 
that of cosmic strings, 
domain walls emit GWs more efficiently.

\begin{figure}[t]
\centering 
\begin{tabular}{l l}
\includegraphics[width=.40\textwidth, bb=0 0 360 351]{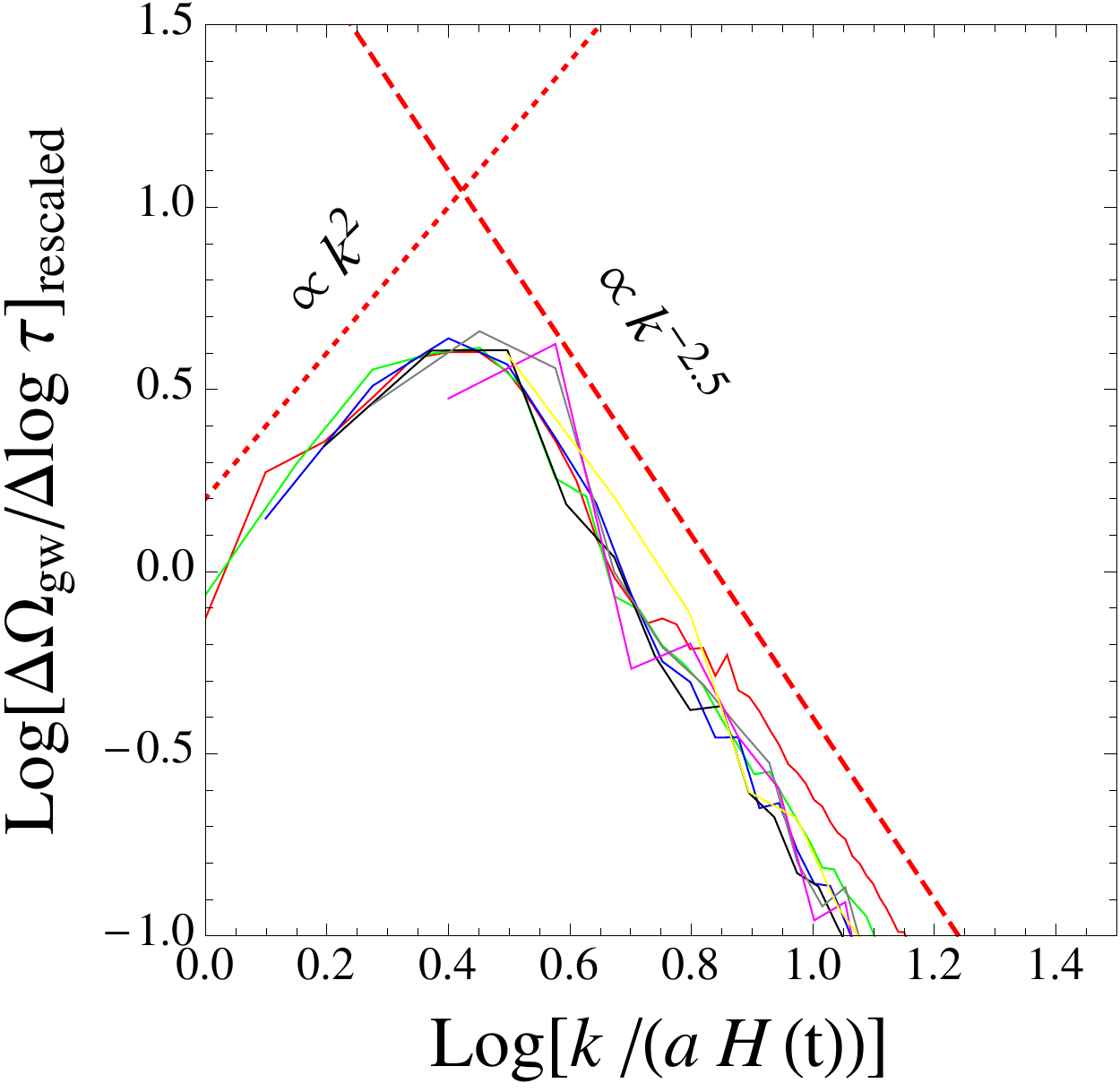} & \quad 
\includegraphics[width=.40\textwidth, bb=0 0 360 351]{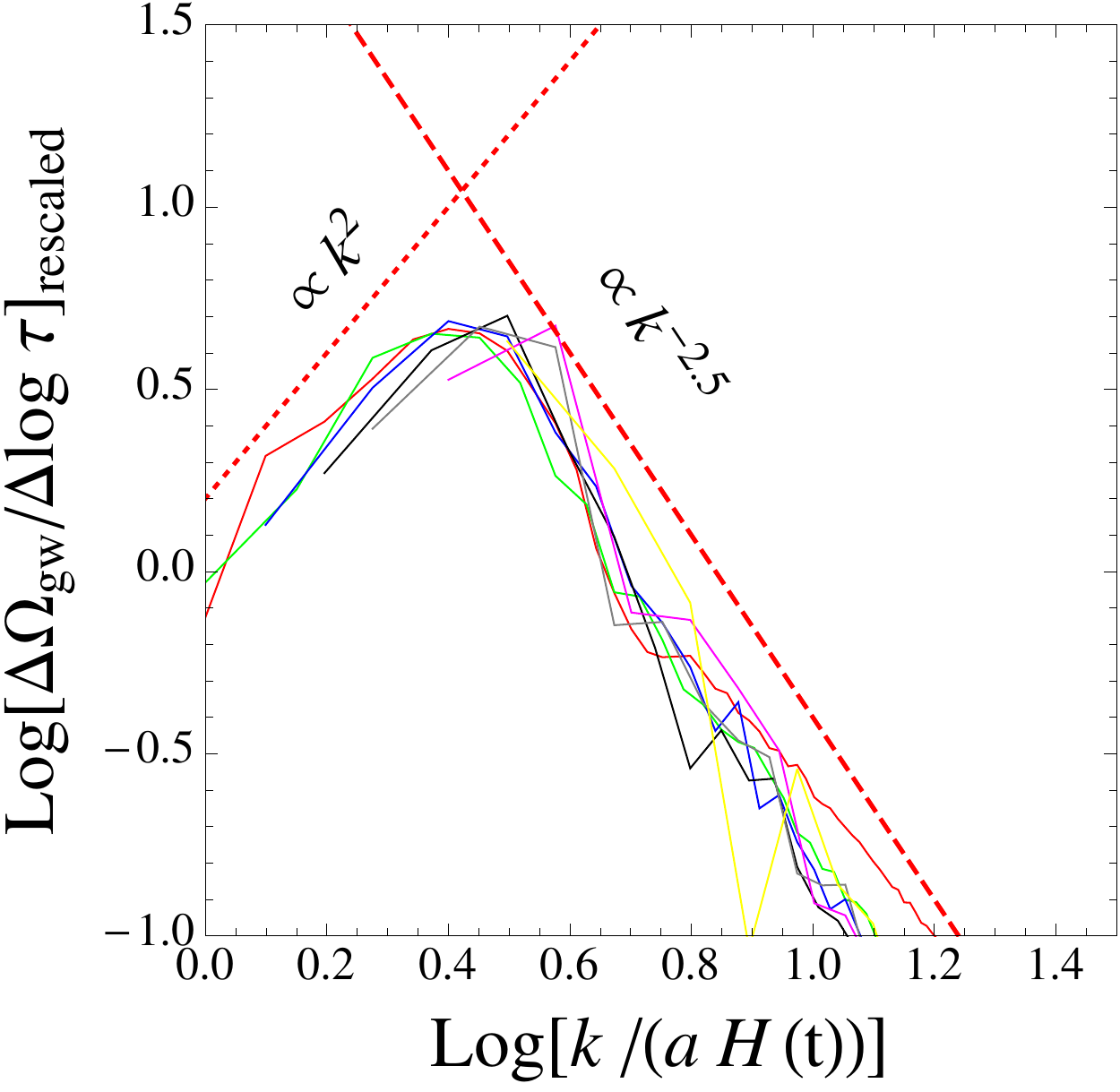} \vspace{0.5cm}\\
\includegraphics[width=.40\textwidth, bb=0 0 360 351]{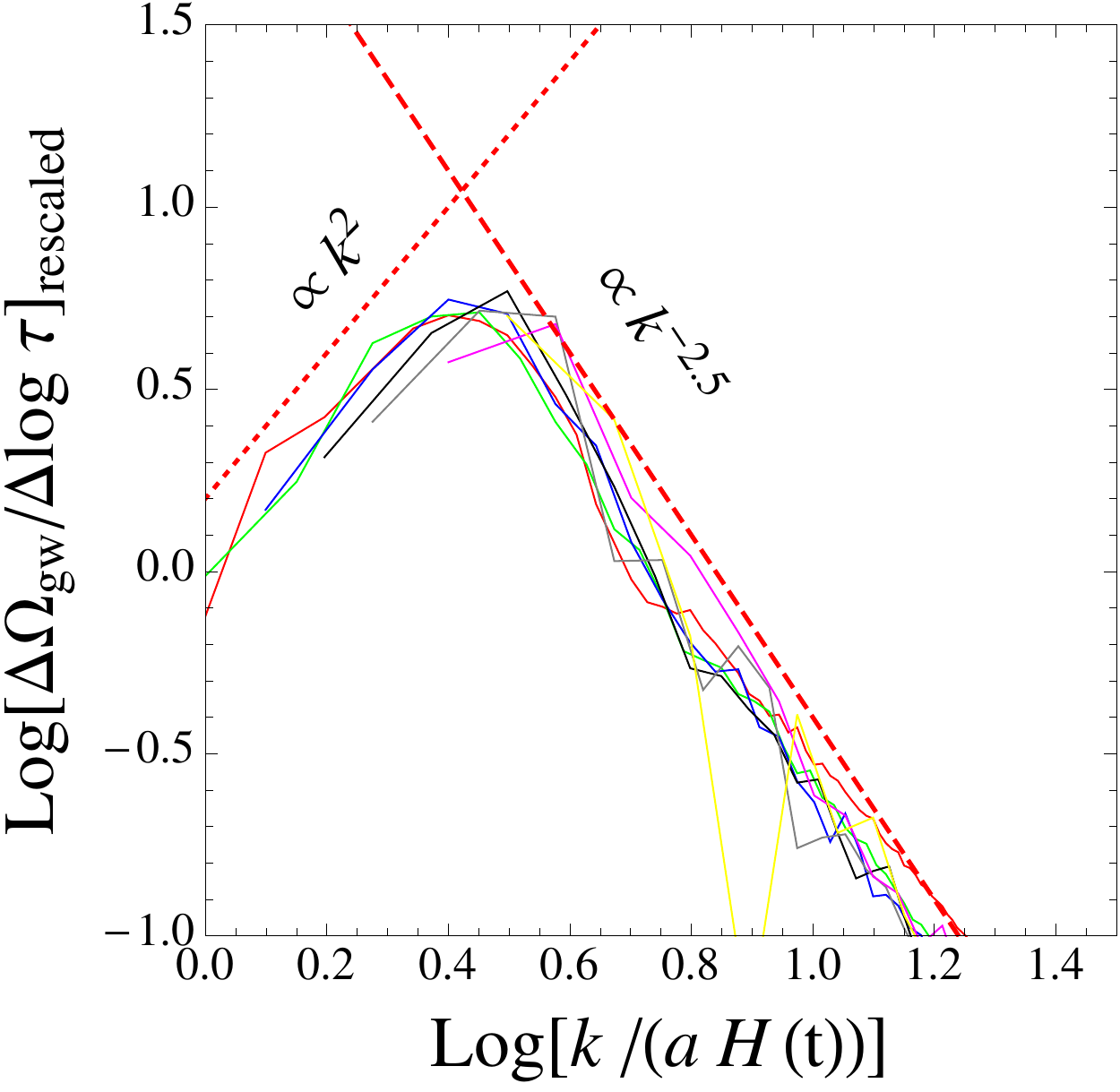} & \quad 
\includegraphics[width=.40\textwidth, bb=0 0 360 351]{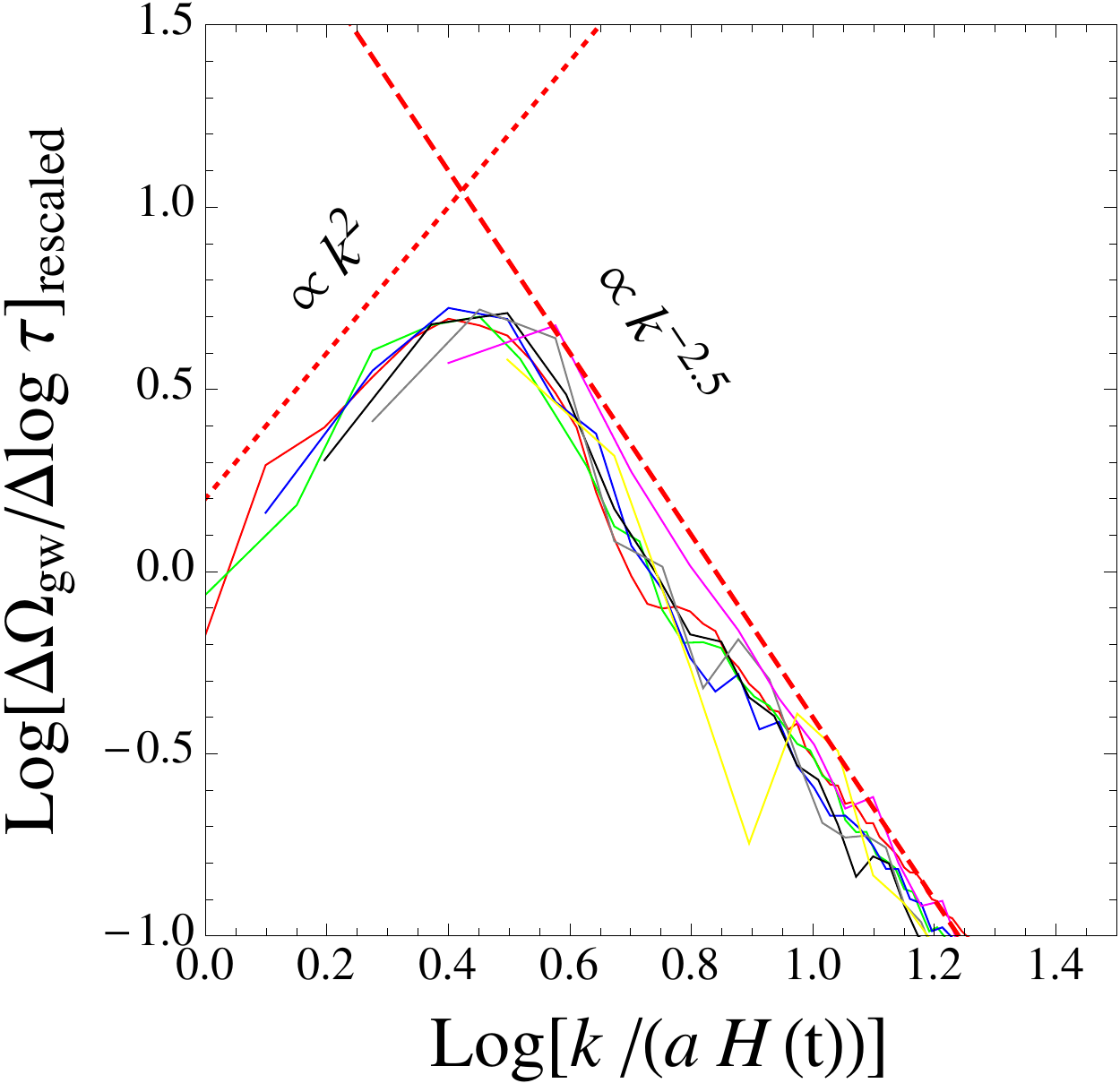} 
\end{tabular}
\caption{
Derivatives for GW spectra in terms of the conformal time. 
The energy density of GWs are rescaled by the factor given by Eq.~(\ref{omega_gw6}) 
with a numerical constant $k_{\rm br}=0.4$, 
which we set to fit our results. 
We plot GW spectra for $\tau a_i H_i = 20, 30, 40, 50, 60, 80,$ and $100$. 
We take $\widetilde{b}_H = 1$ (upper left panel), $3$ (upper right panel), $5$ (lower left panel), 
and $10$ (lower right panel). 
}
  \label{del omega3}
\end{figure}

\begin{figure}[t]
\centering 
\begin{tabular}{l l}
\includegraphics[width=.40\textwidth, bb=0 0 360 351]{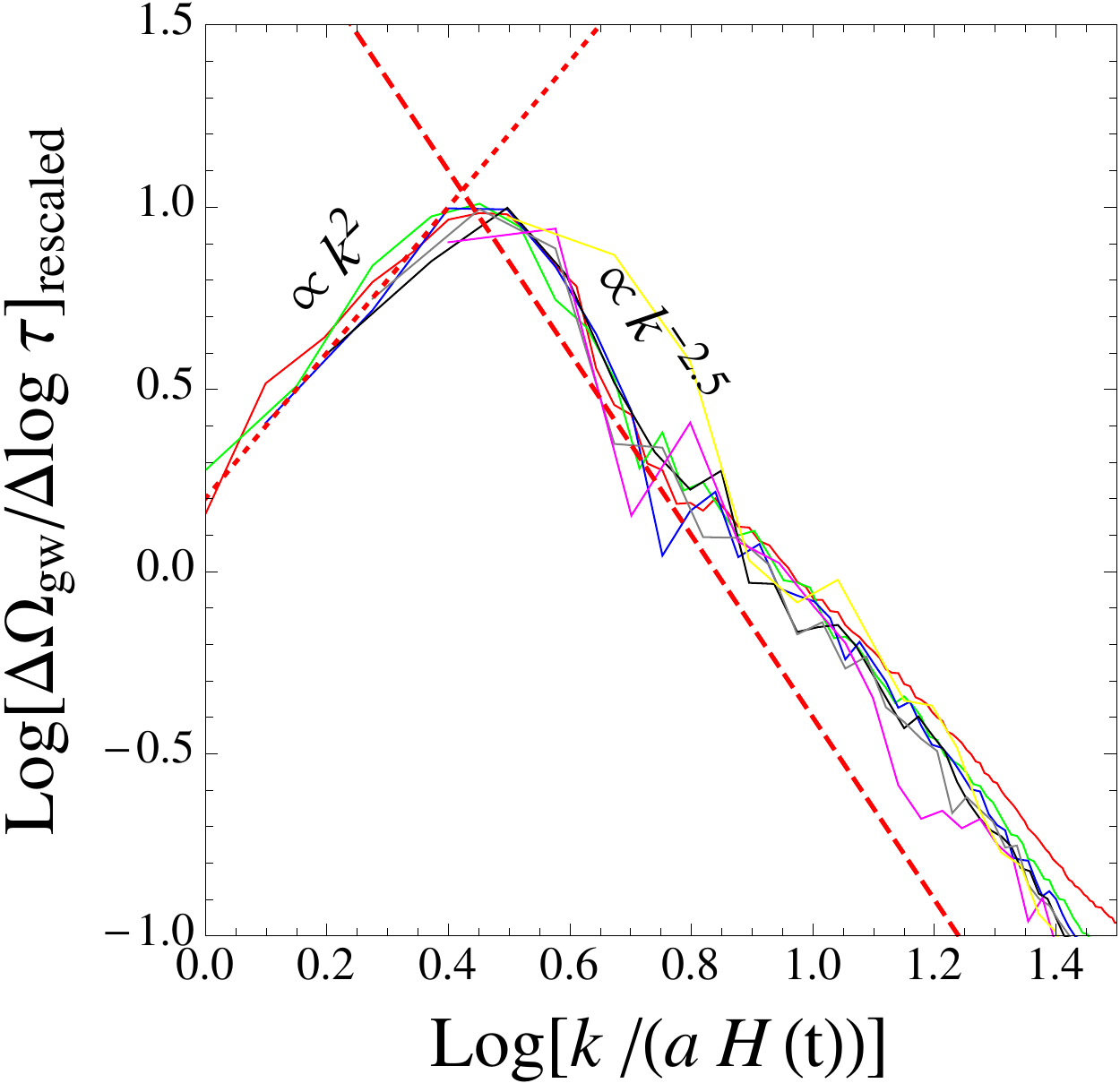} & \quad 
\includegraphics[width=.40\textwidth, bb=0 0 360 351]{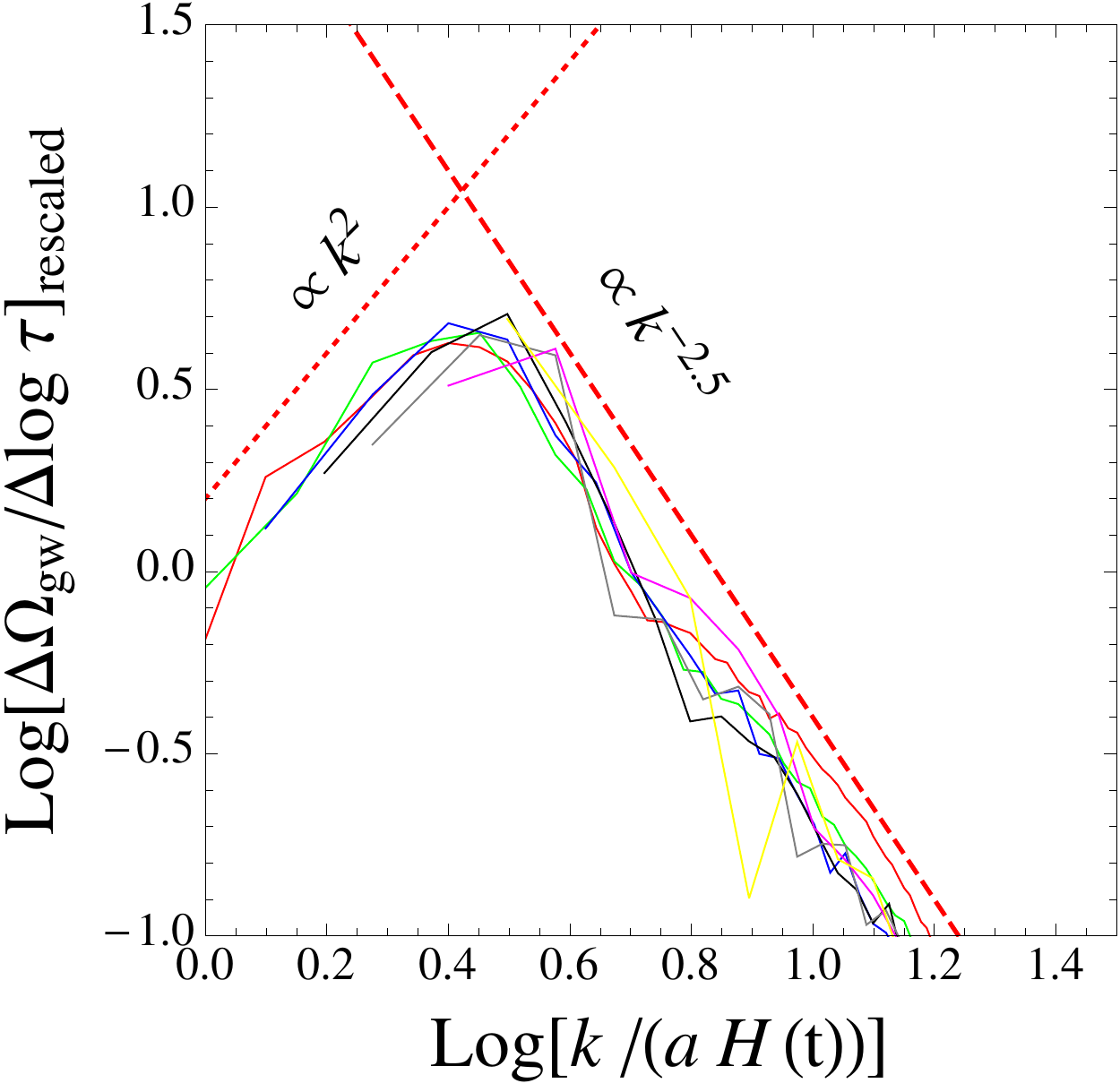} \\
\end{tabular}
\caption{
Same as Fig.~\ref{del omega3} 
but for the cases of $m=4$ and $m'=3$ (left panel), 
and $m=3$ and $m'=2$ (right panel). 
We assume $\widetilde{b}_H = 5$. 
}
  \label{del omega4}
\end{figure}

We plot our numerical results of $\Delta \Omega_{\rm gw} / \Delta \log \tau$ 
in Figs.~\ref{del omega3} and \ref{del omega4}. 
The results are constant in time as expected. 
We can determine the numerical prefactors for 
the emission peak wavenumber and the produced peak energy density of GWs 
from the numerical results 
like 
\beq
 \frac{k^{\Delta}_{\rm peak}}{a(t)} &\simeq& 2.5 H(t), 
 \label{peak4} 
 \\
 \lkk \frac{\Delta \Omega_{\rm gw} }{\Delta \log \tau} \rkk_{\rm peak} 
 &\simeq& 10 \lkk 
  k_{\rm br} 
 b_H C^{m'+2} 
 + 
 \lmk \frac{\abs{c_{H_{\rm osc}}}}{a_H(m-1)} \rmk^{2/(m-2)} 
 \rkk, 
  \label{GW amplitude4}
\\ 
 k_{\rm br} &\simeq& 0.4. 
\eeq 
We find that the numerical prefactors are independent of $m$, $m'$, and $\tau$, 
so that Eqs.~(\ref{peak4}) and (\ref{GW amplitude4}) can be used for any values of $m$, $m'$, and $\tau$.

Figures~\ref{del omega3} and \ref{del omega4} show that 
$\Delta \Omega_{\rm gw} / \Delta \log \tau$ is proportional to 
a certain power of $k$ for large and small wavenumber modes like Eq.~(\ref{large wavenumber}), 
where 
$\alpha \simeq 2.5$ in this case. 
The GW spectra for large wavenumber modes 
in Figs.~\ref{GWspectrum wo U(1)} and \ref{GWspectrum wo U(1)_2} 
can be understood by the same integration as Eq.~(\ref{large scale2}). 
The result of Eq.~(\ref{large scale2}) at $\widetilde{\tau}=50$ is plotted as the black curves 
in Figs.~\ref{GWspectrum wo U(1)} and \ref{GWspectrum wo U(1)_2} 
and well describes the large wavenumber behaviour of GW spectrum. 
Note that 
the GW spectrum is proportional to $k^{-2}$ 
for $k \gg k_{\rm peak}$. 

The gray dotted curves in Figs.~\ref{GWspectrum wo U(1)} and \ref{GWspectrum wo U(1)_2} show that 
simulated super-horizon modes are consistent with the 
spectrum expected from the discussion of causality~\cite{Dufaux:2007pt, Kawasaki:2011vv} 
as explained in Sec.~\ref{interpretation1}. 
The blue dashed lines in Figs.~\ref{GWspectrum wo U(1)} and \ref{GWspectrum wo U(1)_2} 
show the evolution of GW peak wavenumber 
in the case of $m_\phi/H_i = 5 \times 10^{-4}$. 
Since the flat direction starts to oscillate around the potential minimum 
due to its soft mass term, 
domain walls disappear 
at $\tau \simeq \tau_{\rm decay}$.  
The GW peak energy density and frequency at that time 
is given by 
\beq
 \lkk \Omega_{\rm gw} h^2 (\tau_{\rm decay}) \rkk_{\rm peak} 
 &\simeq& 
 \frac{10}{4} \lkk 
 0.4
 b_H C^{m'+2} 
 + 
 \lmk \frac{\abs{c_{H_{\rm osc}}}}{a_H(m-1)} \rmk^{2/(m-2)} 
 \rkk, 
  \label{omega_gw8}
  \\
  \frac{k_{\rm peak}}{a(\tau_{\rm decay})} &\simeq& 3 \frac{\mphi}{\sqrt{\abs{c_{H_{\rm osc}}}}}, 
  \label{peak5}
\eeq
where the numerical prefactors are determined from our numerical results. 
The shape of GW spectrum is in fact freezed out
around $\widetilde{\tau} = \widetilde{\tau}_{\rm decay} \simeq40$, 
which is given by Eq.~(\ref{tau_decay}) with $m_\phi/H_i = 5 \times 10^{-4}$ and $c_{H_{\rm osc}} = -15$.
After that, the peak wavenumber does not change and 
the GW energy density decreases as $\propto a^{-1}$ 
until reheating completes.

In summary, the schematic form of resulting GW spectrum is 
the same as the one described in the previous section, 
even though domain walls form. 
It is impossible to distinguish GW signals 
emitted from cosmic strings and domain walls. 
However, we can obtain information of SUSY scale and reheating temperature 
through the observation of GW signals, as we explain in the next subsection.

\subsection{\label{prediction3}GW signals at present}

The above calculations and discussions 
determine the GW spectrum during the oscillation dominated era. 
In this subsection, we derive 
the GW spectrum at the present epoch, 
taking into account the redshift.

The peak energy density of GWs at $\tau_{\rm decay}$ 
is given by Eq.~(\ref{omega_gw8}) from our numerical simulations. 
Substituting $\lkk \Omega_{\rm gw} h^2 (\tau_{\rm decay}) \rkk_{\rm peak}$ into Eq.~(\ref{present energy density}), 
we obtain 
the present value of GW energy density such as
\beq
 \lkk \Omega_{\rm gw} h^2 (t_0) \rkk_{\rm peak}
  &\simeq& 
  5 \times 10^{-7} 
 \lmk \frac{\abs{c_{H_{\rm osc}}}^{-1/2} \mphi}{10^3 \GeV} \rmk^{-2/3}
 \lmk \frac{T_{\rm RH}}{10^9 \GeV} \rmk^{4/3} \\
 &&\quad \times  \lkk 
 0.4
 b_H C^{m'+2} 
 + 
 \lmk \frac{\abs{c_{H_{\rm osc}}}}{a_H(m-1)} \rmk^{2/(m-2)} 
 \rkk, 
 \label{result omega3}
\eeq
where we use 
Eqs.~(\ref{t_decay}), (\ref{T_RH}), and (\ref{omega_gw8}). 
From Eq.~(\ref{present peak frequency}), 
the GW peak frequency at the present epoch 
is given by 
\beq
 f_0 
 \simeq 700 \text{ Hz} \lmk \frac{\abs{c_{H_{\rm osc}}}^{-1/2} \mphi}{10^3 \GeV} \rmk^{1/3} \lmk \frac{\TR}{10^9 \GeV} \rmk^{1/3}. 
 \label{result f3}
\eeq
Note that the peak frequency is independent of $a_H$, $b_H$, $m$, and $m'$.

Since the GW spectrum is sensitive to what dominates the energy density of the Universe, 
it has information of reheating temperature 
as explained in the previous section. 
In fact, the GW spectrum bends at 
the frequency given by Eq.~(\ref{f_bend}) corresponding to the wavenumber 
around $k \simeq \tau_{\rm RH}^{-1}$~\cite{Seto:2003kc, Nakayama:2008ip, previous work}. 

There are three observables: 
Eqs.~(\ref{f_bend}), (\ref{result omega3}), 
and (\ref{result f3}), 
so that we can in principle obtain three parameters through observation of GW spectrum: 
$\abs{c_{H_{\rm osc}}}^{-1/2} \mphi$, $\TR$, and $M_*/\Mpl$ (up to by a factor of $\mathcal{O}(1)$). 
Let us stress that $\abs{c_{H_{\rm osc}}}^{-1/2} \mphi$ and $\TR$ are not affected by 
higher dimensional operators 
and are independent of the sources of GW signals, such as cosmic strings and domain walls. 
Note that we assume $M_* \ll \Mpl$ 
to avoid 
the effect of the backreaction of GW emission.

\begin{figure}[t]
\centering 
\includegraphics[width=.45\textwidth, bb=0 0 360 339]{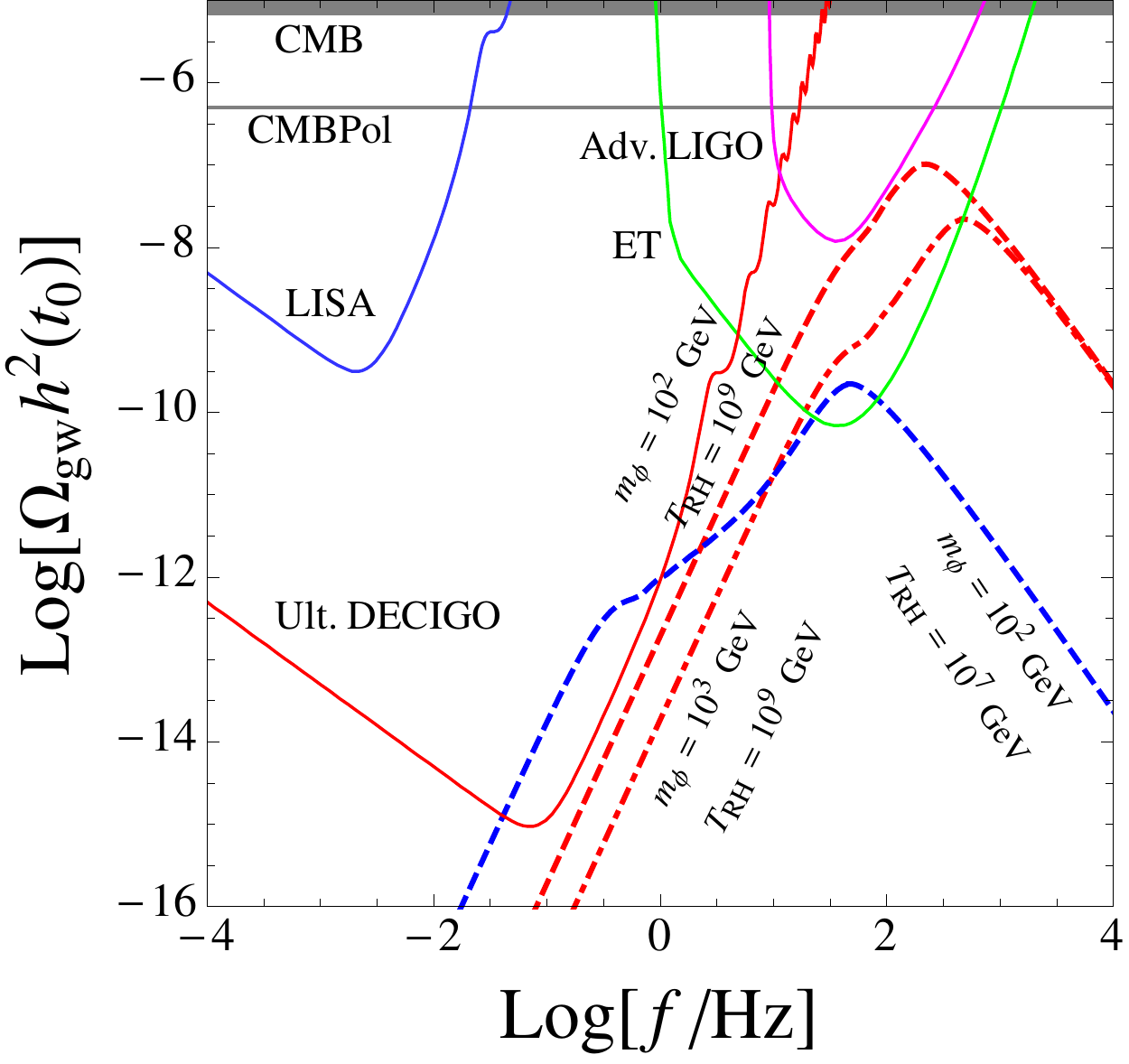} 
\caption{
GW spectra generated by domain walls 
(dash and dot-dashed curves) 
and sensitivities of planned interferometric detectors (solid curves). 
We assume 
$\mphi = 10^2 \GeV$ 
with $\TR = 10^7 \GeV$ (blue dashed curve), 
$\mphi = 10^2 \GeV$ 
with $\TR = 10^9 \GeV$ (red dashed curve), 
and 
$\mphi = 10^3 \GeV$ 
with $\TR = 10^9 \GeV$ (red dot-dashed curve). 
We also assume $c_H' = 3/2$, $a_H' = 1$, $m=4$, and $m'=2$. 
We set the value of $M_*$ as $M_*^2 = \Mpl^2 / 10$. 
}
  \label{detectability3}
\end{figure}

Figure~\ref{detectability3} shows examples of GW spectrum 
predicted by the present mechanism. 
The GW peak energy density and peak frequency are 
given by Eqs.~(\ref{result omega3}) and (\ref{result f3}), which are obtained from our numerical results. 
While the shape of the GW spectrum for low frequency modes are calculated 
from Eqs.~(\ref{A}) and (\ref{B}) with a constant $\TT$ and $\tau_f = \tau_{\rm decay}$, 
that for high frequency modes are calculated from Eq.~(\ref{large scale2}). 
In the figure, we take $\mphi = 10^2 \GeV$ with $\TR = 10^7 \GeV$ (blue dashed curve), 
$\mphi = 10^2 \GeV$ with $\TR = 10^9 \GeV$ (red dashed curve), 
and $\mphi = 10^3 \GeV$ with $\TR = 10^9 \GeV$ (red dot-dashed curve).
To discuss the detectability of GW signals, 
we plot sensitivities of planned interferometric detectors. 
The sensitivity curves are the same as in Fig.~\ref{detectability1},
and details are given in Sec.\,\ref{prediction1}.
We find that 
GW signals would be observed by ET or Ultimate DECIGO.

\section{\label{conclusions}Summary and Conclusions}
We have investigated the dynamics of a flat direction in the early Universe, 
focusing on the case that 
the coefficient of the Hubble induced mass term is positive during inflation 
and is negative after inflation (oscillation dominated era). 
While the flat direction stays at the origin during inflation, 
it obtains a large VEV after inflation due to the negative Hubble induced mass term. 
In many cases, flat directions have $U(1)$ or $Z_n$ symmetries 
and thus topological defects, such as cosmic strings and domain walls, 
form just after the end of inflation. 
When the Hubble parameter decreases down to the soft mass of the flat direction, 
it starts to oscillate around the origin of its potential and topological defects disappear. 
These topological defects disappear well before the BBN epoch 
and they are free from constraints on long-lived topological defects.

The topological defects emit GWs with an emission peak wavenumber corresponding to the Hubble scale at each time 
until they disappear when the Hubble parameter decreases down to the soft mass of the flat direction. 
This means that 
the resulting GW spectrum has a peak wavenumber 
corresponding to the soft mass of the flat direction~\cite{previous work}. 
Therefore, we can obtain the information of SUSY scale through the detection of GW signals. 
In addition, 
since the GW spectrum bends at the wavenumber corresponding to the horizon scale at the end of reheating, 
we can obtain the value of reheating temperature through the observation of the bending frequency. 
Since the GW energy density depends on the energy density of topological defects, 
it is related to the VEV of the flat direction. 
This means that 
the peak energy density of GW spectrum 
and the shape of GW spectrum for high frequency modes 
have information of higher dimensional terms for the flat direction. 
In this paper, we have calculated the GW spectrum emitted from cosmic strings and domain walls 
and have shown that 
the resulting GW spectra actually have the information of SUSY scale, 
reheating temperature, and higher dimensional terms for the flat direction. 
We investigate three cases for higher dimensional potentials for the flat direction; 
the one coming from a nonrenormalizable superpotential, 
the one coming from a $U(1)$ symmetric \Kahler potential, 
and the one coming from $U(1)$ breaking \Kahler potentials. 
Especially in the first case, 
we have found that the resulting GW spectrum for high frequnicy modes 
has information of power-law index for the nonrenormalizable superpotential. 
We have also found that the resulting GW signals would be measured
by future GW observation experiments such as Advanced LIGO, ET, and Ultimate DECIGO.

Here let us comment on the relation between 
the mass of the flat direction and Q-ball formation~\cite{Coleman, 
Qsusy, KuSh, EnMc, KK}. 
Note that $\mphi$ is a function of $\la \phi \ra$ through the renormalization. 
If $\dd \mphi (\phi) / \dd \phi < 0$, 
the flat direction fragments into long-lived non-topological solitons called Q-balls 
soon after the flat direction starts to oscillate around $\phi = 0$ at $t \simeq t_{\rm decay}$. 
This is an obstacle in light of GW detection 
because 
it has been shown that 
GWs are diluted by the existence of Q-ball dominated era~\cite{Chiba:2009zu}. 
In this paper, therefore, 
we assume $\dd \mphi ( \phi ) / \dd \phi > 0$ implicitly. 
This is realized in a wide class of gravity mediation models~\cite{EnMc, Kamada:2014ada}. 
In particular, 
Q-balls never form in SUSY models with 
a hierarchical spectrum between 
gauginos and the other sparticles in the MSSM. 
In models of gauge mediation, 
since the VEV of the flat direction induces the effective mass of gauge fields, 
soft masses are effectively
suppressed as $\propto \phi^{-2}$ for 
$g \abs{\phi} \gtrsim M_s$, where $M_s$ is a messenger scale~\cite{de Gouvea:1997tn, Dvali:1997qv}. 
Thus, the flat direction need to contain a Higgs field 
to obtain a sizable value of $m_{\phi}$ and satisfy $\dd \mphi (\phi) / \dd \phi > 0$~\cite{Fujii:2001dn}. 
If GWs are observed in models of gauge mediation, 
we can identify $\mphi$ as the Higgsino mass parameter $\mu$.

Furthermore, let us comment on the baryon asymmetry 
(see footnote~\ref{footnote5}). 
The flat direction may be kicked by its $A$ term 
along its phase direction after the inflation. 
The phase of the flat direction is randomly distributed through the phase transition. 
Therefore, though the baryon density may be locally generated by the Affleck-Dine mechanism, 
net baryon asymmetry is absent. 
The baryon asymmetry is thus never generated through the dynamics of 
this flat direction. 
To generate the baryon asymmetry, 
we may require another flat direction which 
has a large VEV during inflation and 
realizes the Affleck-Dine baryogenesis. 
The leptogenesis is also viable 
if the gravitino problem can be avoided~\cite{Ibe:2004tg, Ibe:2004gh, Ibe:2011aa, Fukugita:1986hr}.

In the literature, it is known that 
cosmic strings emit Nambu-Goldstone modes efficiently, 
which leads to an effective friction on the dynamics of cosmic strings. 
As a result, cosmic string loops exist for a long time, 
which become more efficient source of GW emission than the cosmic string network 
in the ordinary cosmic strings~\cite{Damour:2001bk, Siemens:2006yp, Kawasaki:2011dp, Kuroyanagi:2012wm}. 
On the other hand, in the present case,
a typical width of cosmic strings is of the order of the Hubble length.
Since the length of cosmic string loops can not be lower than its width,
cosmic string loop formation is suppressed.
This is why the effect of cosmic string loops can be neglected in the present paper. 
Note that the effect of cosmic string loops is incorporated in 
our numerical simulations.

One might wonder if (tachyonic) preheating occurs for the flat direction just after the end of inflation 
and it affects the resulting GW signals~\cite{
Khlebnikov:1997di, Easther:2006gt, Easther:2006vd, GarciaBellido:2007dg, 
GarciaBellido:2007af, Dufaux:2007pt, Dufaux:2008dn}. 
In fact, our numerical calculations incoperate the effect of tachyonic preheating for the flat direction. 
However, let us emphasize that GWs in the observable band are not emitted from preheating 
which occurs when cosmic strings appear, 
but emitted from cosmic strings when they disappear. 
This is 
because the Hubble scale at the time of the preheating is much larger than the one at 
the time when topological defects disappear. 
Since GW signals with such high frequencies are beyond the detectability of future GW observations, 
we neglect GWs emitted from the preheating. 
We can also neglect the GW emission from the field $I$ 
because its typical frequency is far beyond the observable region. 
This is because the characteristic energy scale of that dynamics is the mass of $I$, 
which is of the order of inflation scale. 
The typical GW frequency emitted from that dynamics is 
usually much higher than the observable region of 0.1-1 Hz. 
Furthermore, we consider that the reheating temperature is of the order of or less than $10^{7-9}$ GeV. 
This implies that inflaton decays into radiation perturbatively via Planck suppressed interactions. 
In this case, GW emission is not efficient at the time of reheating.

%
\acknowledgments
This work is supported by the World Premier International Research Center Initiative (WPI Initiative), 
the Ministry of Education, Science, Sports, and Culture (MEXT), Japan (A. K. and M. Y.); 
the Program for Leading Graduate Schools, MEXT, Japan (M. Y.); 
and JSPS Research Fellowships for Young Scientists No.25.8715 (M. Y.).
%

\appendix

\section{\label{inflation}Inflation models}

In this appendix, we briefly review two well-known inflation models in SUSY theories 
to clarify that the field whose $F$ term drives inflation 
is generally different from the field whose oscillation energy dominates the 
energy density of the Universe after inflation. 
We use the Planck unit, that is, we set $\Mpl = 1$ in this appendix. 

\subsection{\label{chaotic inflation}Chaotic inflation model}

A simple chaotic inflation model in supergravity 
has been proposed in Ref.~\cite{Kawasaki:2000yn}, 
where they introduced a shift symmetry to overcome the so-called $\eta$ problem in supergravity. 
They introduced two superfields $X$ and $I$. 
The superfield $I$ has a shift symmetry, which forbids 
the Hubble induced mass term for the imaginary part of its scalar component. 
This ensures the flatness of the potential for the imaginary part, 
so that it can be identified as inflaton. 
In order to introduce a potential for the inflaton, 
a small shift-symmetry breaking term 
is introduced in superpotential like 
\beq
 W = m X I, 
\eeq
where $X$ is another superfield and $m$ is a parameter.

In this appendix, we assume the simplest \Kahler potential 
which satisfies the shift symmetry for the field $I$: 
\beq
 K = \half \lmk I + I^* \rmk^2 + XX^*. 
\eeq
This results in the following potential for the scalar components of the fields $X$ and $I$: 
\beq
 V (I, X) = m^2 e^K \lkk \abs{I}^2 ( 1+ \abs{X}^4) + \abs{X}^2 \lmk 1- \abs{I}^2 
 + ( I + I^*)^2 (1 + \abs{I}^2) \rmk \rkk. 
\eeq
Here, we can see that 
all scalar field except for the imaginary part of $I$ are stabilized around the origin 
due to the exponential factor. 
On the other hand, 
the imaginary part of $I$, that is, the inflaton can take a field value larger than the Planck scale. 
Thus, the inflaton potential 
is simply given by 
\beq
 V \simeq \half m^2 \chi^2, 
 \label{potential for inflaton in SUGRA}
\eeq
where $\chi \equiv \Im I / \sqrt{2}$ is the inflaton. 
In this way, a chaotic inflation model can be constructed in supergravity. 
Here, we should emphasize that 
the potential for inflaton Eq.~(\ref{potential for inflaton in SUGRA}) 
does not come from the $F$ term of the field $I$ 
but from that of the field $X$. 
This results in a very important consequence explained 
in Sec.~\ref{Hubble induced terms}. 
This is also the case in variants of the above chaotic inflation model, 
e.g., the ones proposed in Refs.~\cite{Kallosh:2010ug, Kallosh:2010xz}.

\subsection{\label{hybrid inflation}Hybrid inflation model}

In the simplest hybrid inflation model 
proposed in Ref.~\cite{Copeland:1994vg, Dvali:1994ms}, 
the superpotential is given by 
\beq
 W = \lambda S \bar{\Psi} \Psi - \mu^2 S, 
\eeq
where $S$, $\bar{\Psi}$, and $\Psi$ are superfields, 
and $\lambda$ and $\mu$ are positive parameters satisfying $\lambda \gg \mu$. 
The potential for their scalar components $V(S, \Psi, \bar{\Psi})$ is thus given by 
\beq
 V(S, \Psi, \bar{\Psi}) &=& e^{\abs{S}^2 + \abs{\Psi}^2 + \abs{\bar{\Psi}}^2} 
 \lkk (1-\abs{S}^2 + \abs{S}^4) \abs{\lambda \bar{\Psi} \Psi - \mu^2}^2 \right. \\
 &&\left. + \abs{S}^2 \left\{ \abs{\lambda(1+\abs{\Psi}^2) \bar{\Psi} - \mu^2 \Psi^*}^2 
 + \abs{\lambda (1+ \abs{\bar{\Psi}}^2) \Psi - \mu^2 \bar{\Psi}^*}^2 \right\} \rkk. 
\eeq
When all scalar fields are well below the Planck scale ($\bar{\Psi}, \Psi, S \ll 1$) 
but satisfy the relation of $S \gg 2 \mu / \sqrt{\lambda}$, 
$\bar{\Psi}$ and $\Psi$ are stabilized at the origin of the potential 
due to large effective masses. 
In this case, inflation is driven by a false vacuum energy density of $\mu^4$. 
Driven by 1-loop induced potential for the field $S$, 
it rolls down to the origin of the potential. 
After the field $S$ reaches the critical value of $S_{\rm cr} = 2 \mu / \sqrt{\lambda}$, 
the water-fall fields $\bar{\Psi}, \Psi$ starts to roll down to 
the potential minimum at $\la \bar{\Psi} \Psi \ra = \mu^2/\lambda$. 
Although this simplest model of hybrid inflation in supergravity 
might predict a spectral index inconsistent with the observation, 
many successful hybrid inflation models are considered in the literature 
based on this simplest model. 
Note that 
inflation is driven by the $F$ term of the superfield $S$, 
while the energy density of the Universe is dominated by 
that of oscillation energy of water-fall fields $\bar{\Psi}$ and $\Psi$ after the end of inflation. 
In the main part of this paper, 
the field $X$ is identified with the field $S$ 
while 
$I$ collectively represents water-fall fields 
$\bar{\Psi}$ and $\Psi$.

\section{\label{calculation}Method for the calculation of GWs. }
GWs are emitted by 
cosmic strings~\cite{Vilenkin:1981bx, Caldwell:1991jj, Vachaspati:1984gt, Olmez:2010bi, 
JonesSmith:2007ne, Fenu:2009qf, Figueroa:2012kw} 
and/or domain walls~\cite{Kawasaki:2011vv, Figueroa:2012kw, Hiramatsu:2013qaa} 
between the two phase transitions, 
that is, 
after the end of inflation and before the time of $t_{\rm decay}$. 
In this appendix, we explain how to calculate the spectrum of GWs. 
The calculations are based on Refs.~\cite{Dufaux:2007pt, Kawasaki:2011vv}, 
which is suitable to our situation compared with the method to calculate GW energy density from localized sources 
derived in Ref.~\cite{Weinberg}.

The Fourier transformed transverse-traceless part of the spatial 
metric perturbation $h_{ij}$ 
obeys the following linearized Einstein equation: 
\beq
 h_{ij}'' + 2 \frac{a'}{a} h_{ij}' + k^2 h_{ij} = 16 \pi G T^{TT}_{ij}, 
 \label{eom h}
\eeq
where the prime denotes a derivative with respect to conformal time $\tau$ 
defined by $\dd \tau = a^{-1} \dd t$, 
and 
$\TT$ is the Fourier transformed transverse-traceless part 
of the anisotropic stress. 
When 
the source term $T^{TT}_{ij}$ 
lasts during the interval of $[\tau_{\rm i}, \tau_{\rm f}]$, 
the solution of Eq.~(\ref{eom h}) for $\tau_{\rm f} \lesssim \tau \lesssim \tau_{\rm RH}$ is given by~\cite{Kawasaki:2011vv} 
\beq
 h\ij \tk 
 = 
 A\ij \k \frac{k \tau}{a} j_1 \lmk k \tau \rmk 
 + 
 B\ij \k \frac{k \tau}{a} n_1 \lmk k \tau \rmk,
 \label{h solution}
\eeq
where $j_1$ and $n_1$ are the spherical Bessel and Neumann functions of order one. 
The coefficients $A_{ij}$ and $B\ij$ are given by 
\beq
 A\ij \k &=& - 16 \pi G \int_{\tau_{\rm i}}^{\tau_{\rm f}} \dd \tau' \  \tau' a(\tau') n_1(k \tau') \TT \lmk \tau', {\bf k} \rmk, \label{A}\\
 B\ij \k &=& 16 \pi G \int_{\tau_{\rm i}}^{\tau_{\rm f}} \dd \tau' \  \tau' a(\tau') j_1(k \tau' ) \TT \lmk \tau', {\bf k} \rmk. \label{B}
\eeq
After reheating completes ($\tau \gtrsim \tau_{\rm RH}$), 
the solution is described by the spherical Bessel functions of order zero: 
\beq
 h\ij \tk 
 = 
 A\ij' \k \frac{k \tau}{a} j_0 \lmk k \tau \rmk 
 + 
 B\ij' \k \frac{k \tau}{a} n_0 \lmk k \tau \rmk. 
 \label{h solution2}
\eeq
We match the solution of Eq.~(\ref{h solution}) 
with that of Eq.~(\ref{h solution2}) 
at $\tau = \tau_{\rm RH}$. 
We then obtain $A'$ and $B'$ such as 
\beq
 A\ij' \k &=& 
 -16 \pi G \int_{\tau_{\rm i}}^{\tau_{\rm f}} \dd \tau' \  \tau' a(\tau') 
 f_A ( k \tau') 
 \TT \lmk \tau', {\bf k} \rmk, \label{A}\\
 B\ij' \k &=& 
 16 \pi G \int_{\tau_{\rm i}}^{\tau_{\rm f}} \dd \tau' \  \tau' a(\tau') 
 f_B ( k \tau') 
  \TT \lmk \tau', {\bf k} \rmk, \label{B}
\eeq
with 
\beq
f_A (k \tau') &=& 
 \lkk a_1 n_1(k \tau') - a_2 j_1(k \tau') \rkk, \nonumber\\
f_B (k \tau') &=& 
 \lkk -b_1 n_1(k \tau') + b_2 j_1(k \tau') \rkk, 
 \label{functions}
\eeq
where the coefficients are given by 
\beq
 a_1 &=& \left. x^2 \lkk j_1 (x) \del_x n_0 (x) - n_0 (x) \del_x j_1 (x)  \rkk \right. \vert_{x \to k \tau_{\rm RH}}, 
 \nonumber\\ 
 a_2 &=& \left. x^2 \lkk n_1 (x) \del_x n_0 (x) - n_0 (x) \del_x n_1 (x)  \rkk \right. \vert_{x \to k \tau_{\rm RH}}, 
\nonumber\\
 b_1 &=& - \left. x^2 \lkk j_1 (x) \del_x j_0 (x) - j_0 (x) \del_x j_1 (x)  \rkk \right. \vert_{x \to k \tau_{\rm RH}}, 
\nonumber\\
 b_2 &=& - \left. x^2 \lkk n_1 (x) \del_x j_0 (x) - j_0 (x) \del_x n_1 (x)  \rkk \right. \vert_{x \to k \tau_{\rm RH}}. 
\label{coefficients}
\eeq

The energy density of GWs can be calculated from~\cite{Maggiore:1999vm} 
\beq
 \rho_{\rm gw} = \frac{1}{32 \pi G} \la \dot{h}\ij \lmk t, {\bf x} \rmk \dot{h}\ij \lmk t, {\bf x} \rmk \ra,
 \label{rhogw}
\eeq
where $\la \cdots \ra$ represents an ensemble average for a stochastic background. 
We replace the ensemble average by an average over a volume $V$ of the comoving box. 
This is a good approximation for sub-horizon modes. 
Using the following convention for the Fourier expansion:
\beq
 h\ij ( {\bf x}) = \int \frac{\dd^3 {\bf k}}{(2 \pi)^{3}} h\ij ( {\bf k}) e^{-i {\bf k \cdot x}}, 
\eeq
we rewrite the energy density of GWs as 
\beq
 \rho_{\rm gw} = \frac{1}{32 \pi G a^4} \frac{1}{V} 
 \int \frac{\dd^3 {\bf k}}{(2 \pi)^{3}}
 \frac{k^2}{2} \sum_{i,j} \lmk \abs{A'\ij}^2 + \abs{B'\ij}^2 \rmk. 
 \label{rhogw2}
\eeq
Here we have used $h_{ij}' \simeq k h_{ij}$ 
and taken average over the oscillation period 
since we can observe only sub-horizon modes $k \ll \tau$. 
Also we have averaged over time in Eq.~(\ref{rhogw2}) 
because we are not interested in the resolution of the oscillation of $h\ij (\tau)$ with time. 
We define the GW spectrum as 
\beq
\Omega_{\rm gw} (\tau) 
 &\equiv& 
 \frac{1}{\rho_{\rm tot} \lmk \tau \rmk} \frac{ \dd \rho_{\rm gw} ( \tau) }{\dd \log k}, 
\eeq
where $\rho_{\rm tot} (\tau )$ ($= 3 \M^2 H^2 (\tau)$) is the total energy density of the Universe. 
We can calculate the GW spectrum such as~\cite{Dufaux:2007pt} 
\beq
\Omega_{\rm gw} (\tau) 
 \simeq 
 \frac{k^5}{192 \pi^3 V a^4 H^2} 
 \int \dd \Omega_k
 \sum_{i,j} \lmk \abs{A'\ij }^2 + \abs{B'\ij }^2 \rmk. 
 \label{omega_gw}
\eeq
We can calculate $\TT$ from lattice simulation for the classical dynamics of a scalar field 
and then calculate the value of $A'$ and $B'$ through Eqs.~(\ref{A}) and (\ref{B}). 
Finally, we can obtain the GW spectrum from Eq.~(\ref{omega_gw}).

\section{\label{flavor}Case with flavor symmetry}

In this appendix, we investigate the components of the flat direction in detail. 
In particular, we consider the case with flavor symmetry, 
where topological defects 
may not form when the flat direction obtains a large VEV and phase transition occurs. 
However, 
Nambu-Goldstone (NG) modes are randomly distributed over the Hubble scale through the phase transition, 
so that GWs are emitted from their relaxation dynamics within the Hubble horizon. 
The resulting GW spectrum is qualitatively the same as the ones obtained in the 
main part of this paper. 
This is a realization of so-called self-ordering scalar fields 
in SUSY theories~\cite{Turok:1991qq, Krauss:1991qu, JonesSmith:2007ne, Fenu:2009qf}. 

In the next subsection, we investigate flavor components of the flat direction in detail 
and explain what happens in the case with flavor symmetry. 
Then we briefly review a method suitable to the calculation of GW spectra in this case, 
which has been derived in the literature in the context of self-ordering scalar fields. 
Using this method, we calculate the GW spectrum 
and show the results in Sec.~\ref{results4}.

\subsection{\label{flavor symmetry}Flavor symmetry}

In order to investigate flavor components of flat directions, 
let us consider the $L H_u$ flat direction as an illustration. 
Since the left-handed lepton doublet $L$ contains three flavors, 
the $L H_u$ flat direction has an ambiguity in terms of the flavor direction, 
that is, any flavor directions are $D$-flat directions. 
This implies that the $L H_u$ flat direction 
can be written as 
\beq
\lmk
\begin{array}{llll}
(L)_1^1\\
(L)_2^1\\
(L)_3^1\\
(H_u)^2
\end{array}\rmk
= \frac{1}{\sqrt{2}} \lmk
\begin{array}{llll}
 a_1 &   &  \\
 a_2 \ \ & \cdot \ \ & \cdot \ \ & \cdot \ \ \\
 a_3 &  &   \\
 1 &  & 
\end{array}\rmk 
\lmk
\begin{array}{lll}
\phi \\
\, \cdot \\
\, \cdot \\
\, \cdot
\end{array}\rmk, 
\label{decomposition2}
\eeq 
where the superscripts are SU(2) indices 
and the lower indices are flavor indices. 
Here, $a_i$ satisfy
\beq
 \sum_i \abs{a_i}^2 = 1. 
 \label{a_i}
\eeq
Any combinations of $a_i$ satisfying Eq.~(\ref{a_i}) are D-flat directions. 
This implies that the $L H_u$ flat direction 
has $U(3)$ symmetry 
if we take into account only $D$-flat condition. 
However, Hubble induced terms and higher dimensional superpotentials 
may break that symmetry as we can see in the following discussion.

Next, let us consider Hubble induced mass for the $L H_u$ flat direction.
If we assign R-charges as, for example, $R(Q) = R(L) = R(e^c) = R(u^c) = R(d^c) = 3$ and $R(H_u) = R(H_d) = -4$,%
\footnote{
See footnote~\ref{R-symmetry breaking} for discussion about effects of R-symmetry breaking. 
}
nonrenormalizable K\"{a}helr potentials take the following form: 
\beq
 K =&& \sum_i \abs{(L)_i}^2 + \abs{H_u}^2 + \sum_i  \abs{X}^2 \lmk 1 + \sum_i \frac{c^{ij}_{X}}{M_*^2} (L)_i^* (L)_j 
 + \frac{c^{H_u}_{X}}{M_*^2} \abs{H_u}^2 \rmk \nonumber \\
 &&+ K_{\text{min}} (I) \lmk1 + \sum_i \frac{c^{i j}_I}{M_*^2} (L)_i^* (L)_j  + \frac{c^{H_u}_{I}}{M_*^2} \abs{H_u}^2  \rmk, 
\eeq
where 
$c^{ij}_a$ ($a=X, I$) are Hermitian constant matrices and $c^{H_u}_a$ ($a=X, I$) are real constants.
As explained in Sec.~\ref{Hubble induced terms} and Appendix~\ref{inflation}, 
the minimal \Kahler potential for the field $I$ is given by 
$K_{\text{min}} (I) = (I + I^*)^2/2$ for the case of the chaotic inflation model proposed in Ref.~\cite{Kawasaki:2000yn} 
and 
$K_{\text{min}} (I) = \abs{I}^2$ for the case of hybrid inflation models. 
The $F$ term of $X$ drives inflation 
and satisfies the relation of $\abs{\del W / \del X}^2 \simeq 3 H_{\rm inf}^2 \M^2$. 
The above \Kahler potential 
induces Hubble induced mass terms for fields $\psi_i$ such as 
\beq
 V_{H_{\rm inf}} &=& \sum_i c_{H_{\rm inf}}^{ij} H^2 (L)_i^* (L)_j + c_{H_{\rm inf}}^{H_u} H^2 \abs{H_u}^2, \\
 \label{c_H during inflation 2}
 c_{H_{\rm inf}}^{ij} &=& 3 \left( 1 - c_X^{ij} \frac{\M^2}{M_*^2} \right), \\
 c_{H_{\rm inf}}^{H_u} &=& 3 \left( 1 - c_X^{H_u} \frac{\M^2}{M_*^2} \right), 
\eeq
during inflation. 
After inflation ends, 
the energy density of the Universe is dominated by the oscillation of the field $I$, 
so that Hubble induced mass terms are given by 
\beq
 V_{H_{\rm osc}} &=& \sum_i c_{H_{\rm osc}}^{ij} H^2 (L)_i^* (L)_j + c_{H_{\rm osc}}^{H_u} H^2 \abs{H_u}^2, \\
 \label{c_H 2}
 c_{H_{\rm osc}}^{ij} &=& \frac{3}{2} \lmk 1 - (c_X^{ij}+c_I^{ij}) \frac{\M^2}{M_*^2} \rmk, \\
 c_{H_{\rm osc}}^{H_u} &=& \frac{3}{2} \lmk 1 - (c_X^{H_u}+c_I^{H_u}) \frac{\M^2}{M_*^2} \rmk. 
\eeq
The coefficients $c_a$'s depend on underlying high energy models beyond the cutoff scale $M_*$.
Without any symmetry, we expect that $c_{a}$'s are not related with each other.
This means that Hubble induced mass terms usually break the flavor symmetry, 
which is assumed in the main part of this paper (see footnote~\ref{flavor indices}). 
However, one may assume flavor-universal coefficients such as $c^{ij}_a = c^L_a \delta^{ij}$,
where 
$\delta_{ij}$ is Kronecker's delta,
just like the case in the constrained MSSM, for instance. 
In this case, 
the $L H_u$ flat direction has $U(3)$ symmetry.

Nonrenomalizable terms should stabilize the flat direction as we discussed in Sec~\ref{higher dimensional terms}.
It can take a form of 
\beq
 V_{\rm NR} = a^{{i_1}{j_1}\dots{i_{m-n}}{j_{m-n}}}_{H} H^2 (L)_{i_1}^* (L)_{j_1}\dots(L)_{i_{m-n}}^* (L)_{j_{m-n}} \abs{H_u}^{2n-2} / \M^{2m-4},
 \eeq 
where the coefficient $a_H$ is of the order of $(\M/M_*)^{2m-2}$ and may also break flavor symmetry in general. 
If we consider, for example, the case of $a^{{i_1}{j_1}\dots{i_{m-n}}{j_{m-n}}}_{H} = a_H \delta^{{i_1}{j_1}}\dots\delta^{i_{m-n}j_{m-n}}$,%
\footnote{
This assumption is inconsistent with the observations of neutrino oscillations~\cite{Agashe:2014kda}. 
In this appendix, however, we take the $L H_u$ flat direction as an example to take advantage of 
its simple flavor structure. 
} 
the superpotential respects $U(3)$.

Suppose that the Hubble induced mass term is flavor-universal 
and the higher dimensional superpotential has the $U(3)$ symmetry. 
As we can see from Eqs.~(\ref{c_H during inflation 2}) and (\ref{c_H 2}), 
the coefficient of the Hubble induced mass term during inflation, 
$c_{H_{\rm inf}}$, is generally different from the one 
during the oscillation dominated era, $c_{H_{\rm osc}}$~\cite{previous work}. 
We consider the case that 
the coefficient of the Hubble induced mass term is positive during inflation 
and is negative after inflation.
In this case, 
the flat direction stays at the origin during inflation 
and has a large VEV after inflation, 
that is, the phase transition occur just after inflation. 
At this phase transition, 
the $U(3)$ symmetry is broken to 
$U(2)$ symmetry.
In this case, topological defects do not form at this phase transition. 
However, there is another source of GWs.
Since NG models have no correlation over distances larger than the Hubble radius due to the causality, 
they are 
randomly distributed over the Hubble scale. 
Therefore, 
the NG modes obtain gradient energy.
The gradient energy turns into the kinetic energy and decreases inside the Hubble radius.
As a result, the modes with wavenumber around the Hubble scale ($k \sim a H(t)$) 
dominantly emits GWs~\cite{Turok:1991qq, Krauss:1991qu, JonesSmith:2007ne, Fenu:2009qf}. 
Since the NG modes outside the Hubble radius grow through the nonlinear couplings,
the peak frequency of the resultant GW signals correspond to the Hubble scale 
when the flat direction start to oscillate around the origin at $t \simeq t_{\rm decay}$ (Eq.~(\ref{t_decay})).
This is another source of GWs which gives a contribution to the GW signal. 
Note that 
it is possible that, say, there is $O(N) \times U(1)$ symmetry and is broken down to $O(N-1)$ symmetry. 
In this case, topological defects also form 
and the resulting GW signal is given by the sum of both contributions. 
In the following subsections, we calculate 
the GW energy density and spectrum from the above mechanism.

Hereafter we focus only on the global symmetry of the flat direction for the following reasons.
The gauge fields may be randomly distributed at the phase transition.
We can take the unitary gauge if the topological defects are not formed.
Since, the broken gauge field obtains the effective mass of $g \la \phi \ra$, 
the field configuration begins to oscillate and 
becomes trivial ($A^\mu = 0$) soon after the phase transition. 
For the unbroken gauge symmetry, the corresponding gauge field remains massless. 
Here we should note that while long wavelength modes of such gauge fields remain constant until 
they enter the horizon, the NG modes 
grow through the nonlinear term~\cite{Fenu:2009qf}.
Therefore, GW emission from the local symmetry can be neglected.

\subsection{Analytical calculation of GW spectra for self-ordering scalar fields}

Here we explain 
a method suitable to the calculation of GW spectra emitted from 
self-ordering scalar fields. 
We consider the case with $O(N)$ symmetry for simplicity.
Note that $L H_u$ flat direction considered in the previous section has $O(6)$ symmetry rather than $U(3)$ symmetry 
if gauge and Yukawa couplings are ignored.
This is similar to $O(4)$ custodial symmetry in the Standard Model.
We expect that the results do not change qualitatively for the case with other symmetries 
because the discussion in the previous subsection hardly depends on the detail of the manifold 
where NG bosons reside. 

Instead of replacing the ensemble average by an average over a volume of the comoving box 
in Eq.~(\ref{rhogw}), 
here we just rewrite it by assuming 
\beq
 \la \TT (\tau_1,  {\bf k}) T_{ij}^{{\rm TT}*} ( \tau_2 ,  {\bf k}' ) \ra \equiv 
 (2 \pi)^3 \delta^{3}(  {\bf k} -  {\bf k}' ) \Pi ( k, \tau_1, \tau_2), 
\eeq
and then obtain 
the energy density spectrum of GWs in the radiation dominated era after reheating as 
\beq
\Omega_{\rm gw} (\tau) 
 &\simeq& 
 \frac{2 \pi k^5}{3 (2 \pi)^{3} a^4 H^2 \M^4} 
 \int_{\tau_{\rm i}}^{\tau_{\rm f}} \dd \tau_1 \  \tau_1 a(\tau_1) 
 \int_{\tau_{\rm i}}^{\tau_{\rm f}} \dd \tau_2 \  \tau_2 a(\tau_2) \nonumber\\
 &\times&
\lkk f_A (k \tau_1) 
 f_A (k \tau_2) 
+ f_B (k \tau_1) 
 f_B (k \tau_2) \rkk 
 \Pi \lmk k, \tau_1, \tau_2 \rmk. 
 \label{omega_gw9}
\eeq

To calculate the spectrum of GWs, 
we quote the result of $\Pi (k, \tau_1, \tau_2)$ 
calculated in the literature~\cite{Turok:1991qq, JonesSmith:2007ne, Fenu:2009qf}, 
where they have considered a global $O(N)$ symmetric scalar field 
with a quartic potential. 
While we consider higher power potentials for flat directions, 
their result is applicable to the present case. 
This is because the potential for scalar fields is irrelevant to NG bosons 
after the scalar fields obtains VEV. 
Within the large $N$ approximation, 
they have obtained 
\beq
 \Pi ( k, \tau_1, \tau_2) 
 &=& 
 \int \frac{\dd^3 q}{(2 \pi)^3} q^4 \lkk 1- ( \hat{\bf k} \cdot \hat{\bf q})^2 \rkk^2 \nonumber\\
 &\times& 
 \mathcal{P}^{ab} (q, \tau_1, \tau_2) \mathcal{P}^{ab} ( \abs{\bf{k}-\bf{q}}, \tau_1, \tau_2 ), 
 \label{Pi_k}\\
 \mathcal{P}^{ab} (k, \tau_1, \tau_2) 
 &=& 
 \frac{2835}{16} \pi^3 (\tau_1\tau_2)^{3/2} 
 \frac{J_3 (k \tau_1) J_3 ( k \tau_2)}{(k \tau_1)^3 (k \tau_2)^3} \frac{\delta_{ab}}{N} \la \phi \ra^2, 
 \nonumber\\
 \label{power}
\eeq
where $J_3 (x)$ is the Bessel function of order 3.

\subsection{\label{results4}Results for self-ordering scalar fields}

When the Hubble parameter decreases down to the mass of the flat direction $m_\phi$, 
the flat direction starts to oscillate around the origin of the potential.
After that, 
the field value of the flat direction decreases with time due to the expansion of the Universe 
such as $\phi  \propto a^{-3/2}$. 
After the phase transition at the end of inflation 
and until they start to oscillate around the origin at $t \simeq t_{\rm decay}$, 
the dynamics of NG modes emit GWs continuously. 
We obtain the GW spectrum by evaluating Eqs.~(\ref{omega_gw9}), (\ref{Pi_k}), and (\ref{power}) numerically.

We obtain GW peak energy density and peak wavenumber at $t \simeq t_{\rm decay}$ such as 
\beq
 \Omega_{\rm gw} h^2 (\tau_{\rm decay}) &\simeq& 0.14 N^{-1} \lmk \frac{\la \phi \ra}{\M} \rmk^4, \label{omega_gw10}\\
 \frac{k_{\rm peak}}{a(\tau_{\rm decay})} &\simeq& 2.3 \frac{\mphi}{\sqrt{\abs{c_{H_{\rm osc}}}}}. 
\eeq
Note that we assume $M_* / \Mpl \ll 1$ so that 
we can neglect the backreaction of the flat direction dynamics on the cosmic expansion.
Substituting the above formulas into Eqs.~(\ref{present energy density}) and (\ref{present peak frequency}), 
we obtain the present value of the GW peak energy density such as 
\beq
 \Omega_{\rm gw} h^2 (t_0) 
 \simeq 
 3 \times 10^{-8} N^{-1} 
 \lmk \frac{\abs{c_{H_{\rm osc}}}^{-1/2} \mphi }{10^3 \GeV} \rmk^{-2/3} \lmk \frac{\TR}{10^9 \GeV} \rmk^{4/3} 
 \lmk \frac{\la \phi \ra}{\M} \rmk^{4}, 
\eeq
and 
the present value of peak frequency $f_0$ such as
\beq
 f_0 
\simeq 
550 \text{ Hz} \lmk \frac{\abs{c_{H_{\rm osc}}}^{-1/2} \mphi}{10^3 \GeV} \rmk^{1/3} \lmk \frac{\TR}{10^9 \GeV} \rmk^{1/3}.
\eeq

Since a typical length scale of the dynamics of NG modes 
is the Hubble length, 
the emission peak wavenumber of GWs is estimated as 
the Hubble scale.
In addition, 
the GW energy density of Eq.~(\ref{omega_gw9}) is proportional to 
$\la \phi \ra^4$. 
These observations imply that 
the resulting GW spectrum emitted from the dynamics of NG modes 
is qualitatively the same as the ones obtained in Secs.~\ref{W ne 0}, \ref{W=0}, and \ref{W=0 without U(1)}, 
depending on higher dimensional potentials for the flat direction. 
In particular, since the GW spectrum is sensitive to what dominates the energy density of the Universe, 
it has information of reheating temperature 
as explained in those sections. 
In fact, the GW spectrum bends at the frequency 
given by Eq.~(\ref{f_bend}) corresponding to the wavenumber 
around $k \simeq \tau_{\rm RH}^{-1}$~\cite{Seto:2003kc, 
Nakayama:2008ip, previous work}.

\begin{figure}[t]
\centering 
\includegraphics[width=.45\textwidth, bb=0 0 360 339]{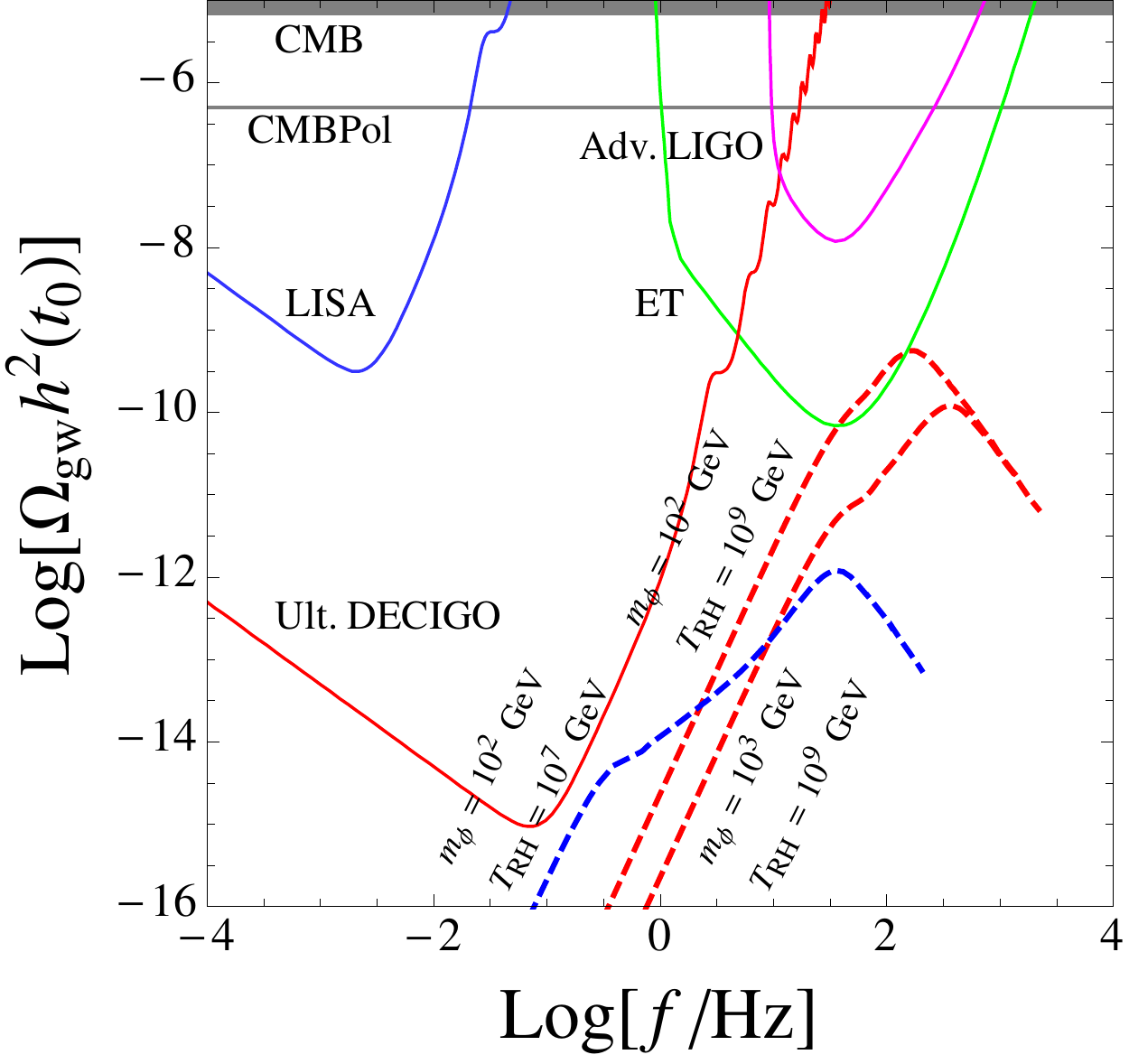} 
\caption{
GW spectra generated by the dynamics of NG modes in the case with $O(6)$ flavor symmetry 
(dash and dot-dashed curves) 
and sensitivities of planned interferometric detectors (solid curves). 
We assume 
$\mphi = 10^2 \GeV$ 
with $\TR = 10^7 \GeV$ (blue dashed curve), 
$\mphi = 10^2 \GeV$ 
with $\TR = 10^9 \GeV$ (red dashed curve), 
and 
$\mphi = 10^3 \GeV$ 
with $\TR = 10^9 \GeV$ (red dot-dashed curve). 
We also assume $c_{H_{\rm osc}}= -15$ and $\la \phi \ra^2 = \Mpl^2 / 10$. 
}
  \label{detectability4}
\end{figure}

In summary, the schematic form of resulting GW spectrum is 
similar to the one described in the main part of this paper, 
even if there is a large flavor symmetry such as $O(N)$.
Figure~\ref{detectability4} shows examples of GW spectrum 
generated by the self-ordering scalar fields. 
We assume that $\la \phi \ra = M_*$ is constant until $t = t_{\rm decay}$ as in Secs.~\ref{W=0}. 
The GW spectra are 
calculated 
from Eqs.~(\ref{omega_gw9}), (\ref{Pi_k}), and (\ref{power}). 
In the figure, we take $\mphi = 10^2 \GeV$ with $\TR = 10^7 \GeV$ (blue dashed curve), 
$\mphi = 10^2 \GeV$ with $\TR = 10^9 \GeV$ (red dashed curve), and $\mphi = 10^3 \GeV$ with $\TR = 10^9 \GeV$ 
(red dot-dashed curve).
To discuss the detectability of GW signals, 
we plot sensitivities of planned interferometric detectors. 
The sensitivity curves are the same as in Fig.~\ref{detectability1},
and details are given in Sec.\,\ref{prediction1}.
Detecting these GW signals seems somewhat challenging in the current design of Advanced LIGO, ET, and Ultimate DECIGO
unlike those from topological defects in Secs.~\ref{W ne 0}, \ref{W=0}, and \ref{W=0 without U(1)}.
However, let us stress that if we can detect these GW signals from somehow, for example, through longer observation time $T \gg 1$~yr,
they provide us invaluable information of $\abs{c_{H_{\rm osc}}}^{-1/2} \mphi$, $\TR$, and $M_*/\Mpl$ 
(up to by a factor of $\mathcal{O}(1)$).

\end{document}